\documentclass[aps,prx,preprint,onecolumn,citeautoscript,footinbib,superscriptaddress,
eqsecnum]{revtex4-1}  
\usepackage{amsmath,amssymb,bm,bbm} 
\usepackage{graphicx}  
\usepackage{color} 
\usepackage[dvipsnames]{xcolor}
\usepackage[papersize={8.5in,11in}]{geometry}
\usepackage[colorlinks=true]{hyperref}  
\usepackage{ulem}
\usepackage[mathscr]{euscript}
\usepackage[section]{placeins}
\usepackage{comment}
\hypersetup{
    bookmarks=true,         % show bookmarks bar?
    unicode=false,          % non-Latin characters 
    pdftoolbar=true,        % show Acrobat
    pdfmenubar=true,        % show Acrobat 
    pdffitwindow=false,     % window fit to page when opened
    pdfstartview={FitH},    % fits the width of the page to the window
    pdfsubject={},   % subject of the document
    pdfcreator={},   % creator of the document
    pdfproducer={}, % producer of the document
    pdfkeywords={} {} {}, % list of keywords
    pdfnewwindow=true,      % links in new window
    colorlinks=true,       % false: boxed links; true: colored links
    linkcolor=magenta, %red,          % color of internal links (change box color with linkbordercolor)
    citecolor=blue,        % color of links to bibliography
    filecolor=magenta,      % color of file links
    urlcolor=blue           % color of external links
} 

\usepackage{amsfonts}
\usepackage{upgreek}
\usepackage{slashed}
\usepackage{latexsym}

\newcommand{\beq}{\begin{equation}}
\newcommand{\eeq}{\end{equation}}
\def\bea{\begin{eqnarray}}
\def\eea{\end{eqnarray}}
\newcommand{\nn}{\nonumber \\}

\usepackage{float}
\usepackage[caption = false]{subfig}

\newcommand{\lam}{\lambda} % Constraint
\newcommand{\kin}{s_0} % f-elec energy
\newcommand{\kinr}{s} % renormalized f-elec energy
\newcommand{\gc}{g_0} % bare fermionic bath coupling
\newcommand{\gcr}{g} % renormalized fermionic bath coupling
\newcommand{\gam}{\gamma_0} % bare Bosonic bath coupling
\newcommand{\gamr}{\gamma} % renormalized Bosonic bath coupling

\newcommand{\fd}{f^{\dagger}} % f creation
\newcommand{\fa}{f} % f annihilation
\newcommand{\bd}{b^{\dagger}} % b creation
\newcommand{\ba}{b} % b annihilation
\newcommand{\ca}{\psi_{k \alpha}} % c annihilation
\newcommand{\pa}{\phi_{a}} % \phi
\newcommand{\charg}{q} % Charge
\newcommand{\No}{\aleph} %Constant: \gamma_{E} -2 \log (2)-\psi ^{(0)}\left(\frac{3}{2}\right) 
\newcommand{\LS}{\Lambda_{S}}  
\newcommand{\LP}{\Lambda_{c}}

\newcommand{\up}{^}

\newcommand{\betg}{\beta (g)} % beta function of g 
\newcommand{\betgam}{\beta (\gamma)} % beta function of \gamma 
\newcommand{\bets}{\beta (s)} % beta function of s

\newcommand{\rb}{\bar{r}}
\newcommand{\ep}{\epsilon}

\newcommand{\zf}{Z_{f}}
\newcommand{\zb}{Z_{b}}
\newcommand{\zg}{Z_{g}}
\newcommand{\zp}{Z_{\phi}}
\newcommand{\zgam}{Z_{\gamma}}
\newcommand{\rgs}{\mu}

\newcommand{\sdd}{\frac{S_{d}}{2\widetilde{S}_{d+1}}}

\newcommand{\iw}{i\omega}
\newcommand{\inu}{i\nu}

\newcommand{\lo}{L_{0}}
\newcommand{\lop}{L'_{0}}
\newcommand{\lopp}{L''_{0}}

\newcommand{\Lgam}{L_{\gamma}}
\newcommand{\Lg}{L_{g}}

\newcommand{\la}{L_{1}}
\newcommand{\lap}{L'_{1}}
\newcommand{\lapp}{L''_{1}}

\newcommand{\lb}{L_{2}}
\newcommand{\lbp}{L'_{2}}
\newcommand{\lbpp}{L''_{2}}

\newcommand{\lc}{L_{3}}
\newcommand{\lcp}{L'_{3}}
\newcommand{\lcpp}{L''_{3}}

\newcommand{\Da}{D_{1\phi}}
\newcommand{\Db}{D_{2\phi}}
\newcommand{\Dc}{D_{3\phi}}

\newcommand{\Dap}{D'_{1\psi}}
\newcommand{\Dbp}{D'_{2\psi}}
\newcommand{\Dcp}{D'_{3\psi}}

\newcommand{\Dapp}{D''_{1\psi}}
\newcommand{\Dbpp}{D''_{2\psi}}
\newcommand{\Dcpp}{D''_{3\psi}}

\newcommand{\po}{P_{0}}
\newcommand{\Pgam}{P_{\gamma}}
\newcommand{\Pg}{P_{g}}

\newcommand{\Pa}{P_{1}}
\newcommand{\pap}{P'_{1}}
\newcommand{\papp}{P''_{1}}

\newcommand{\pb}{P_{2}}
\newcommand{\pbp}{P'_{2}}
\newcommand{\pbpp}{P''_{2}}

\newcommand{\pc}{P_{3}}
\newcommand{\pcp}{P'_{3}}
\newcommand{\pcpp}{P''_{3}}

\usepackage{braket}
\usepackage{comment}
\usepackage{slashed}

\renewcommand{\vec}[1]{\boldsymbol{#1}}

\begin{document}

\preprint{\href{https://arxiv.org/abs/1912.08822}{arXiv:1912.08822}}

\title{Deconfined critical point in a doped random\\ quantum Heisenberg magnet}

\author{Darshan G. Joshi}
\affiliation{Department of Physics, Harvard University, Cambridge MA 02138, USA}

\author{Chenyuan Li}
\affiliation{Department of Physics, Harvard University, Cambridge MA 02138, USA}

\author{Grigory Tarnopolsky}
\affiliation{Department of Physics, Harvard University, Cambridge MA 02138, USA}

\author{Antoine Georges}
\affiliation{
Center for Computational Quantum Physics, Flatiron Institute, New York, NY 10010 USA
}
\affiliation{Coll{\`e}ge de France, 11 place Marcelin Berthelot, 75005 Paris, France}
\affiliation{CPHT, CNRS, {\'E}cole Polytechnique, IP Paris, F-91128 Palaiseau, France}
\affiliation{DQMP, Universit{\'e} de Gen{\`e}ve, 24 quai Ernest Ansermet, CH-1211 Gen{\`e}ve, Suisse}

\author{Subir Sachdev}
\affiliation{Department of Physics, Harvard University, Cambridge MA 02138, USA}

\date{\today
\\
\vspace{0.4in}}

\begin{abstract}
We describe the phase diagram of electrons on 
%a cluster of $N \rightarrow \infty$ sites, 
a fully connected lattice with random hopping, 
subject to a random Heisenberg spin exchange interactions between any pair of sites 
and a constraint of no double occupancy. 
%Double electron occupancy is prohibited on all sites. 
A perturbative renormalization group analysis yields a critical point with fractionalized excitations at a non-zero critical value $p_c$ of the hole doping $p$ 
away from the half-filled insulator. We compute the renormalization group to two loops, but some exponents are obtained to all loop order. We argue that the critical point $p_c$ is flanked by confining phases: a disordered Fermi liquid with carrier density $1+p$ for $p>p_c$, %at higher doping, 
and a metallic spin glass with carrier density $p$ for $p<p_c$. %at lower doping. 
Additional evidence for the critical behavior is obtained from a large $M$ analysis of a model which extends the SU(2) spin symmetry to SU($M$).
We discuss the relationship of the vicinity of this deconfined quantum critical point to key aspects of cuprate phenomenology. 
\end{abstract}

\maketitle
\tableofcontents

\section{Introduction}
\label{sec:intro}

Much evidence has now accumulated for a fundamental transformation in the ground state of the cuprate superconductors near optimal hole doping $p=p_c$ \cite{CPLT18,Vishik2012,He14,Fujita14,Badoux16,Loram01,Loram19,Michon18,Bourges18,Shen19,CPana1,CPana2,Julien19}. The transformation is primarily associated with a change in the mobile carrier density from to $p$ to $1+p$ as $p$ increases across $p_c$. There are indications of various broken symmetries for $p<p_c$, including charge and bond density waves \cite{He14,Fujita14}, spin glass order \cite{CPana1,CPana2,Julien19}, orbital currents \cite{Bourges18} and nematic order \cite{Fujita14}. However, it appears that the restoration of the broken symmetry cannot be the driving mechanism for a quantum phase transition at $p=p_c$: the broken symmetries are weak and differ among the cuprates, and the transport \cite{Badoux16}, thermodynamic \cite{Loram01,Loram19,Michon18}, electronic \cite{Vishik2012,He14,Fujita14,Shen19}, and spin dynamics \cite{CPana1,CPana2,Julien19} signatures are strong. 

\begin{figure}[tb]
\begin{center}
\includegraphics[height=3.75in]{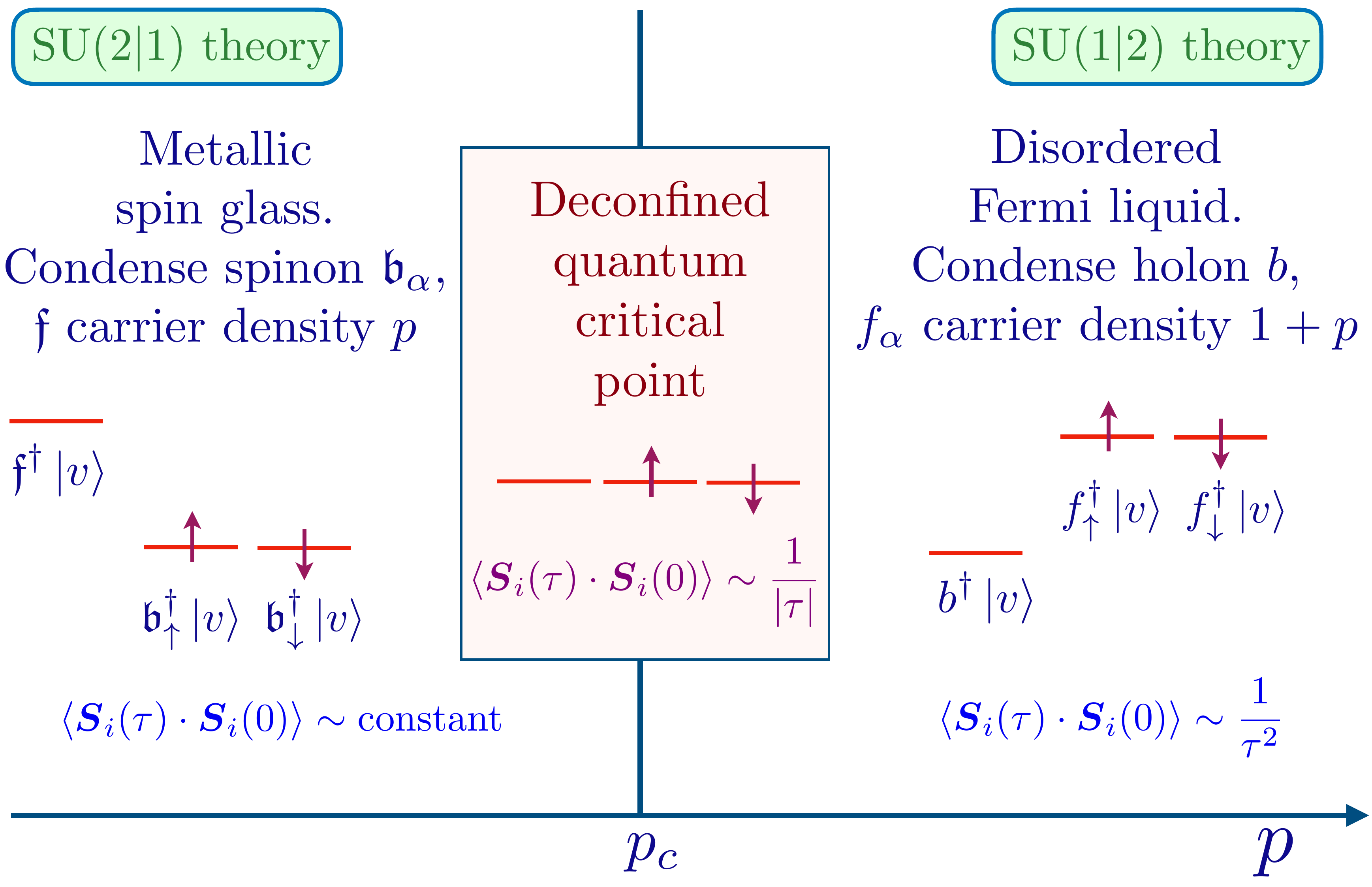} 
\end{center}
\caption{Proposed phase diagram of $H_{tJ}$ in (\ref{HtJ}) as function of hole doping $p$. The three states on each site are realized either by fermionic spinons ($f_\alpha$) and bosonic holons ($b$) with operators as in (\ref{defcS1}), or
by bosonic spinons ($\mathfrak{b}_\alpha$) and fermionic holons ($\mathfrak{f}$) with operators as in (\ref{defcS2}); $\left|v \right\rangle$ is the holon and spinon vacuum. The three states are nearly degenerate at $p=p_c$, which implies $p_c = 1/3$ at zeroth order.
The critical theory
can be described by both operator representations and exhibits an exact fermion-boson duality. Away from the critical point, the lower energy state is chosen to be bosonic, and that boson is condensed: so the spinons $\mathfrak{b}_\alpha$ condense for $p<p_c$, and the holons $b$ condense for $p>p_c$. The spin correlations of SY \cite{SY92}, decaying as $1/|\tau|$, are realized at $p=p_c$. In the impurity model $H_{\rm imp}$ in (\ref{Himp}), increasing $p$ corresponds to increasing $s_0$.}
\label{fig:phasediag}
\end{figure}

This paper will study a model with all-to-all randomness (see (\ref{HtJ}) below) which exhibits a deconfined quantum critical point \cite{senthil1} with many similarities to the mysterious cuprate phenomenology. Our model has a quantum critical point at $p=p_c$ with fractionalized excitations, separating metallic states with carriers densities of $p$ and $1+p$ (see Fig.~\ref{fig:phasediag}). The overdoped state is a conventional disordered Fermi liquid, the underdoped `pseudogap' phase with carrier density $p$ has spin glass order, but the quantum critical point is described by fractionalized excitations. Our critical theory is distinct from a Landau-Hertz-Millis-type theory \cite{SRO1995,Sengupta95} of the quantum fluctuations of the spin glass order in a metal; the latter theory has no fractionalization at criticality and does not exhibit a change in carrier density across the transition. Moreover, our $p=p_c$ critical theory is expected to maximally chaotic \cite{Maldacena:2016hyu}, similar to the Sachdev-Ye-Kitaev (SYK) \cite{SY92,kitaev2015talk} models, and unlike the Landau theories \cite{Mao:2019xvt}.

Our results provide a simple rationale for the existence of a quantum phase transition in correlated metals with `Mottness'. Broken symmetries are not essential, and only play a secondary role. At low doping, we have fermionic `holons' of density $p$, moving in a background of condensed bosonic `spinons' (see Fig.~\ref{fig:phasediag}). At higher doping, we have condensed bosonic holons, so that the fermionic spinons behave like a Fermi liquid of hole-like carrier density $1+p$. This statistical transmutation, and corresponding transformation in the many-body state, is accomplished by a strongly coupled deconfined critical point which exhibits a boson-fermion duality. Note that, because of the presence of the Higgs-like condensates on both sides of the critical point, there is no true fractionalization for either $p>p_c$ or $p<p_c$.

There have been discussions \cite{SSV05} of deconfined critical points between a magnetic metal with `small' Fermi surfaces, and a non-magnetic heavy Fermi liquid with a `large' Fermi surface (a review of related ideas is in Ref.~\onlinecite{ColemanSi01}). However, to date, no tractable realization of this scenario has been found in non-random systems. Our results show that a similar scenario is naturally realized in models with random couplings. We also note the study of Ref~\onlinecite{Haule07}, which found an evolution between small and large Fermi surface across optimal doping in a plaquette dynamical mean field theory.

Our model is in the class of SYK models \cite{SY92,kitaev2015talk} with random all-to-all couplings, which have been extensively exploited recently for descriptions of strange metals and the quantum information theory of black holes. Specifically, we generalize the insulating 
random Heisenberg magnet originally studied in Ref.~\onlinecite{SY92} to metallic states 
of a $t-J_{ij}$ model at non-zero doping, along the lines of Ref.~\onlinecite{PG98}. 
We consider electrons, annihilated by $c_{i \alpha}$, spin $\alpha = \uparrow, \downarrow$ on sites $i=1 \ldots N$ with 
double occupancy prohibited $\sum_\alpha c_{i \alpha}^\dagger c_{i \alpha}^{} \leq 1$. The Hamiltonian is the familiar $t$-$J$ model with
\beq
H_{tJ} = \frac{1}{\sqrt{N}} \sum_{i \neq j=1}^N t_{ij}
P \, c_{i\alpha}^\dagger c_{j \alpha}^{} P\, + \frac{1}{\sqrt{N}} \sum_{i < j=1}^N J_{ij} \vec{S}_i \cdot \vec{S}_j - \mu \sum_{i} c_{i \alpha}^\dagger c_{i \alpha}^{} \label{HtJ}
\eeq
where $P$ is the projection on non-doubly occupied sites, $\mu$ is the chemical potential and $\vec{S}_i = (1/2) c_{i \alpha}^\dagger \vec{\sigma}_{\alpha\beta} c_{i \beta}$ is the spin operator on site $i$, with $\vec{\sigma}$ the Pauli matrices. The complex hoppings $t_{ij}= t_{ji}^\ast$ and the real exchange interactions $J_{ij}$ are independent random numbers with zero mean and mean-square values $\overline{ |t_{ij}|^2} = t^2$
and $\overline{J_{ij}^2} = J^2$. We will work at a variable hole density $p$, defined by
\beq
\frac{1}{N} \sum_{i} \left\langle c_{i \alpha}^\dagger c_{i \alpha}^{} \right\rangle = 1-p\,.
\label{doping}
\eeq

The insulating $p=0$ case of $H_{tJ}$ was studied in Ref.~\onlinecite{SY92}, and in Ref.~\onlinecite{PG98} for non-zero $p$, after generalizing the SU(2) spin symmetry to SU($M$) and taking the large $M$ limit (see Appendix~\ref{app:largeM}). A gapless critical ground state was found \cite{SY92} at large $M$ for $p=0$. However, subsequent numerical studies \cite{MJR02,MJR03} of the insulating SU($M=2$) case found a spin glass ground state, and such insulating spin glass states had also been examined 
in the large $M$ limit \cite{GPS00,GPS01}.
At non-zero $p$, the particular large $M$ limit in Ref.~\onlinecite{PG98} predicted a Fermi liquid ground state for all $p>0$.
We argue here that for the physical SU(2) case, the Fermi liquid only appears above a critical doping $p>p_c$, and that there is a metallic critical state very similar to the critical state of Ref.~\onlinecite{SY92} at $p=p_c$.
Our study was motivated by the observation of such critical spin correlations in metals in numerical studies of multi-orbital Hubbard models \cite{Werner18,Werner19}, and at the metal-insulator transition of a disordered Hubbard model at half-filling ($p=0$) \cite{Cha19}. 

It is convenient to describe the 3 states on each site of the $t$-$J$ model by spinon and holon operators. 
We can choose the spinons to be either fermions or bosons, and the holon to have the opposite statistics: the final results should be the same for all physical observables. With fermionic spinons $f_\alpha$ and bosonic holons $b$, the 3 physical states are $f_\alpha^\dagger \left| v \right\rangle$ and $b^\dagger \left| v \right\rangle$ (where $\left| v \right\rangle$ is the spinon and holon vacuum, and we drop site indices), see Fig.~\ref{fig:phasediag}. Then the operators 
\beq
c_\alpha = b^\dagger f_\alpha^{} \quad, \quad \vec{S} = \frac{1}{2} f_\alpha^\dagger \vec{\sigma}_{\alpha\beta} f_\beta^{} \quad , \quad V = b^\dagger b^{} + \frac{1}{2}f^\dagger_\alpha f_\alpha^{}
\label{defcS1}
\eeq
are all physical observables which realize the superalgebra SU($1|2$) \cite{Wiegmann88} (see Appendix~\ref{app:super}). We are interested in the 3-dimensional representation of physical states obeying 
\beq 
f_\alpha^\dagger f_\alpha^{} + b^\dagger b = 1\,;
\label{const1}
\eeq
Hence, the physical states are invariant under the U(1) gauge transformation $f_\alpha \rightarrow f_\alpha e^{i \phi}$, $b \rightarrow be^{i \phi}$, while individual spinon and holon excitations carry U(1) gauge charges.

Alternatively we can use a representation with bosonic spinons $\mathfrak{b}_\alpha$ and fermionic holons $\mathfrak{f}$. Now the gauge-invariant operators are
\beq
c_\alpha = \mathfrak{f}^\dagger \mathfrak{b}_\alpha \quad, \quad \vec{S} = \frac{1}{2} \mathfrak{b}_\alpha^\dagger \vec{\sigma}_{\alpha\beta} \mathfrak{b}_\beta \quad , \quad V = \mathfrak{f}^\dagger \mathfrak{f}^{} + \frac{1}{2}\mathfrak{b}^\dagger_\alpha \mathfrak{b}_\alpha^{}\,.
\label{defcS2}
\eeq
This realizes the same superalgebra SU($2|1) \equiv$ SU($1|2$) as (\ref{defcS1}), and the same 3-dimensional representation is obtained by the constraint
\beq 
\mathfrak{b}_\alpha^\dagger \mathfrak{b}_\alpha^{} + \mathfrak{f}^\dagger \mathfrak{f} = 1\,.
\label{const2}
\eeq
Note that while we find it convenient to refer to the superalgebra, there will be no supersymmetry in our results: $H_{tJ}$ does not commute with all SU($1|2$) generators.

We can now describe the structure of our main results illustrated in Fig.~\ref{fig:phasediag}. We find a deconfined critical point $p=p_c$ at which the 3 spinon and holon states are nearly degenerate. Assuming all three states are equally probable at criticality, we obtain a critical density $p_c=1/3$. Indeed, $p_c=1/3$ is the zeroth order result of our renormalization group (RG) analysis, as we show in Appendix~\ref{app:particle}; however, there are non-universal higher order corrections to the value of $p_c$.
We can formulate this critical theory using {\it either} of the representations in (\ref{defcS1}) and (\ref{defcS2}). This exact fermion-boson duality is an elementary analog of the fermion-boson duality of 2+1 dimensional field theories describing the deconfined critical point of the square lattice antiferromagnet \cite{Wang2017,Karch16}; as in the 2+1 dimensional theories, we will find that a Wess-Zumino-Witten \cite{Witten83,Wiegmann88,Tanaka05,Senthil06,LeeSachdev15} term ($\mathcal{S}_B$ in (\ref{Z}) and (\ref{SB2}) below) plays a central role in the criticality. 

Away from the critical point, there is a runaway RG flow to states in which either the spinon or holon states are lower in energy. As illustrated in Fig.~\ref{fig:phasediag}, we argue that this RG flow implies that we should now {\it choose\/} between the representations in (\ref{defcS1}) and (\ref{defcS2}) so that the lower energy state is a boson, and we should condense that boson. For $p>p_c$, the holon states are lower in energy; so we should choose (\ref{defcS1}) and condense $b$---this breaks the U(1) gauge symmetry, and we obtain a disordered Fermi liquid in which the $f_\alpha$ behave like electrons of density $1-p$ (which is the same as holes of density $1+p$ for the underlying Hubbard model). Conversely, for $p<p_c$, the spinon states are lower in energy; so we should choose (\ref{defcS2}) and condense $\mathfrak{b}_\alpha$---this again breaks the U(1) gauge symmetry, and we obtain a metallic spin glass in which the $\mathfrak{f}$ behave like holes of density $p$. Note that as $p \rightarrow 0$, this metallic spin glass becomes the insulating spin glass found in earlier studies \cite{GPS00,GPS01,MJR02,MJR03}. 
The large-$M$ limit considered in Ref.~\onlinecite{PG98} captures the disordered Fermi liquid, but does not capture the metallic spin-glass phase and the deconfined critical point at 
a non-zero $p=p_c$.

We will describe the nature of the infinite volume ($N\rightarrow\infty$) limit of $H_{tJ}$ in Section~\ref{sec:largeN}, and map possible critical states of the large $N$ limit to quantum impurity models in Section~\ref{sec:qimp}.
The RG analysis of the impurity model appears in Section~\ref{sec:rg}, where we obtain critical exponents of the deconfined critical point to two-loop order; the anomlaous dimensions of the electron and spin operators are obtained to all-loop order. We turn to the phases flanking the critical point in Section~\ref{sec:away}, and summarize our results in Section~\ref{sec:conc}. The appendices contain various technical details; in particular, the RG equations for a generalized model with SU($M$) spin symmetry appear in Appendix~\ref{app:RGM}, and the large $M$ analysis appears in Appendix~\ref{app:largeM}.

\section{Large volume limit}
\label{sec:largeN}

The limit of large volume ($N \rightarrow \infty$) of $H_{tJ}$ is obtained by the methods described in Refs.~\cite{SY92,GPS00,GPS01} for the insulating model at $p=0$.
We introduce field replicas in the path integral, and average over $t_{ij}$ and $J_{ij}$. 
At the $N=\infty$ saddle point, the problem reduces to a single site
problem, with the fields carrying replica indices. 
The replica structure is important in the spin glass phase \cite{GPS00,GPS01}, which we will explore in subsequent work. 
In the interests of simplicity, we drop the replica indices here as they play no significant role in the critical theory and the RG equations. 
Within the imaginary time path integral formalism (with $\tau\in[0,1/T]$, with $T$ the temperature), 
the solution of the model involves a local single-site effective action which reads:
\begin{eqnarray}
\mathcal{Z} &=& \int \mathcal{D} c_\alpha (\tau) e^{-\mathcal{S}-\mathcal{S_\infty}} \nonumber \\
\mathcal{S} &=& \int d \tau \left[c_\alpha^\dagger (\tau)  \left(\frac{\partial}{\partial \tau}-\mu \right)  c_\alpha (\tau) \right] 
 + t^2 \int d\tau d \tau' R (\tau - \tau') c_\alpha^\dagger (\tau) c_\alpha (\tau')  \nonumber \\
&~&~~ - \frac{J^2}{2} \int d\tau d \tau' Q (\tau - \tau') \vec{S} (\tau) \cdot \vec{S} (\tau') \,, \label{Zc}
\end{eqnarray}
In this expression, $\mu$ is the chemical potential determined to satisfy (\ref{doping}) and $\mathcal{S_\infty}$ is the action associated with the constraint 
of no double occupancy ($U=\infty$). 
Decoupling the path integral introduces fields analogous to $R$ and $Q$ which are initially off-diagonal in the spin SU(2) indices. 
We have assumed above that the large-volume  limit is dominated by the saddle point in which
spin rotation symmetry is preserved on the average, and so $R$ and $Q$ were taken to diagonal in spin indices.
The path integral $\mathcal{Z}$ is a functional of the fields $R(\tau)$ and $Q(\tau)$, and we define its correlators 
\begin{eqnarray}
\overline{R}(\tau - \tau') &=& -  \left\langle c^{}_{\alpha} (\tau) c^{\dagger}_\alpha (\tau') \right\rangle_\mathcal{Z} \nonumber \\
\overline{Q} (\tau - \tau') &=&  \frac{1}{3} \left\langle \vec{S} (\tau) \cdot \vec{S} (\tau') \right\rangle_\mathcal{Z} 
\end{eqnarray}
%Finally, the saddle-point equations of the large $N$ limit require that we impose the self-consistency conditions
In the thermodynamic ($N\rightarrow\infty$) limit, %and in the phase without spin-glass ordering, 
the solution of the model is obtained by imposing the two self-consistency conditions:  
\beq
R(\tau) = \overline{R}(\tau) \quad, \quad Q(\tau) = \overline{Q} (\tau).
\label{selfcons}
\eeq
These equations and the mapping to a local effective action are part of the extended dynamical mean-field theory framework (EDMFT), which becomes 
exact for random models on fully connected lattices~\cite{Sengupta95}.
They can also be viewed as an EDMFT approximation to the non-random $t$-$J$ model~\cite{Smith2000,Haule02,Haule03,Haule07}. 
%We note that equations closely related to the large volume saddle point equations above were obtained in a non-random model \cite{Haule02,Haule03}.
To make contact with notations often used in the (E)DMFT literature, we note that $t^2 R(\tau-\tau')$ is the self-consistent `hybridization function' (dynamical mean-field) 
$\Delta(\tau-\tau')$.  

This path-integral representation can be formulated in the 
%We present the single site imaginary time ($\tau$) path integral in the 
SU($1|2$) representation (\ref{defcS1}) as:
\begin{eqnarray}
\mathcal{Z} &=& \int \mathcal{D} f_\alpha (\tau) \mathcal{D} b (\tau) \mathcal{D} \lambda (\tau) e^{-\mathcal{S}_B - \mathcal{S}_{tJ}} \nonumber \\
\mathcal{S}_B &=& \int d \tau \left[ f_\alpha^\dagger (\tau)  \frac{\partial}{\partial \tau}  f_\alpha (\tau) + b^\dagger (\tau)  \frac{\partial}{\partial \tau} b (\tau) 
+ i \lambda (\tau) \left( f_\alpha^{\dagger} (\tau) f_\alpha^{} (\tau) + b^\dagger (\tau) b^{} (\tau) - 1 \right)\right] \nonumber \\
\mathcal{S}_{tJ} &=& \int d \tau \, s_0 \, f_\alpha^{\dagger} (\tau) f_\alpha^{} (\tau) 
+ t^2 \int d\tau d \tau' R (\tau - \tau') f_\alpha^\dagger (\tau) b(\tau) b^\dagger(\tau') f_\alpha (\tau')  \nonumber \\
&~&~~ - \frac{J^2}{2} \int d\tau d \tau' Q (\tau - \tau') \vec{S} (\tau) \cdot \vec{S} (\tau') \,, \label{Z}
\end{eqnarray} 
where  $\vec{S}(\tau)$ is to be represented by (\ref{defcS1}). 
Here $\mathcal{S}_B$ is the kinematic Berry phase ({\it i.e.\/} the Wess-Zumino-Witten term \cite{Witten83}) 
of the SU($1|2$) superspin at each site \cite{Wiegmann88}, 
$\mathcal{S}_{tJ}$ is the action containing the terms arising from $H_{tJ}$, %$T$ is the temperature, 
$\lambda$ is the Lagrange multiplier imposing Eq.~(\ref{const1}) and the chemical potential $s_0$ is determined to satisfy (\ref{doping}), which now becomes
\beq
\left\langle b^\dagger b^{} \right\rangle_{\mathcal{Z}} = p\,.
\label{bbp}
\eeq
Note that $\mathcal{Z}$ is a U(1) gauge theory, and under the U(1) gauge transformation
$\lambda \rightarrow \lambda - \partial_\tau \phi$.

Let us also present the exactly equivalent formulation of the large $N$ saddle point in terms of the SU($2|1$) superspin. Now the Berry phase $\mathcal{S}_B$ in (\ref{Z}) is replaced by 
\beq
\mathcal{S}_B = \int d \tau \left[ \mathfrak{b}_\alpha^\dagger (\tau)  \frac{\partial}{\partial \tau}  \mathfrak{b}_\alpha (\tau) + \mathfrak{f}^\dagger (\tau)  \frac{\partial}{\partial \tau} \mathfrak{f} (\tau) + i \lambda (\tau) \left( \mathfrak{b}_\alpha^{\dagger} (\tau) \mathfrak{b}_\alpha^{} (\tau) + \mathfrak{f}^\dagger (\tau) \mathfrak{f}^{} (\tau) - 1 \right)\right]\,,
\label{SB2}
\eeq
while $\mathcal{S}_{tJ}$ has the same form, apart from representing $c_\alpha (\tau)$ and $\vec{S}(\tau)$ by (\ref{defcS2}), and replacing the $s_0$ term by
$s_0 \mathfrak{b}_\alpha^\dagger (\tau) \mathfrak{b}_\alpha (\tau)$. The density constraint determining $s_0$ 
in (\ref{bbp}) is replaced by
\beq
\left\langle \mathfrak{f}^\dagger \mathfrak{f}^{} \right\rangle_{\mathcal{Z}} = p\,. \label{ffp}
\eeq

Appendix~\ref{app:largeM} analyzes the path integral (\ref{Z}) using a large $M$ expansion in an approach which generalizes the SU(2) spin symmetry to SU($M$); a similar large $M$ method has been used previously for a Hubbard model \cite{Florens02,Florens04,Florens13} and other phases of a disordered $t$-$J$ model \cite{Fu2018}.

The body of the paper will focus on an RG analysis of $\mathcal{Z}$. This is performed by expressing it in terms of an auxiliary quantum impurity problem, which we will now set up.

\subsection{Mapping to a quantum impurity problem}
\label{sec:qimp}

In our RG analysis, we find it useful to consider the path integral as a functional of the fields $R(\tau)$ and $Q(\tau)$ with an arbitrary time dependence, and to defer imposition of the self-consistency conditions in (\ref{selfcons}). As we are looking for critical states, we assume that these fields have a power-law decay in time with
\begin{equation}
Q(\tau) \sim \frac{1}{|\tau|^{d-1}} \quad, \quad R(\tau) \sim \frac{\mbox{sgn}(\tau)}{|\tau|^{r+1}}\,. 
\label{QRpower}
\end{equation}
where, for now, $d$ and $r$ are arbitrary numbers determining exponents, which will only be determined after imposing (\ref{selfcons}). Our analysis exploits the freedom to choose $d$ and $r$: we will show that a systematic RG analysis of the path integral $\mathcal{Z}$ is possible to all orders in $\epsilon$ and $\rb$, where
\beq
\epsilon = 3 -d \quad, \quad \rb = \frac{1-r}{2}\,\,\,;\,\,\,
Q(\tau) \sim \frac{1}{|\tau|^{2-\epsilon}} \quad, \quad R(\tau) \sim \frac{\mbox{sgn}(\tau)}{|\tau|^{2-2\rb}}\,. 
\eeq
The analysis assumes $\epsilon$ and $\rb$ are of the same order, and expands order-by-order in homogeneous polynomials in $\epsilon$ and $\rb$.
Such RG analyses were carried out in Refs.~\cite{SBV1999,VBS2000,SS2001} 
for an insulating spin model in which $t=0$, 
and by Fritz and Vojta~\cite{VojtaFritz04,FritzVojta04,FritzThesis} for a pseudogap Anderson impurity model in which $J=0$ (see also Refs.~\cite{Si1993a,Si1993b}); we note that the $\rb$ expansion of Refs.~\cite{VojtaFritz04,FritzVojta04,FritzThesis} was in good agreement with numerical studies \cite{Ingersent98}.  We will combine these analyses here, and our notations here for $\epsilon$ and $\rb$ follow these earlier works.

We proceed by decoupling the last two terms in $\mathcal{S}$  by introducing 
fermionic ($\psi_\alpha$) and bosonic ($\phi_a$, $a = x,y,z$) fields respectively, and then the 
path integral $\mathcal{Z}$ reduces to a quantum impurity problem. The `impurity' is a SU($1|2$) superspin
realizing the 3 states on each site of the $t$-$J$ model, and this is coupled to a `superbath' of the $\psi_\alpha$ and $\phi_a$ excitations. The quantum impurity problem is specified by
the Hamiltonian
\begin{eqnarray}
H_{\rm imp} && = (s_0 + \lambda) f_\alpha^\dagger f_\alpha + \lambda \, b^\dagger b + g_0 \left( f^\dagger_\alpha b \, \psi_\alpha (0) + \mbox{H.c.} \right) + \gamma_0 f_\alpha^\dagger \frac{\sigma^a_{\alpha\beta}}{2} f_\beta \, \phi_a (0) \nonumber \\
&&~~~~~~~ + \int |k|^r dk \, k \, \psi_{k\alpha}^\dagger \psi_{k \alpha} + \frac{1}{2} \int d^d x \left[ \pi_a^2 + (\partial_x \phi_a)^2 \right] \,. \label{Himp}
\end{eqnarray}
For completeness let us also explicitly present the Hamiltonian using a SU($2|1$) impurity of bosonic spinons and fermionic holons
\begin{eqnarray}
H_{\rm imp} && = (s_0 + \lambda) \mathfrak{b}_\alpha^\dagger \mathfrak{b}_\alpha + \lambda \, \mathfrak{f}^\dagger \mathfrak{f} + g_0 \left( \mathfrak{b}^\dagger_\alpha \mathfrak{f} \, \psi_\alpha (0) + \mbox{H.c.} \right) + \gamma_0 \mathfrak{b}_\alpha^\dagger \frac{\sigma^a_{\alpha\beta}}{2} \mathfrak{b}_\beta \, \phi_a (0) \nonumber \\
&&~~~~~~~ + \int |k|^r dk \, k \, \psi_{k\alpha}^\dagger \psi_{k \alpha} + \frac{1}{2} \int d^d x \left[ \pi_a^2 + (\partial_x \phi_a)^2 \right] \,. \label{Himp2}
\end{eqnarray}
We note several features of $H_{\rm imp}$, which apply equally to (\ref{Himp}) and (\ref{Himp2}):
\begin{itemize}
\item The bosonic bath is realized by a free massless scalar field in $d$ spatial dimensions,
as in Refs.~\cite{SBV1999,VBS2000,SS2001}. The field $\pi_a$ is canonically conjugate to the field $\phi_a$. 
The impurity spin $\vec{S}$ couples to the value of $\phi_a$ at the spatial origin, $\phi_a (0) \equiv 
\phi_a (x=0, \tau)$. It is easy to verify that upon integrating out $\phi_a$ from $H_{\rm imp}$, we obtain the $J$ term in $\mathcal{S}_{tJ}$, with $Q(\tau)$ obeying (\ref{QRpower}).
\item The fermionic bath is realized by free fermions $\psi_{k \alpha}$ with energy $k$ and a `pseudogap' 
density of states $\sim |k|^r$. The impurity electron operator $c_\alpha$ is coupled to $\psi_\alpha (0)  \equiv \int |k|^r dk \, \psi_{k \alpha}$. Integrating out $\psi_{k \alpha}$ from $H_{\rm imp}$ yields the $t$ term in $\mathcal{S}_{tJ}$, with $R(\tau)$ obeying (\ref{QRpower}).
\item We have replaced the path integral over the Lagrange multiplier $i\lambda$ in $\mathcal{S}_B$ by a constant real $\lambda$. The constraint (\ref{const1}) can be conveniently and exactly imposed by the Abrikosov method of sending $\lambda \rightarrow \infty$ \cite{SS2001,VojtaFritz04,FritzVojta04,FritzThesis}, as we will see in Section~\ref{sec:rg}. So the consequences of $\mathcal{S}_B$ will be accounted for exactly, and that is also the case for the alternative analysis in Appendix~\ref{app:RGM}, where $\mathcal{S}_B$ is accounted for by the exact implementation of the superalgebras.
\item The two formulations of $H_{\rm imp}$ in (\ref{Himp}) and (\ref{Himp2}) are equivalent, and lead to identical RG equations. This is because, ultimately, the quantum dynamics depends only upon the superspin algebra and the representation of the superspin on each site, and these are the same for SU($2|1$) formulation by (\ref{defcS1},\ref{const1}) and SU($1|2$) formulation by (\ref{defcS2},\ref{const2}). An explicit derivation of the one-loop RG equations using only the superspin algebra and representation appears in Appendix~\ref{app:RGM}.
\item The model is now characterized by 3 couplings constants, $s_0$, $\gamma_0$, and $g_0$, we will derive the RG equations for these couplings in Section~\ref{sec:rg}. The coupling of the superspin to the fermionic bath is $g_0$, and to the bosonic bath is $\gamma_0$: we will see that the RG flow of these couplings is marginal for small $\epsilon$ and $\rb$, and they are attracted to a deconfined critical point. 
\item The coupling $s_0$ acts like a `Zeeman field' on the superspin, which breaks the degeneracy between the spinon and holon states. The flow of $s_0$ is strongly relevant at the deconfined critical point, and this drives the system into one of the two phases in Fig.~\ref{fig:phasediag}.
\end{itemize}
We note that impurity models with both fermionic and bosonic baths have been considered earlier by Sengupta \cite{Sengupta00}, and by Si and collaborators \cite{SiNature,Si2002,Si2013}, but not for the `superspin' case with significant particle-hole asymmetric charge fluctuations on the impurity site. Specifically, we fully project out doubly occupied states, while keeping holon states low energy, and these features are crucial to the structure of our critical theory, as in Refs.~\cite{VojtaFritz04,FritzVojta04,FritzThesis}.
Also, the effect of a Zeeman field in an impurity spin model with a bosonic environment was studied in Refs.~\cite{Huang16,Whitsitt17,Chen18} in the context of the superfluid-insulator transition.

%%%%%%%%%%%%%%%%%%%%%%%%%%%%%%%%%%%%%%%%%%%%%%%%%%%%%%%%%%%%%%%%%%%%%%
\section{RG analysis}
\label{sec:rg}

This section will present the RG analysis of the impurity model defined by (\ref{Himp}). 
The RG analysis will initially not account for the self-consistency conditions (\ref{selfcons}). We will apply them later in Section~\ref{sec:anom}.

We will employ the SU($1|2$) superspin formulation, although essentially the same analysis can be applied to the SU($2|1$) superspin, with exactly the same results. 
A key feature of the computation is that we impose the local constraint (\ref{const1}) exactly. This implemented by the Abrikosov method of taking the $\lambda \rightarrow \infty$ limit, as described in earlier analyses \cite{SS2001,VojtaFritz04,FritzVojta04,FritzThesis}.

An alternative approach to obtain the RG equations generalizes the method of Refs.~\cite{SBV1999,VBS2000} for SU(2) spins to superspins in either SU($2|1$) or SU($1|2$). This method utilizes only gauge-invariant information contained in the superspin algebra and its representation, and is presented in Appendix~\ref{app:RGM}. The RG equations are identical to those obtained in this section.

At the tree-level, we can identify the scaling dimensions from (\ref{Himp}):
\begin{align}
&\text{dim} [\fa] = \text{dim} [\ba] = 0 \,, ~~~~ \text{dim} [\ca] = -\frac{1+r}{2} = -\text{dim} [\psi_{\alpha} (0)]   \,, \nonumber \\
&\text{dim} [\gc] = \frac{1-r}{2} \equiv \rb \,, ~~~~ \text{dim} [\gam] = \frac{3-d}{2} \equiv \frac{\ep}{2} \,, ~~~~
\text{dim} [\pa] = \frac{d-1}{2} \,.
\label{eq:sca_dim}
\end{align}
This establishes $r=1$ and $d=3$ as upper critical dimensions. %Since $0<r<1$, we have $0<\rb<1/2$.

We define the following renormalized fields and couplings,
\begin{align}
&\fa_{\alpha} = \sqrt{\zf} \fa_{R \alpha} \,, ~~~ \ba = \sqrt{\zb} \ba_{R} \,, ~~~ \gc = \frac{\rgs\up{\rb} \zg}{\sqrt{\zf \zb}} \gcr  \,, ~~~ %\nonumber \\
\gam = \frac{\rgs\up{\ep/2} \zgam}{\zf \sqrt{\zp \widetilde{S}_{d+1}}} \gamr \,, ~~~ \pa = \sqrt{\zp} \phi_{R a} \,,
\label{eq:renorm_fact}
\end{align}
where $\widetilde{S}_d = \Gamma (d/2-1)/(4 \pi\up{d/2})$.
The renormalization factors are to be obtained from self-energy and vertex corrections, as we will show below. We will work with $\kin=0$ and subsequently derive the flow away from it. Also, we work at zero temperature, i.e., $\beta \rightarrow \infty$.

\subsection{Fermion and boson self energy}

\begin{figure}[t]
\centering
\subfloat[]{\includegraphics[width=0.15\textwidth]{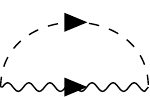}} ~~~~~~
\subfloat[]{\includegraphics[width=0.15\textwidth]{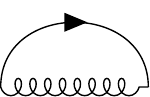}} ~~~~~~
\subfloat[]{\includegraphics[width=0.15\textwidth]{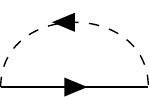}} ~~~~~~
\subfloat[]{\includegraphics[width=0.2\textwidth]{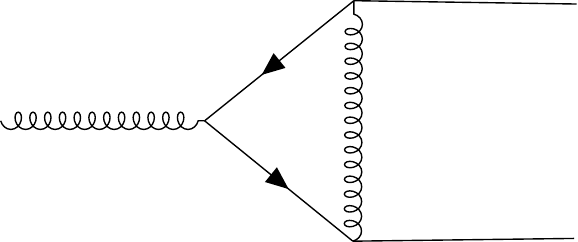}}
\caption{One-loop diagrams for self energy and vertex corrections. Fermion self-energy diagrams are shown in (a) and (b), boson self energy is shown in (c), and $\gam$ vertex correction is shown in (d). In these diagrams, solid line is for $f$ propagator,  dashed line is for $\psi$ propagator, wavy is for $b$ propagator, and spiral is for $\phi$ propagator.}
\label{fig:dia}
\end{figure}

Interaction terms in our action lead to self-energy corrections to the fermion and boson propagator. The corresponding Feynman diagrams at one-loop level are shown in Fig. \ref{fig:dia} (a)-(c), while Feynman diagrams for self-energy to two-loop order are shown in Figs. \ref{fig:dia2f} and \ref{fig:dia2b}. Here we show explicit result for one-loop diagrams and refer to Appendix 
\ref{app:2loop} for details regarding two-loop calculations.

%%%%%%%%%%%%%%%%
%%%%%%%%%%%%%%%%

We first calculate the fermionic self energy at one-loop level. At this level there are no diagrams which mixes the vertices corresponding to the bosonic bath coupling and the fermionic bath coupling. Here we have two relevant diagrams and we quote the self-energy below: %From Fig. \ref{fig:dia}(a),
\begin{align}
\label{eq:se_f1}
\Sigma\up{f}_{\ref{fig:dia} (a)} &= -\gc\up{2} \frac{1}{\beta} \sum_{\iw_{n}} \int_{-\infty}\up{\infty} dk \frac{|k|\up{r}}{\iw_{n} -k} 
\frac{1}{\inu - \iw_{n} - \lam} = \gc\up{2} \int_{-\infty}\up{\infty} dk \frac{|k|\up{r} \theta(k)}{\inu -\lam -k} ~~~~~~
(\text{Recall $\lam \rightarrow \infty$}) \nonumber \\
&= \gc\up{2} \pi  \csc (\pi  r) (\lam-i \nu )^r
%\nonumber \\
= A_{\mu} (\inu-\lam) \gcr\up{2} \left[-\frac{1}{2\rb} + i\pi \right] \hspace{0.25\textwidth}  \,, \\
%%%%%%
%\end{align}
%Similarly, from Fig. \ref{fig:dia}(b)
%\begin{align}
\label{eq:se_f2}
\Sigma\up{f}_{\ref{fig:dia} (b)} &= 
\gam\up{2} \frac{3}{4} \frac{1}{\beta} \sum_{\iw_{n}} \int \frac{d\up{d}k}{(2\pi)\up{d}} \frac{1}{\omega_{n}\up{2} +k\up{2}} 
\frac{1}{\inu + \iw_{n} - \lam} 
= \gam\up{2} \frac{3}{4} \frac{S_{d}}{2} \int_{0}\up{\infty} dk \frac{k\up{d-2} }{\inu -\lam -k} \nonumber \\
&= \gam\up{2} \frac{3}{4} \frac{S_{d}}{2} \pi  \csc (\pi (d-2)) (\lam-i \nu )^{-2+d}
%\nonumber \\
= B_{\mu} \frac{3}{4} \gamr\up{2} (\inu-\lam) 
\left[-\frac{1}{\ep} + \frac{1}{2} (\No + 2 i \pi ) \right]  \,,
\end{align}
where we wrote $\int d\up{d}k/(2\pi)\up{d} = {S_{d}} \int dk k\up{d-1}$, $S_d = {2}/({\Gamma(d/2) (4\pi)\up{d/2}})$, 
\begin{align}
\label{eq:Amu}
A_{\mu} &= \mu\up{2\rb} (\inu-\lam)\up{-2\rb} \frac{\zg\up{2}}{\zf \zb} \,, \\
\label{eq:Bmu}
B_{\mu} &= \mu\up{\ep} (\inu-\lam)\up{-\ep} \frac{\zgam\up{2}}{\zf\up{2} \zp} \,, \\
\label{eq:n0}
\No &=2(\gamma_{E} -1) = -0.845569\ldots \,,
\end{align}
with $\gamma_{E}$ being the Euler's constant.

For the bosonic self energy there is only one diagram (Fig. \ref{fig:dia}(c)) at the one-loop level. The self energy is evaluated as follows:
\begin{align}
\Sigma\up{b}_{\ref{fig:dia} (c)} &= 2 \gc\up{2} \frac{1}{\beta} \sum_{\iw_{n}} \int_{-\infty}\up{\infty} dk \frac{|k|\up{r}}{\iw_{n} -k} 
\frac{1}{\inu + \iw_{n} - \lam} = \gc\up{2} \int_{-\infty}\up{\infty} dk \frac{|k|\up{r} \theta(k)}{\inu -\lam -k} ~~~~~~
(\text{Recall $\lam \rightarrow \infty$}) \nonumber \\
&= 2 \gc\up{2} \pi  \csc (\pi  r) (\lam-i \nu )^r
%\nonumber \\
= A_{\mu} (\inu-\lam) 2 \gcr\up{2} \left[-\frac{1}{2\rb} + i\pi \right] \hspace{0.25\textwidth}  \,.
\label{eq:se_b1}
\end{align}
A factor of $2$ is due to the spin index of internal $f$ and $\psi$-line. 

The expressions for $\Sigma\up{f}_{\ref{fig:dia} (a)}$ and $\Sigma\up{f}_{\ref{fig:dia} (c)}$ agree with those in Refs.~\cite{VojtaFritz04,FritzVojta04,FritzThesis}, while that for $\Sigma\up{f}_{\ref{fig:dia} (b)}$ agrees with that in Ref.~\onlinecite{SS2001}.
Similarly, the self-energy diagrams at two-loop level are evaluated in a straightforward manner,
as shown in Appendix~\ref{app:2loop}.

%%%%%%%%%%%%%%%%%%%%%%%%%%%%%%%%%%%%%%%%%%%%%%%%%%%%%%%%%%%%%%%%%%%%%%%%%

\subsection{Vertex correction}

There is no one-loop vertex correction to the fermionic bath coupling $\gc$. However, it does acquire corrections at two-loop level and the corresponding diagrams are shown in Fig. \ref{fig:dia2vg}. The bosonic bath coupling $\gam$ has vertex corrections both at the one-loop and two-loop level. The one-loop diagram is shown in Fig. \ref{fig:dia}(d), while the two-loop diagrams are shown in Fig. \ref{fig:dia2vgam}. Here we explicitly evaluated the one-loop vertex correction to $\gam$,
\begin{align}
\label{eq:vertex_gam}
\Gamma\up{\gamma}_{\ref{fig:dia} (d)} &= (-\frac{1}{4}) \gam\up{3} \frac{1}{\beta} \sum_{i\omega_{1n}} \int \frac{d\up{d}k_{1}}{(2\pi)\up{d}}
\frac{1}{\omega_{1n}\up{2} + k_{1}\up{2}} \frac{1}{i\Omega_{1n}+i\omega_{1n}-\lam} \frac{1}{i\Omega_{2n}+i\omega_{1n}-\lam} 
\nonumber \\
&= (-\frac{1}{4}) \gam\up{3} \int \frac{d\up{d}k_{1}}{(2\pi)\up{d}}
\frac{1}{2k_{1}} 
\frac{1}{i\Omega_{1n}-k_{1}-\lam} \frac{1}{i\Omega_{2n}-k_{1}-\lam} \nonumber \\
%&=\gam B_{\mu} \gamr\up{2} (-\frac{1}{4}) 
%\left[ \frac{1}{\ep} -1 + \frac{1}{2} \left(-\No -2 i \pi \right) \right]
%\nonumber \\
&= -\gam B_{\mu} \gamr\up{2} \frac{1}{4} 
\left[ \frac{1}{\ep} -1 + \frac{1}{2} \left(-\No -2 i \pi \right) \right] \,.
\end{align}
This expression agrees with that in Ref.~\onlinecite{SS2001}.
We can similarly evaluate the two-loop level corrections and these are quoted in the Appendix \ref{app:2loop}. 

%%%%%%%%%%%%%%%%%%%%%%%%%%%%%%%%%%%%%%%%%%%%%%%%%%%%%%%%%%%%%%%%%%%%%%%%%

\subsection{Beta functions and fixed points}
\label{sec:betafunctions}

We now demand the cancellation of poles in the expression for the renormalized vertex and the $f/b$ Green's functions at the external frequency, $\inu -\lam=\rgs$. This leads to the following expressions of the renormalization factors. Note that $\zp=1$ exactly, owing to the absence of bulk interaction terms such as $\phi\up{4}$. For the rest we have,
\begin{align}
\label{eq:zf2}
\zf &= 1 - \frac{\gcr\up{2}}{2\rb} - \frac{3\gamr\up{2}}{4\ep} 
-\frac{15 \gamr^4}{32 \ep^2}+\frac{3 \gamr^4 }{8 \ep} -\frac{\gcr^4}{4 \rb^2}+\frac{\gcr^4}{2 \rb}
-\frac{3 \gcr^2 \gamr^2}{8 \ep \rb} 
+\frac{3 \gcr^2 \gamr^2 }{8 \rb (\ep+2 \rb)} 
+\frac{3 \gcr^2 \gamr^2 (2+\No)}{8 (\ep+2 \rb)} 
\,, \\
%&\text{where $N=\gamma_{E} -\log (4)-\psi ^{(0)}\left(\frac{3}{2}\right)$, with $\gamma_{E}$ being EulerGamma} \nonumber \\
%%%%%
\label{eq:zb2}
\zb &= 1 - \frac{\gcr\up{2}}{\rb} 
-\frac{\gcr^4}{4 \rb^2}+\frac{\gcr^4}{2 \rb}
-\frac{3 \gcr^2 \gamr^2}{4 \ep \rb}
+\frac{3 \gcr^2 \gamr^2}{2 \ep (\ep+2 \rb)}
+\frac{3 (2-\No)  \gcr^2 \gamr^2}{4 (\ep+2\rb)}
\,, \\
%%%%
\label{eq:zg2}
\zg &= 1 + \frac{\gcr\up{4}}{4\rb} \,, \\
%%%%
\label{eq:zgam2}
\zgam &= 1 + \frac{\gamr\up{2}}{4\ep} 
+\frac{9 \gamr^4}{32 \ep^2} -\frac{\gamr^4}{8 \ep} +\frac{\gcr^2 \gamr^2}{4 \ep \rb}
-\frac{\gcr^2 \gamr^2}{4 \rb (\ep+2\rb)}
- \frac{\gcr^2 \gamr^2 \No}{4 (\ep + 2\rb)}
\,.
\end{align}

Using Eqns. (\ref{eq:beta1}) and (\ref{eq:beta2}), we obtain the beta functions as follows:
\begin{align}
\label{eq:beta2g}
\betg &=  - \rb \gcr + \frac{3}{2} \gcr^3 + \frac{3}{8} \gcr \gamr^2 
- \gcr^5 + \frac{3}{16} \gcr^3 \gamr^2 \No - \frac{9}{8} \gcr^3 \gamr^2   - \frac{3}{8} \gcr \gamr^4     \\
%%%%%%%%%%%
\label{eq:beta2gam}
\betgam &=  - \frac{\ep}{2} \gamr +  \gamr^3  + \gcr^2 \gamr - \gamr^5
- \frac{5}{8} \gcr^2 \gamr^3 \No - \frac{3}{4} \gcr^2 \gamr^3  - 2 \gcr^4 \gamr
\end{align}

We can find the fixed points to two-loop order by setting the beta functions to zero. This gives us four fixed points 
$(\gcr*\up{2},\gamr*\up{2})$:
\begin{align}
\label{eq:fp1_2}
FP_{1} &= (0,0) \,, \\
\label{eq:fp2_2}
FP_{2} &= \left(0, \frac{\ep}{2} + \frac{\ep\up{2}}{4} \right) \,, \\
\label{eq:fp3_2}
FP_{3} &= \left( \frac{2\rb}{3} + \frac{8}{27} \rb\up{2}, 0 \right)  \,, \\
\label{eq:fp4_2}
FP_{4} &= \left( -\frac{\ep}{6}+\frac{8\rb}{9} + \ep\up{2} \left[ -\frac{25}{324} + \frac{\No}{24} \right] 
+ \rb\up{2} \left[ -\frac{304}{729} + \frac{8 \No}{27} \right] + \ep \rb \left[ \frac{119}{243} - \frac{5 \No}{18} \right],
\right. \nonumber \\
&\left. ~~\frac{2\ep}{3}-\frac{8\rb}{9} + \ep\up{2} \left[ \frac{40}{81} - \frac{\No}{9} \right] 
+ \rb\up{2} \left[ \frac{1600}{729} - \frac{64 \No}{81} \right] + \ep \rb \left[ -\frac{416}{243} + \frac{20 \No}{27} \right]
 \right) \,.
\end{align}

\begin{figure}[t]
\centering
\includegraphics[width=0.5\textwidth]{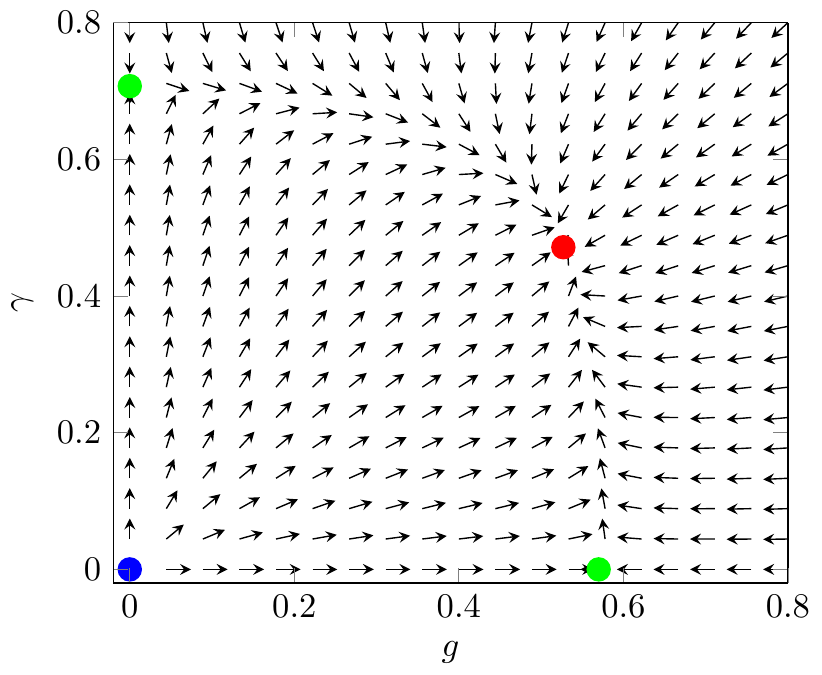}
\caption{One loop RG flow in the $\gcr - \gamr$ plane plotted for $\ep = 1$ and $\rb=0.5$. The red point is the stable fixed point ($FP_{4}$), green points are the saddle ($FP_{2}$ and $FP_{3}$) and blue point is the unstable Gaussian fixed point ($FP_{1}$). Note that the flow in the $\kinr$ direction is always relevant (not shown).}
\label{fig:rg_g_gam}
\end{figure}

The stability of the fixed points can be analyzed by looking at the eigenvalues of the stability matrix. We find that the Gaussian fixed point is always unstable. Importantly, we find that the non-trivial fixed point, $FP_{4}$, with $\gcr* \ne 0$ and $\gamr* \ne 0$ is stable for a range of values in the parameter space of $\ep$ and $\rb$. In Fig. \ref{fig:rg_g_gam} we plot the RG flow in the $\gcr-\gamr$ plane at one-loop level and show the different fixed points.

These fixed points corresponds to the underlying $t$-$J$ model at a non-zero doping density $p$. However, the precise value of $p$ depends upon high energy details, and cannot be deduced from the fixed point couplings, as we discuss in Appendix~\ref{app:particle}.
%%%%%%%%%%%%%%%%
%%%%%%%%%%%%%%%%

%%%%%%%%%%%%%%%%%%%%%%%%%%%%%%%%%%%%%%%%%%%%%%%%%%%%%

\subsection{Anomalous dimensions of $f_\alpha$ and $b$}

With the beta function at hand, it is straight forward to calculate the anomalous dimension of the fermion and boson propagators, defined as follows:
\begin{equation}
\label{eq:adim2}
\eta_{f} = \left. \rgs \frac{d \ln \zf}{d \rgs} \right|_{FP} \,, ~~~~~ \eta_{b} = \left. \rgs \frac{d \ln \zb}{d \rgs} \right|_{FP}\,.
\end{equation}
Note that these anomalous dimensions are gauge-dependent, and not physically observable. We have defined them in the gauge $\lambda = \mbox{constant}$. 
In terms of the coupling constants,
\begin{align}
\label{eq:etaf2}
\eta_{f} &= \gcr\up{2} + \frac{3}{4} \gamr\up{2} - 2 \gcr\up{4} - \frac{3}{4} \gamr\up{4} 
- \frac{3}{4} \gcr\up{2} \gamr\up{2} \left( 1 + \frac{\No}{2} \right) \,, \\
%%%%
\label{eq:etab2}
\eta_{b} &= 2 \gcr\up{2} - 2 \gcr\up{4} - \frac{3}{4} \gcr\up{2} \gamr\up{2} (2-\No) \,.
\end{align}
At the fixed points, we obtain the following expressions for the anomalous dimension,
\begin{align}
\label{eq:adim1_2}
&FP_1: \eta_{f} = 0 \,,~~~~ \eta_{b} = 0 \,, \\
\label{eq:adim2_2}
&FP_2: \eta_{f} = \frac{3}{8}\ep %- \frac{3}{16} \ep\up{3} - \frac{3}{64} \ep\up{4} 
\,,~~~~ \eta_{b} = 0 ~~~~~ \text{$\eta_{f}$ does not recieve correction at two loop } \,, \\
\label{eq:adim3_2}
&FP_3: \eta_{f} = \frac{2}{3}\rb -\frac{16}{27}\rb\up{2} %- \frac{64}{81}\rb\up{3} - \frac{128}{729}\rb\up{4}  
\,,~~~~ 
\eta_{b} = \frac{4}{3}\rb - \frac{8}{27}\rb\up{2} %- \frac{64}{81}\rb\up{3} - \frac{128}{729}\rb\up{4} 
\,, \\
\label{eq:adim4_2}
&FP_4: \eta_{f} = \frac{1}{3}\ep + \frac{2}{9}\rb - \frac{\ep\up{2}}{81} - \frac{256}{729}\rb\up{2} + \frac{32}{243} \ep \rb \,,~~~~ 
\eta_{b} = -\frac{1}{3}\ep + \frac{16}{9}\rb - \frac{7}{162}\ep\up{2} - \frac{896}{729}\rb\up{2} + \frac{112}{243} \ep \rb \,.
\end{align}

%%%%%%%%%%%%%%%%%%%%%%%%%%%%%%%%%%%%%%%%%%%%%%%%%%%%%%%%%%%%%%%%%%%%%%%%%

\subsection{Anomalous dimensions of the electron and spin operators}
\label{sec:anom}

We now calculate the anomalous dimensions $\eta_c$ and $\eta_S$ of the physical and gauge-invariant composite operators,
the electron $c_\alpha$ and the spin $\vec{S}$ specified in (\ref{defcS1}), defined such that 
\begin{equation}
\overline{R} (\tau) \sim \frac{\mbox{sgn}(\tau)}{|\tau|^{\eta_c}}
\quad , \quad \overline{Q}(\tau)
\sim \frac{1}{|\tau|^{\eta_S}}\,,
\end{equation}
at large $\tau$.
We will show that it is possible to determine these anomalous dimensions to {\it all orders} in the $\ep$ and $\rb$ expansions, as was also the case in previous analyses \cite{SBV1999,VBS2000,SS2001,VojtaFritz04,FritzVojta04,FritzThesis}.

To compute these scaling dimensions, we add source terms to the action
\begin{equation}
\label{eq:S_comp2}
\mathcal{S}_c = \frac{1}{\beta}  \sum_{i \omega_n} \left( 
\LS \fd_{\alpha} \frac{\sigma\up{a}_{\alpha \beta}}{2} \fa_{\beta} 
+ \LP [\fd_{\alpha} \ba + H.c.] \right)\,.
\end{equation}
Within the field-theoretic RG scheme, we have
\begin{equation}
\label{eq:lam_renorm2}
\LS = \frac{Z_{ff} \Lambda_{S,R}}{\zf} \,,~~~~~ \LP = \frac{Z_{fb} \Lambda_{c,R}}{\sqrt{\zf \zb}} \,.
\end{equation}
The composite operators $\hat{S} = \fd_{\alpha} ({\sigma\up{a}_{\alpha \beta}}/{2}) \fa_{\beta}$ and 
$c_{\alpha}\up{\dagger} = \fd_{\alpha} \ba$ are renormalized as follows:
\begin{equation}
\label{eq:sc_renorm}
\hat{S} = \sqrt{Z_{S}} \hat{S}_{R} \,,~~~~~ c = \sqrt{Z_{c}} c_{R} \,.
\end{equation}
It turns out that the diagrams contributing to the vertex corrections to $\LS$ and $\LP$ are exactly those we encountered while evaluating $\zgam$ and $\zg$ respectively. Thus we have,
\begin{equation}
\label{eq:zs_zc}
Z_{S} = \left( \frac{\zf}{\zgam} \right)\up{2} \,,~~~~~ Z_{c} = \frac{\zf \zb}{\zg\up{2}} \,.
\end{equation}
It is these identities which enable use to compute the anomalous dimensions exactly.
We evaluate the required anomalous dimensions as,
\begin{equation}
\label{eq:etasc2}
\eta_{S} = \frac{d \ln Z_{S}}{d \ln \mu} \,, ~~~~~ \eta_{c} = \frac{d \ln Z_{c}}{d \ln \mu} \,.
\end{equation}
We can now make an exact statement for $\eta_{S}$ for fixed points with $\gamr \neq 0$. From Eqns. (\ref{eq:etasc2}) and 
(\ref{eq:zs_zc}) we obtain,
\begin{equation}
\eta_{S} = \frac{2}{\zf \zgam} \bigg[  
\left( \zgam \frac{\partial \zf}{\partial \gcr} - \zf \frac{\partial \zgam}{\partial \gcr} \right) \betg 
+ \left( \zgam \frac{\partial \zf}{\partial \gamr} - \zf \frac{\partial \zgam}{\partial \gamr} \right) \betgam 
\bigg]  \,.
\end{equation}
Substituting the above equation in Eqn. (\ref{eq:beta2}), we obtain,
\begin{equation}
\frac{\ep}{2} \gamr \zgam \zf - \gamr \eta_{S} \frac{\zf \zgam}{2} + \betgam \zf \zgam = 0 \,,
\end{equation}
which leads to 
\begin{equation}
\label{eq:etas_eq}
\gamr \eta_{S} = \gamr \ep + 2 \betgam \,.
\end{equation}
The first term on the r.h.s. of (\ref{eq:etas_eq}) arises from the tree-level scaling dimension, while the second term contains potential corrections higher order in $\epsilon$. However,
at the fixed point where $\gamr = \gamr\up{*} \neq 0$, we have $\beta (\gamr\up{*}) =0$ and 
therefore, 
\beq
\eta_{S} = \ep\,, \quad \mbox{to all orders in $\ep$ and $\rb$.}
\label{etaSep}
\eeq
The same value of $\eta_S$ is also obtained in the large $M$ expansion in (\ref{Deltafe}) and (\ref{etaSDeltaf}).

Similarly, using Eqns. (\ref{eq:etasc2}) and (\ref{eq:zs_zc}) in combination with Eqn. (\ref{eq:beta1}) we obtain the following relation:
\begin{equation}
\label{eq:etac_eq}
\gcr \eta_{c} = 2 \rb \gcr + 2 \betg \,.
\end{equation}
Thus at the fixed point, $\beta (\gcr\up{*}) =0$, such that $\gcr\up{*} \neq 0$, we obtain 
\beq
\eta_{c} = 2 \rb\,, \quad \mbox{to all orders in $\ep$ and $\rb$.}
\label{etacrb}
\eeq
The same value of $\eta_c$ is also obtained in the large $M$ expansion in (\ref{e5}) and (\ref{etaclargeM}).

We can now state the main result of this subsection: at the non-trivial fixed point $FP_{4}$ ($\gcr\up{*} \neq 0, \gamr\up{*} \neq 0$) we have $\eta_{S} = \ep$ and 
$\eta_{c} = 2\rb$ to all orders in $\ep$ and $\rb$. 

Finally, we can impose the self-consistency condition, Eq. (\ref{selfcons}). This implies equating the exponents in Eq. (\ref{QRpower}) to the anomalous dimensions found above, {\it i.e.\/}, $\eta_{S}=d-1$ and $\eta_{c}=r+1$. Solving these equations requires using values of $\epsilon$ and $\rb$ which are of order unity, but because Eqs.~(\ref{etaSep}) and (\ref{etacrb}) are valid to all orders, we can use such values with confidence. Solving these self-consistency conditions we obtain our exact results for $\eta_c$ and $\eta_S$
\beq
\eta_S = 1 \quad, \quad \eta_c = 1\,, \label{etaSetac}
\eeq
at the self-consistent values $\epsilon = 1$ and $\rb=1/2$.
These anomalous dimensions imply the critical correlators at $FP_4$
\beq
\left\langle \vec{S} (\tau) \cdot \vec{S} (0) \right\rangle \sim \frac{1}{|\tau|}
\quad, \quad \left\langle c_\alpha^{} (\tau) c_{\alpha}^\dagger (0) \right\rangle \sim
\frac{\mbox{sgn}(\tau)}{|\tau|} \,, \label{Gc3}
\eeq
as indicated in Fig.~\ref{fig:phasediag}. We note that the fixed point $FP_{4}$ in (\ref{eq:fp4_2}) is stable when evaluated at 
$\ep=1$ and $\rb=1/2$, although formally we cannot trust our expansion for the stability exponent at such large values of $\ep$ and $\rb$.

%%%%%%%%%%%%%%%%%%%%%%%%%%%%%%%%%%%%%%%%%%%%%%%%%%%%%%
%%%%%%%%%%%%%%%%%%%%%%%%%%%%%%%%%%%%%%%%%%%%%%%%%%%%%%

%%%%%%%%%%%%%%%%%%%%%%%%%%%%%%%%%%%%%%%%%%%%%%%%%%%%%%%%%%%%%%%%%%%%%%%%%

\section{Moving away from the critical point}
\label{sec:away}

The RG flow equations presented in Section~\ref{sec:betafunctions} have a fixed point $FP_4$ which realizes the deconfined critical point of Fig.~\ref{fig:phasediag}. This fixed point has one relevant direction, corresponding to the on-site energy $s$ which distinguishes the local energy of the spinon and holon states. As $s$ flows to $+\infty$, the holon state is lower in energy, corresponding to the $p>p_c$ region of the phase diagram in Fig.~\ref{fig:phasediag}. Conversely, as $s$ flows to $-\infty$, the spinon states are lower in energy, corresponding to the $p<p_c$ region of the phase diagram.

We can now exploit the choice of superspin representations between SU($1|2$)
and SU($2|1$) to understand the fate of the RG flow. The energy of the ground state is minimized if we maximize the occupation of the lower energy state, and this is achieved if we choose the representation in which this lower energy state is bosonic. This implies that we should choose SU($1|2$) for $p>p_c$ and SU($2|1$) for $p<p_c$, as indicated in Fig.~\ref{fig:phasediag}.
We now describe the structure of these theories in turn

\subsection{Overdoped region}
\label{sec:overdoped}

In the SU($1|2$) theory for $p>p_c$, we condense the boson $b$. With $\langle b \rangle \neq 0$, we have from (\ref{defcS1}), $c_\alpha \sim f_\alpha$, and the hopping term $t_{ij}$ in $H_{tJ}$ in (\ref{HtJ}) reduces to a renormalized hopping term for the $f_\alpha$ spinons. Indeed, the resulting theory for the $f_\alpha$ fermions is similar to that studied by Parcollet and Georges \cite{PG98}, and more recently in SYK-like extensions \cite{Balents2017,Zhang2017,Chowdhury2018,Patel2017,Patel2019}.

A description of this phase, far from the $p=p_c$ critical point, can be obtained from $H_{tJ}$ in (\ref{HtJ}) by taking the large $M$ limit of Ref.~\onlinecite{PG98}: we consider spinons $f_\alpha$ with $\alpha = 1 \ldots M$, the bosons $b$ do not acquire any additional index (so we are considering a SU($1|M$) superspin), and we rescale $t^2 \rightarrow t^2/M^2$ and $J^2 \rightarrow J^2/M$. (Note that this large $M$ limit is distinct from that described in our Appendix~\ref{app:largeM}, where the bosons do acquire an additional index, and we rescale $t^2 \rightarrow t^2/M$ and $J^2 \rightarrow J^2/M$.)

In the large $M$ limit of Ref.~\onlinecite{PG98}, the $b$ bosons can be replaced by their condensate $\langle b \rangle = b_0 \sqrt{M}$. The $f_\alpha$ fermions have an effective hopping of strength $t \, b_0^2 $. As shown in Ref.~\onlinecite{PG98}, the theory behaves a like a disordered Fermi liquid below a coherence scale 
\beq
E_c = \frac{(t b_0^2)^2}{J}.
\eeq
Note that this disordered Fermi liquid has a hole carrier density of $1+p$. This follows from $c_\alpha \sim f_\alpha$, and the density of $f_\alpha$ fermions obtained from (\ref{const1}) and (\ref{bbp}).

As we approach the critical point, with $p \searrow p_c$, we expect $E_c$ and $\langle b \rangle$
to both vanish algebraically. However, we do not expect the large $M$ theory of Ref.~\onlinecite{PG98} to properly describe the approach to the critical point: in this large $M$ theory, we obtain an insulating
state as $\langle b \rangle$ vanishes, whereas our $p=p_c$ critical point is metallic. Indeed, as $b$ is gauge-charged field, the value of $\langle b \rangle$ is not a gauge-invariant quantity which can be directly compared between different approaches. However, the crossover scale $E_c$ is better defined, and
we can deduce the behavior of $E_c$ near $p=p_c$ by the RG analysis of Section~\ref{sec:rg}. We expect
\beq
E_c \sim |p-p_c|^{1/\lambda_s}
\eeq
where $\lambda_s$ is the relevant RG eigenvalue with which $s$ flows away from the $FP_4$ fixed point, specified in (\ref{lambdas}).

\subsection{Pseudogap region}
\label{sec:pseudogap}

For $p<p_c$, we use the SU($2|1$) theory, and condense the $\mathfrak{b}_\alpha$ spinons to obtain spin glass order, as described in Refs.~\cite{GPS00,GPS01}. The presence of the mobile $\mathfrak{f}$ fermions will make this a metallic spin glass with carrier density $p$, as determined by (\ref{ffp}).

We need to extend the insulating spin glass theory of Refs.~\cite{GPS00,GPS01} to the metallic spin glass, and this will be studied in greater detail in future work. Here we note that a systematic description appears possible in the large $M$ limit of a SU($M|M'$) formulation noted in Appendix~\ref{app:larger}, where the $\mathfrak{f}$ fermions acquire an additional `orbital' label $\ell = 1 \ldots M'$, and we keep $k=M'/M$ fixed in the large $M$ limit. This should be contrasted from the SU($M'|M$) theory described in Appendix~\ref{app:largeM}, in which the $b_\ell$ bosons are assumed to not condense.

\subsection{Specific heat}
\label{sec:heat}

This section discusses qualitative features of the specific across the phase transition at $p=p_c$.

Right at $p=p_c$, the critical theory is expected \cite{GPS01} to have a non-vanishing extensive entropy $S_0$ as $T \rightarrow 0$. This follows from the similarity of the random insulating magnet \cite{SY92}, and many other models in the SYK class.

\begin{figure}
\begin{center}
\includegraphics[height=2.75in]{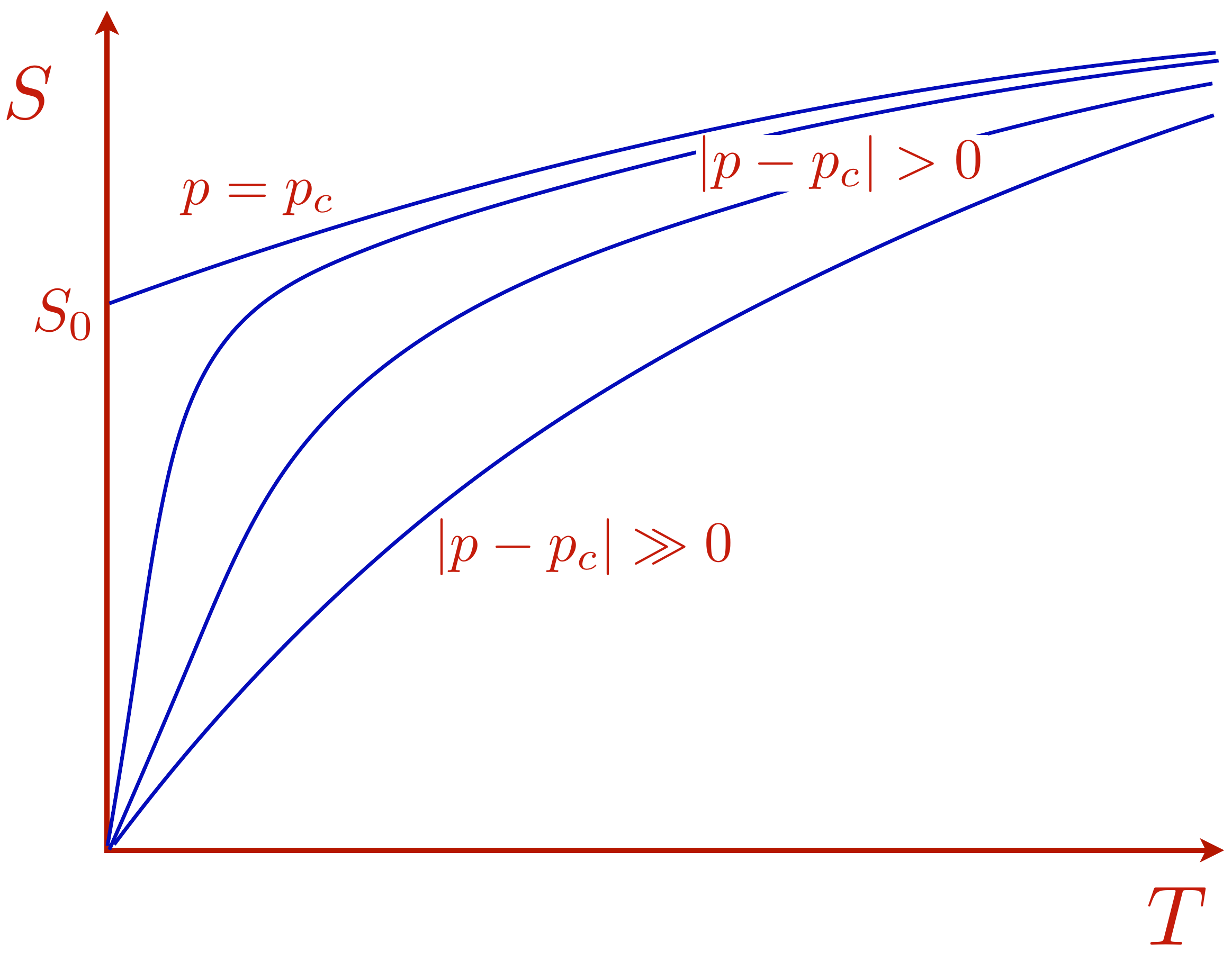} 
\end{center}
\caption{Schematic plot of the temperature dependence of the entropy $S$ of $H_{tJ}$. 
At $p=p_c$, there is a non-vanishing extensive entropy at zero temperature $S_0$. 
%Away from $p=p_c$, there is divergence in 
The linear-in-$T$ coefficient of the specific heat $\gamma = \lim_{T \rightarrow 0} C/T = \lim_{T \rightarrow 0} dS/dT$ 
diverges as we approach $p_c$.}
\label{fig:entropy}
\end{figure}

Away from $p=p_c$, we expect that the entropy follows the behavior of the critical point at temperatures above the coherence scale $E_c$, before vanishing lineary with $T$ at temperatures below $E_c$, as shown in Fig.~\ref{fig:entropy}. 
We can therefore estimate that the linear-in-$T$ coefficient of the specific heat $C$ is
\beq
\gamma = \lim_{T \rightarrow 0} \frac{C}{T} = \lim_{T \rightarrow 0} \frac{dS}{dT} \sim  \frac{S_0}{E_c}
\eeq
So we expect $\gamma$ to diverge as $|p-p_c|^{-1/\lambda_s}$ in the infinite range model $H_{tJ}$. It is notable that this behavior resembles experimental observations \cite{Michon18}.

\section{Conclusions}
\label{sec:conc}

We have presented a RG analysis (in Section~\ref{sec:rg}) of the phase diagram of the $t$-$J$ model in (\ref{HtJ}) with random and all-to-all hopping and spin exchange. The predictions of this analysis are presented in Fig.~\ref{fig:phasediag}: a deconfined critical point which separates two phases without fractionalization: a `pseudogap' metallic spin glass phase with carrier density $p$, from a disordered Fermi liquid with carrier density $1+p$. The change in the carrier density across the critical point, and the behavior of the specific heat across the critical point implied by the entropy in 
Fig.~\ref{fig:entropy} are in good qualitative accord with experimental observations \cite{Badoux16,Loram01,Loram19,Michon18}.
The SYK-like structure of the deconfined critical point connects naturally with a class of theories with linear-in-$T$ resistivity in the strange metal \cite{PG98,Balents2017,Zhang2017,Chowdhury2018,Patel2017,Patel2019}; a linear-in-$T$ resistivity was also found in numerical studies \cite{Haule02,Haule03} of lattice models without disorder described by equations closely related to those of the large $M$ limit of Appendix~\ref{app:largeM}. And we note that there is a recent report of spin glass correlations in the pseudogap phase \cite{Julien19}, extending earlier impurity-induced observations \cite{CPana1,CPana2}.

It is useful to compare the structure of $H_{tJ}$ in (\ref{HtJ}) with that of SYK lattice models \cite{Balents2017,Zhang2017,Chowdhury2018,Patel2017}. The SYK models have a random 4-fermion interaction term, and a random 2-fermion hopping term of strength $t$, but no on-site fermion constraint. At the lowest energies, the 2-fermion term is always relevant and drives the system away from SYK criticality to a Fermi liquid state. In the present $t$-$J$ model analysis, the presence of the local constraint (\ref{const1}) or (\ref{const2}) is crucial; in terms of the fractionalized particles in (\ref{defcS1}) or (\ref{defcS2}), {\it both\/} the $t$ and $J$ terms are 4-particle terms. Consequently, they can balance each other, and allow for a critical SYK-like state to exist at a critical $p=p_c$ all the way down to $T=0$. At $p=p_c$, the 3 states of the $t$-$J$ model obeying the constraint (\ref{const1}) or (\ref{const2}) are nearly degenerate in our perturbative renormalization group analysis. Consequently, we find $p_c = 1/3$ at zeroth order (see Fig.~\ref{fig:phasediag}).

Appendix~\ref{app:largeM} presents a low energy analysis of the large $M$ saddle point of $H_{tJ}$ obtained by generalizing the SU(2) symmetry to SU($M$). The main difference from the RG analysis is that a critical {\it phase} appears in the low energy and large $M$ theory, rather than a critical point. However, the exponents of the gauge-invariant spin and electron operators in the appropriate large $M$ phase are the {\it same\/} as those obtained by the renormalization group analysis of the deconfined critical point to all orders in $\epsilon$ and $\rb$: compare (\ref{Deltafe},\ref{etaSDeltaf}) and (\ref{e5},\ref{etaclargeM}) with  (\ref{etaSep}) and (\ref{etacrb}). We also note that the similarity between Appendix~\ref{app:largeM} and the SYK equations indicates that our critical theory obeys maximal chaos \cite{Maldacena:2016hyu}.

A useful perspective on our renormalization group analysis provided by viewing the 3 states on each site as different orientations of a SU($1|2$) or SU($2|1$) superspin \cite{Wiegmann88}. In the large $N$ limit of a $N$-site $t$-$J$ model with random and all-to-all interactions, we obtain an impurity model in which the superspin is coupled to both fermionic and bosonic baths self-consistently. In addition, there is a `Zeeman field' acting on the superspin which breaks the degeneracy between the spinon and holon states; this field is a strongly relevant perturbation at the deconfined critical point, and drives the system into the two phases flanking the critical point. (See Refs.~\cite{Huang16,Whitsitt17,Chen18} for analogous RG studies of a relevant ``Zeeman field'' at an impurity site (represented by a SU(2) spin) in a Bose-Hubbard model at the superfluid-insulator transition, which are in good agreement with numerics.) In the overdoped phase, the holon state has a lower energy (see Fig.~\ref{fig:phasediag}): so we use the SU($1|2$) superspin in which the holon is a boson, and condense it to obtain a disordered Fermi liquid, analogously to Ref.~\onlinecite{PG98}. 
Conversely, in the underdoped phase, the spinon states have a lower energy, so we use the SU($2|1$) superspin in which the spinons are bosons, and condense them to obtain a metallic spin glass. 

Despite our use of the superspin terminology, the model studied and the renormalization group equations are not supersymmetric: the fermionic $c_\alpha$ generators in (\ref{defcS1}) and (\ref{defcS2}) do not commute with the Hamiltonian. We can extend $H_{tJ}$ by including off-site interations with the generator $V$ in (\ref{defcS1}) and (\ref{defcS2}), and this will include density-density interactions. With this extension, we have the possibility of charge glass order in the pseudogap, and critical density fluctuations (likely similar to those observed in Refs.~\onlinecite{Mitrano18,Husain19}) at possibly supersymmetric fixed points. (Supersymmetric $t$-$J$ models have been studied in one dimension without disorder \cite{Sutherland75,Blatter90,Sarkar91,Essler92,Czech03}.)

Finally, we comment on the extent to which a model with all-to-all randomness can be mapped to the cuprates. Randomness is present in the experimental systems, and also serves important simplifying purposes in our theoretical analysis. Moreover, certain approximations to models without randomness lead to closely related saddle point equations \cite{Smith2000,Haule02,Haule03,Haule07}.  Several of the broken symmetries do not exist in the random model, and subtle questions \cite{SSST18} about the structure of the Fermi surface in momentum space can be avoided. However, the central issues of carrier density, fractionalization, emergent gauge charges, and associated quantum phase transitions remain well defined even in the presence of randomness. Given the recent spin glass observations \cite{Julien19}, and as we noted in Section~\ref{sec:pseudogap}, it will be useful to study the metallic pseudogap state, and the interplay between the spin glass order and charge transport. 
A possible approach is the extend the large $M$ theory of Refs.~\cite{GPS00,GPS01} to include fermionic holons, as well as numerical studies for $M=2$.   

\subsection*{Acknowledgements}
%********************************************************

We thank P.~Cha, J.C. Seamus Davis, M.-H. Julien, E.-A.~Kim, G.~Kotliar, O.~Parcollet, N.~Seiberg, A.~Sengupta, Z.-X. Shen, Qimiao Si, T.~Senthil, L.~Taillefer, M.~Vojta, and Ya-Hui Zhang for valuable discussions.
This research was supported by the National Science Foundation under Grant No. DMR-1664842 and U.S. Department of Energy under Grant DE-SC0019030.
G.T.  acknowledges support from the MURI grant W911NF-14-1-0003 from ARO and by DOE grant DE-SC0007870. 
The Flatiron Institute is a division of the Simons Foundation.

\appendix

\section{Superalgebras}
\label{app:super}

The operators in (\ref{defcS1}) obey the commutation and anti-commutation relations
\bea
[S^a, S^b] = i \epsilon_{abc} S^c \quad &,& \quad 
\{c_\alpha , c_\beta \} = 0 \quad , \quad 
   \{c_\alpha , c_\beta^\dagger\} = \delta_{\alpha\beta} V +  \sigma^a_{\alpha \beta} S^a \nonumber \\ 
 ~  [ S^a , c_\alpha ] = - \frac{1}{2} \sigma^a_{\alpha\beta} c_\beta \quad &,&
 ~  [ S^a , c_\alpha^\dagger ] = \frac{1}{2} \sigma^a_{\beta\alpha} c_\beta^\dagger \quad , \quad
 ~ [S^a , V] = 0 \nonumber \\
 ~ [V, c_\alpha ] = \frac{1}{2} c_\alpha \quad &,& \quad
 ~ [V, c_\alpha^\dagger ] = -\frac{1}{2} c_\alpha^\dagger\,. \label{super}
\eea
The constraint (\ref{const1})
commutes with all operators of the superalgebra. Imposing this constraint yields the fundamental representation of SU($1|2$).

Alternatively, we can use the operators in (\ref{defcS2}).
These realize the SU($2|1$) algebra, and it can be verified that these operators also obeys the superalgebra in Eq.~(\ref{super}). The constraint projecting to the fundamental representation is now
(\ref{const2}).

\subsection{Larger symmetries}
\label{app:larger}

We consider the model for general $M$ and $M'$, with the electron operator 
\beq
c_{\ell \alpha} = b_{\ell}^\dagger f_{\alpha} \,, \label{defc}
\eeq
acquiring both spin ($\alpha = 1 \ldots M$) and `orbital' ($\ell = 1 \ldots M'$) indices.
We define the SU($M$) spin operator
\beq
S^a = f_\alpha^\dagger T^a_{\alpha\beta} f_\beta \label{STM}
\eeq
where the matrices $T^a$ obey
\begin{align}
\textrm{tr}(T^{a}T^{b}) = \frac{1}{2}\delta^{ab}, \quad T^{a}T^{a} = \frac{M^{2}-1}{2M} \cdot \textbf{1}\,, \quad T^{a}_{\alpha\beta}T^{a}_{\gamma\delta}=\frac{1}{2}
\Big(\delta_{\alpha\delta}\delta_{\beta \gamma}-\frac{1}{M}\delta_{\alpha\beta}\delta_{\gamma\delta}\Big)\,.
\label{SUM}
\end{align} 
The operators $c_{\ell \alpha}$, $S^a$, the operator
\beq 
V = \frac{1}{M}  f_\alpha^\dagger f_\alpha + \frac{1}{M'}  b_\ell^\dagger b_\ell^{}\,,
\eeq
and the operators $b_\ell^\dagger T^a_{\ell \ell'} b_{\ell'}$ are the $(M+M')^2 -1$ generators of the superalgebra SU($M'|M$).  

A general constraint fixing the representation is
\beq
f_\alpha^\dagger f_\alpha + b_\ell^\dagger b_\ell^{} = P\,, \label{constraintM}
\eeq
with $P$ a positive integer, and our interest in the case $P=M/2$ which realizes the representation in which the SU($M$) subalgebra is self-conjugate. Note that the fundamental representation of the superalgebra is $P=1$, but this does not lead to a convenient large $M$ limit.

We can also consider a bosonic spinon and fermionic holon decomposition for general $M$, $M'$
\beq
c_{\ell \alpha} = \mathfrak{f}^\dagger_\ell \mathfrak{b}_\alpha \label{defc2}
\eeq
The analogous steps will lead to a realization of the SU($M|M'$) superalgebra, which is identical to the 
SU($M'|M$) superalgebra. However, the constraint
\beq
\mathfrak{b}_\alpha^\dagger \mathfrak{b}_\alpha + \mathfrak{f}_\ell^\dagger \mathfrak{f}_\ell^{} = P \label{constraintM2}
\eeq
now leads to a {\it different representation\/} of the superalgebra from (\ref{constraintM}) for $P \neq 1$ \cite{RS89}. So the models defined by (\ref{defc}) and (\ref{defc2}) are in general different, although they are the same for the $M=2$, $M'=1$ case of interest to us. Note also that in all cases the Hamiltonian contains an exchange interaction involving the SU($M$) generators only, and not the SU($M'$) generators.

\subsection{Operator expectation values}

We will compute the only RG equations for the SU($M'|M$) theory in Appendix~\ref{app:RGM} following the method in Appendix C of Ref.~\onlinecite{VBS2000}.
After using the identity (\ref{SUM}), the computations in Appendix~\ref{app:RGM}  can be reduced to the following operator traces.

First, let us compute the dimension, $\mathcal{D}(M,M',P)$, of the superspin Hilbert space. To compute this, it is useful to compute the grand-canonical partition function, while ignoring the constraint (\ref{constraintM}).
\beq
\mathcal{Z}(z) = \mbox{Tr} \, z^{f_\alpha^\dagger f_\alpha + b_\ell^\dagger b_\ell^{}} = \frac{(1 + z)^M}{(1-z)^{M'}}
\eeq
where $z$ is the common fugacity. 
Then we can impose the constraint (\ref{constraintM}), and the dimension of the Hilbert space is given by the coefficient of $z^{P}$ in the series expansion of $\mathcal{Z}(z)$, or equivalently
\beq
\mathcal{D}(M,M',P) = \oint_{|z|=c < 1} \frac{dz}{2 \pi i} \frac{1}{z^{P+1}} \mathcal{Z} (z)
\eeq
Some sample values from Mathematica are
\beq
\mathcal{D}(2,1,1) = 3 \quad, \quad \mathcal{D}(6,6,3) = 292 \quad, \quad \mathcal{D}(20,14,10) = 553844224 \quad , \quad \ldots
\eeq
Now we can compute the general expectation value
\bea
\mathcal{I}_{m,m'}  &\equiv& \left\langle \left(f_\alpha^\dagger f_\alpha\right)^m \left( b_\ell^\dagger b_\ell \right)^{m'} \right\rangle  \\
&=& \frac{1}{\mathcal{D}(M, M',P)} \oint_{|z|=c < 1} \frac{dz}{2 \pi i} \frac{1}{z^{P+1}} \,
\left[ \left( z \frac{d}{dz} \right)^{m} (1+z)^M \right] \left[ \left( z \frac{d}{dz} \right)^{m'} \frac{1}{(1-z)^{M'}} \right] \nonumber \,.
\eea
Note $\mathcal{I}_{0,0} = 1$.

The values for $M=2$, $P=1$, and $M'=1$ case of interest to us are simple:
\bea
\mathcal{I}_{m,0} &=& \frac{2}{3}, \quad m \geq 1; \quad
\mathcal{I}_{0,m'}  = \frac{1}{3}, \quad m' \geq 1; \quad
\mathcal{I}_{m,m'} = 0, \quad m \geq 1 ~{\rm and}~ m' \geq 1 \,.
\eea
This simplicity is the reason  Section~\ref{sec:rg} was able to compute the RG equations using Feynman diagrams and the Abrikosov method.

Some other random values
\bea
\mathcal{I}_{2,3}  = \frac{342}{73} , \quad \mathcal{I}_{7,4}  = \frac{3384}{73}, &\quad& \mbox{for $M=6$, $M'=6$, $P=3$;} \nonumber \\
  \mathcal{I}_{7,4}  = \frac{238531161015}{6698} , \quad  \mathcal{I}_{3,9}  = \frac{3197447102115}{6698}, &\quad& \mbox{for $M=20$, $M'=14$, $P=10$.}
\eea

\section{RG equations for general $M$, $M'$}
\label{app:RGM}

This appendix generalizes the method of Refs.~\cite{SBV1999,VBS2000} for SU(2) spins to superspins in SU($M'|M)$. This method utilizes only gauge-invariant information contained in the superspin algebra and its representation; thus the Berry phase $\mathcal{S}_B$ (see (\ref{Z}) and (\ref{SB2})) of the supergroup \cite{Wiegmann88} is exactly accounted for by the commutation and anti-commutation relations. The RG equations obtained here reduce to those of Section~\ref{sec:rg} at $M=2$, $P=1$, $M'=1$.

We consider here the Hamiltonian
\begin{eqnarray}
H_{\rm imp} && = g_0 \left( c^\dagger_{\ell \alpha} \, \psi_{\alpha\ell} (0) + \mbox{H.c.} \right) + \gamma_0 S^a \, \phi_a (0) \nonumber \\
&&~~~~~~~ + \int |k|^r dk \, k \, \psi_{k\alpha \ell}^\dagger \psi_{k \alpha \ell} + \frac{1}{2} \int d^d x \left[ \pi_a^2 + (\partial_x \phi_a)^2 \right] \,, \label{HimpM}
\end{eqnarray}
where $c_{\ell \alpha}$ and $S^a$ are defined in (\ref{defc}) and (\ref{STM}), and we impose exactly the constraint (\ref{constraintM}). Recall, the indices $\alpha,\beta = 1 \ldots M$, $\ell = 1 \ldots M'$, and $a= 1 \ldots M^2 -1$.

The setup of the renormalization factors in the present perturbation theory is somewhat different from (\ref{eq:renorm_fact}). We now write, using the operators defined in (\ref{defc}) and (\ref{STM}), 
\begin{equation}
\label{eq:renorm_fact_m}
S\up{a}=\sqrt{Z_{S}} S\up{a}_{R}\,, ~~~ c_{\ell \alpha}=\sqrt{Z_{c}}c_{R,\ell\alpha} \,, ~~~
\gam= \frac{\mu\up{\ep/2} \widetilde{Z}_{\gamma}}{\sqrt{Z_{S} \widetilde{S}_{d+1}}} \gamr \,, ~~~
\gc= \frac{\mu\up{\rb} \widetilde{Z}_{g}}{\sqrt{Z_{c} \Gamma(r+1)}} \gcr \,.
\end{equation}
The renormalization constants $Z_S$ and $Z_c$ are the same as those defined in (\ref{eq:sc_renorm}), but we will now compute them in a different manner. The notation of our renormalization constants also differs from that in Ref.~\onlinecite{SS2001}, and we provide a translation in Table~\ref{tab:translate}.
\begin{table}[h]
\begin{tabular}{c|c}
Ref.~\onlinecite{SS2001}~ & ~Present paper \\
\hline
$Z_h$ & $Z_f$ \\
$\widetilde{Z}_\gamma$ & $Z_\gamma$ \\
$Z$ & $Z_\phi = 1$ \\
$Z'$ & $Z_S$ \\
$Z_\gamma$ & $\widetilde{Z}_\gamma = 1$ \\
\hline
\end{tabular}
\caption{Correspondence between the notations of Ref.~\onlinecite{SS2001} and the present paper.}
\label{tab:translate}
\end{table}
Unlike Ref.~\onlinecite{SS2001}, we do not have bulk interactions of the bosonic bath field $\phi$, and hence we have $Z_\phi = 1$, as we noted in Section~\ref{sec:betafunctions}. For the same reasons, it was argued in Refs.~\cite{SBV1999,VBS2000,SS2001} that (in our notation) $\widetilde{Z}_\gamma=1$ in the absence of bulk interactions. The reasoning extends also to the fermionic bath, and so we have $\widetilde{Z}_\gamma = 1$ and $\widetilde{Z}_g =1$ exactly.
These identities can also be understood from the statement below (\ref{eq:sc_renorm}) that the vertex corrections $\Lambda_{S,c}$ arise from the same diagrams as $Z_{\gamma,g}$.
We will now compute $Z_S$ and $Z_c$ by renormalizing the two-point correlators of $S^a$ and $c_{\ell \alpha}$, and this will sufficient to obtain the needed beta functions.

\subsection{Spin correlator}

This subsection evaluates the spin correlator, $\langle O_{1} \rangle \equiv \langle S\up{a}(\tau) S\up{a}(0) \rangle$, and its renormalization will yield $Z_{S}$. We follow the strategy of Ref.~\onlinecite{VBS2000}:
use time-ordered perturbation theory to expand the correlator in powers of $g_0$ and $\gamma_0$, insert the two-point correlators of the bulk fields, and then explicitly evaluate the traces over the $S^a$ and $c_{\ell \alpha}$ operators using the superspin algebra described in Appendix~\ref{app:super}. This effectively exactly evaluates the path integral over the Berry phase $\mathcal{S}_B$ in (\ref{Z}).

\begin{figure}[t]
\centering
\subfloat[]{\includegraphics[width=0.12\textwidth]{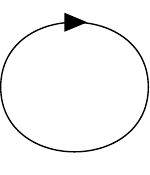}} ~~~
\subfloat[]{\includegraphics[width=0.12\textwidth]{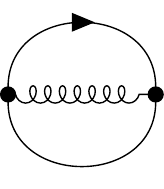}} ~~~
\subfloat[]{\includegraphics[width=0.12\textwidth]{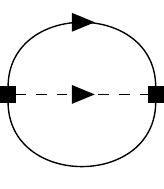}} ~~~
\subfloat[]{\includegraphics[width=0.12\textwidth]{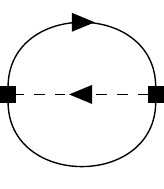}}
\caption{Diagrams contributing to the denominator, $D$ (Eq. (\ref{eq:d})), of $\langle O_{1} \rangle = \langle S\up{a}(\tau) S\up{a}(0) \rangle$. The oriented line denotes the trajectory of the SU($M'|M$) superspin in imaginary time, a filled circle is a $\gam$ vertex, and a filled square is a $\gc$ vertex. The spiral curve denotes the $\phi$ propagator, and the dashed curve represents the $\psi$ propagator.}
\label{fig:denom}
\end{figure}

\begin{figure}[t]
\centering
\subfloat[]{\includegraphics[width=0.12\textwidth]{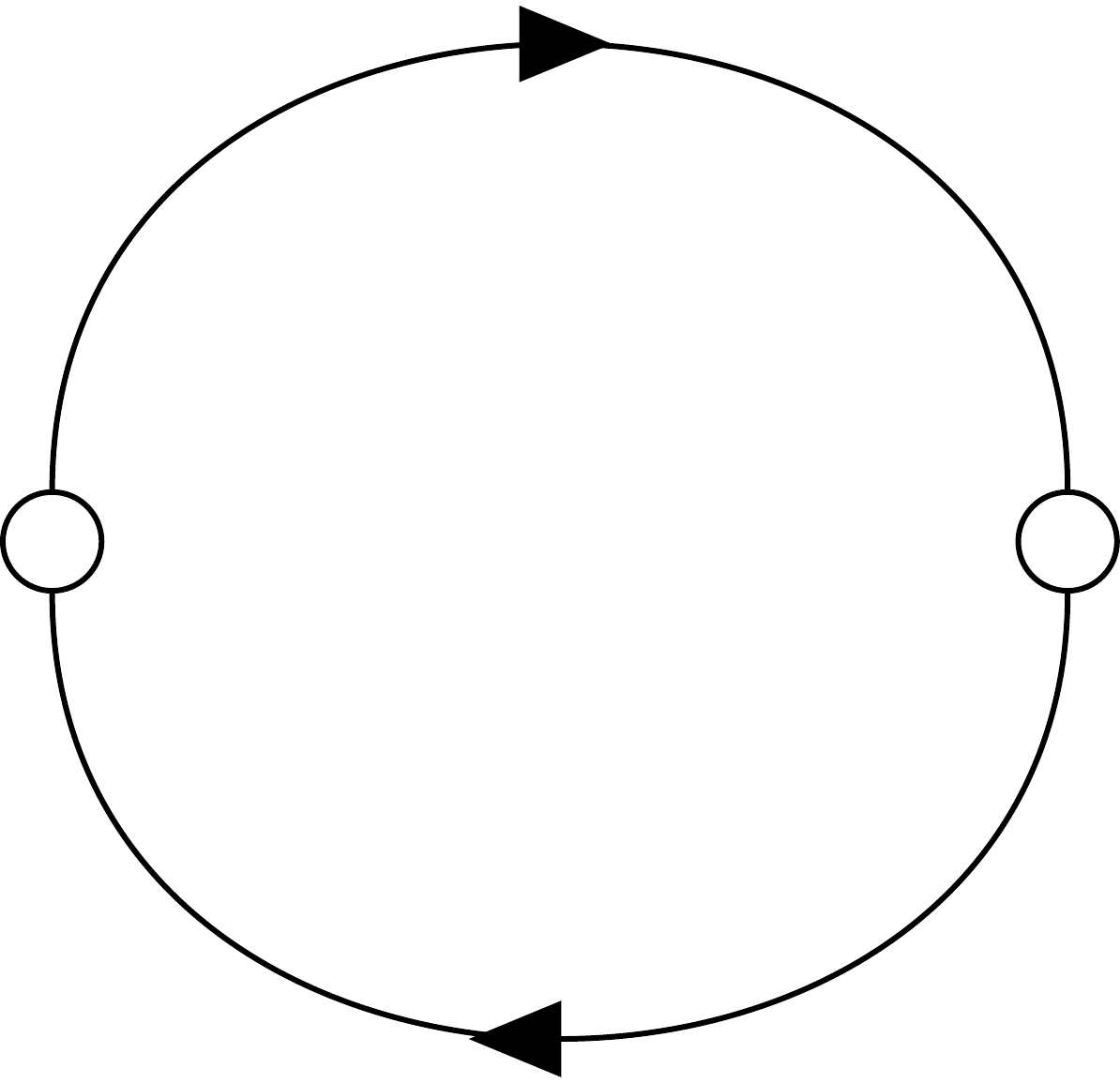}} ~~~
\subfloat[]{\includegraphics[width=0.12\textwidth]{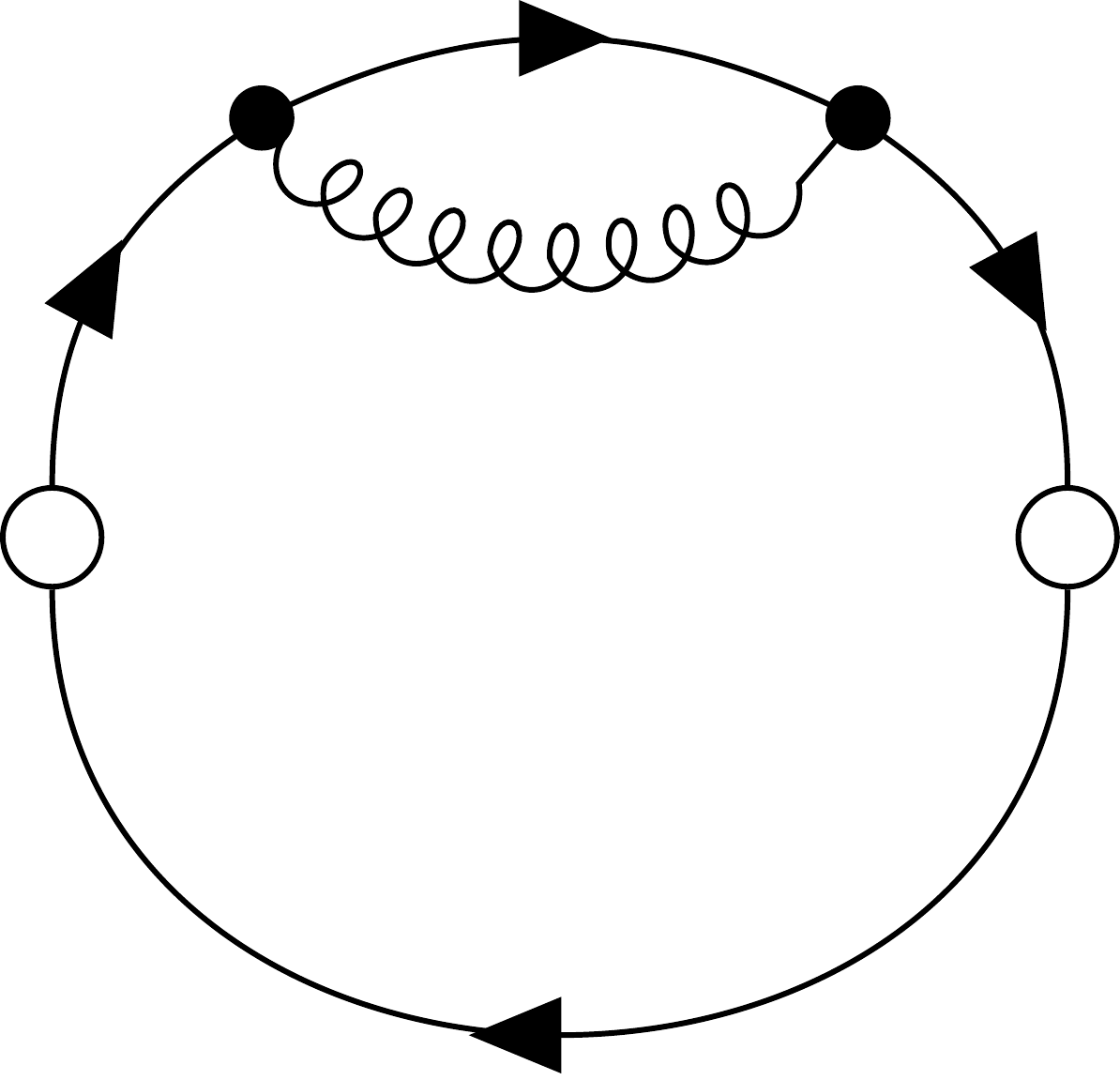}} ~~~
\subfloat[]{\includegraphics[width=0.12\textwidth]{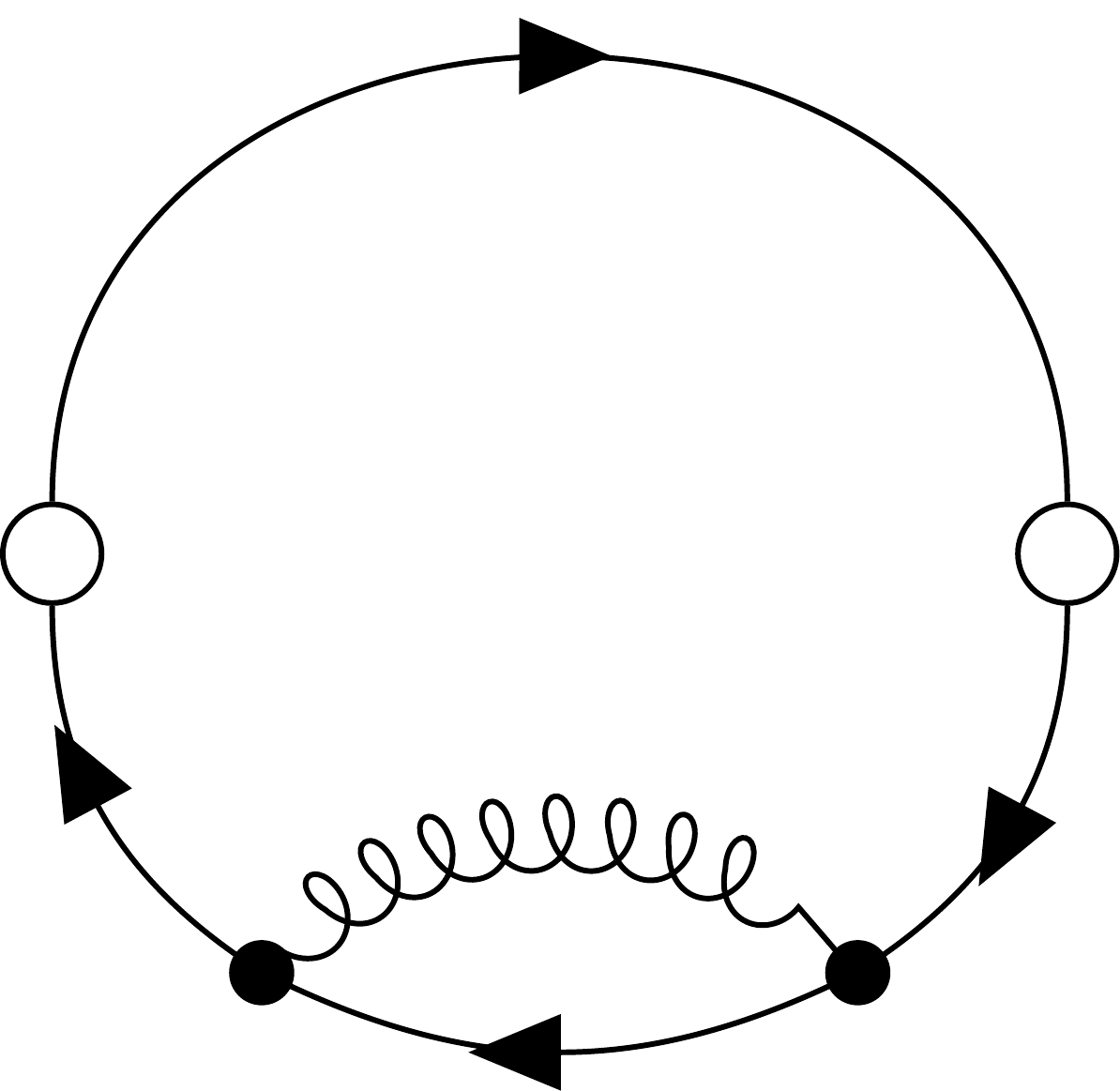}} ~~~
\subfloat[]{\includegraphics[width=0.12\textwidth]{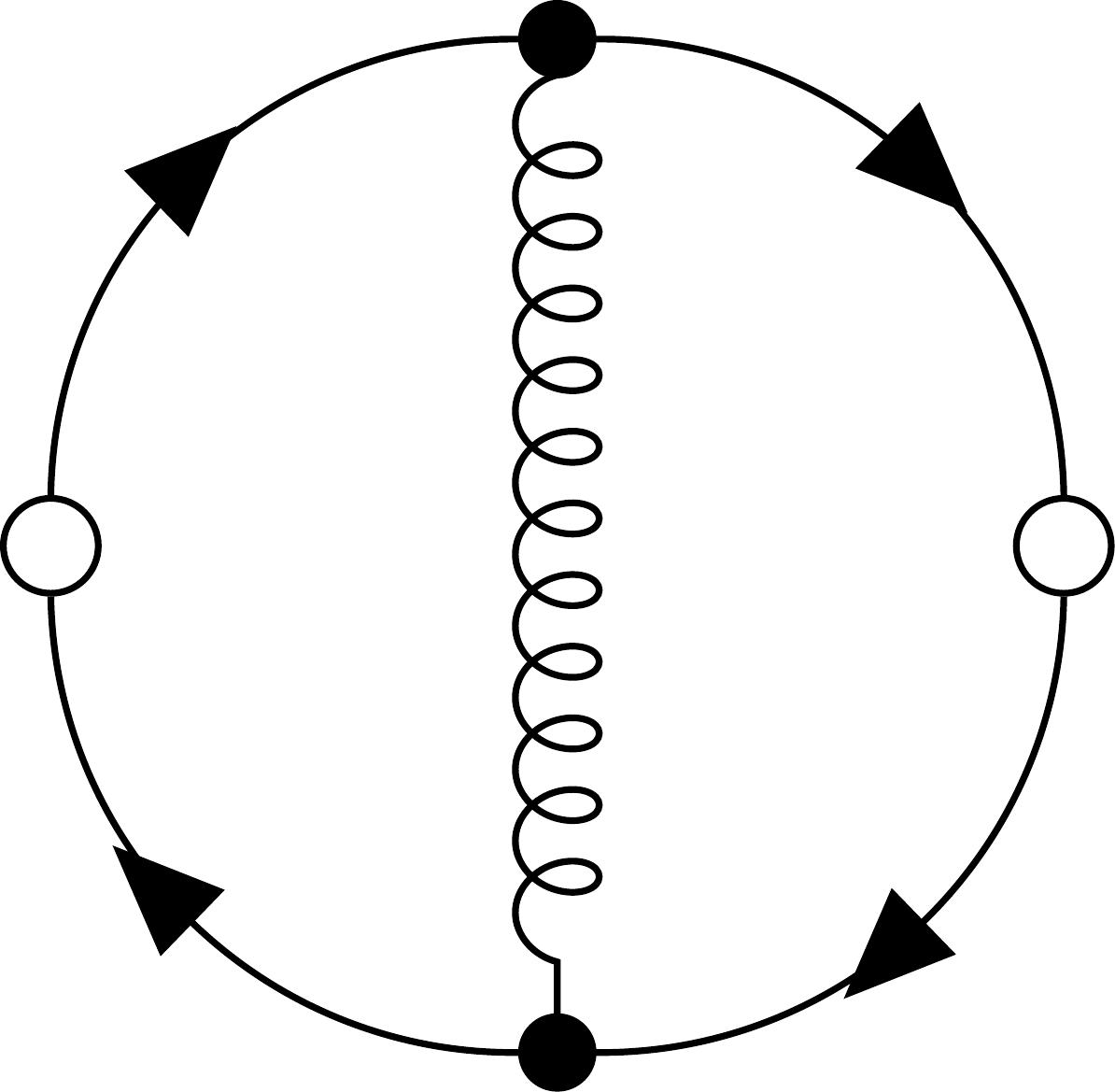}} ~~~
\subfloat[]{\includegraphics[width=0.12\textwidth]{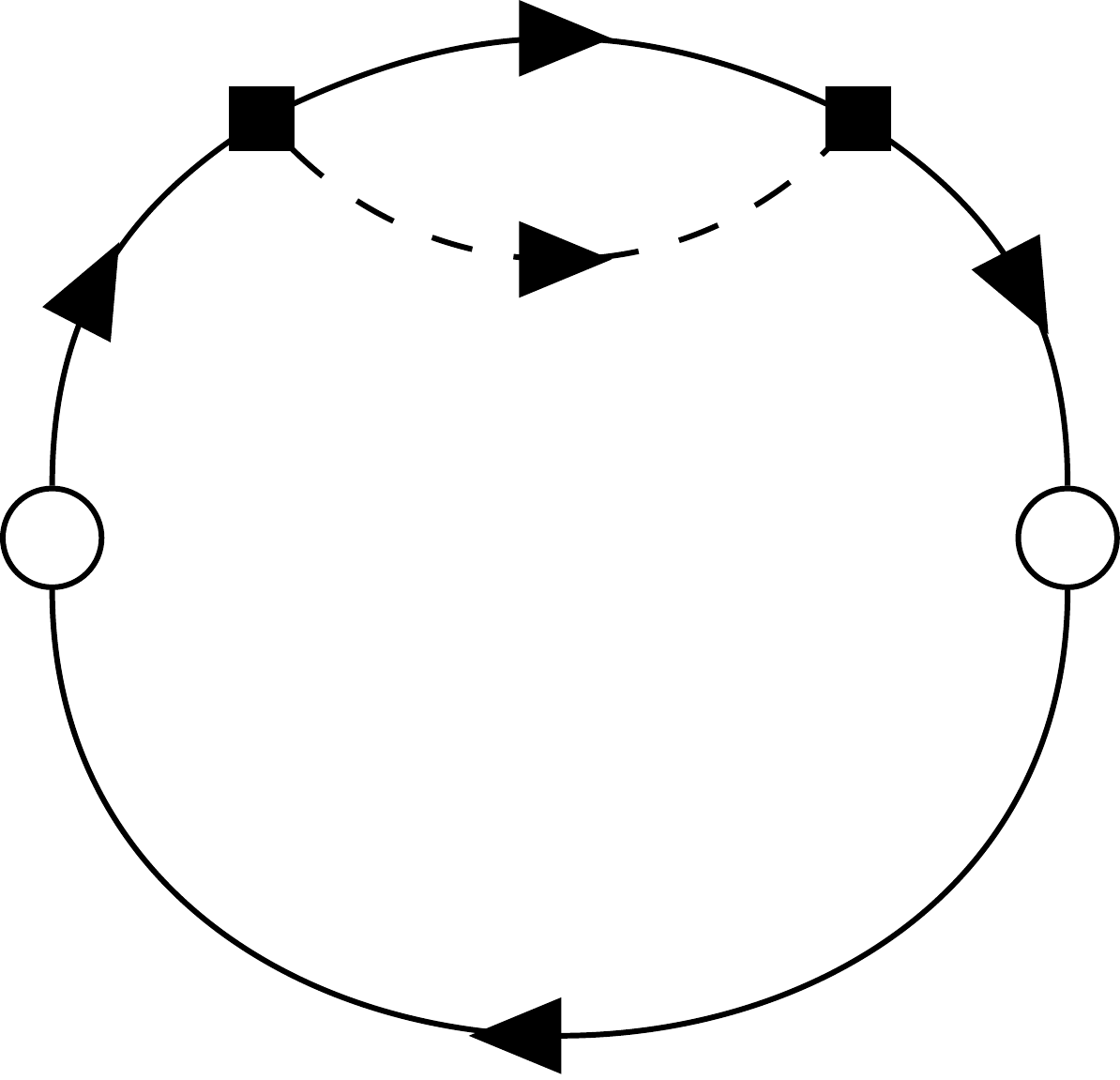}} \\
\subfloat[]{\includegraphics[width=0.12\textwidth]{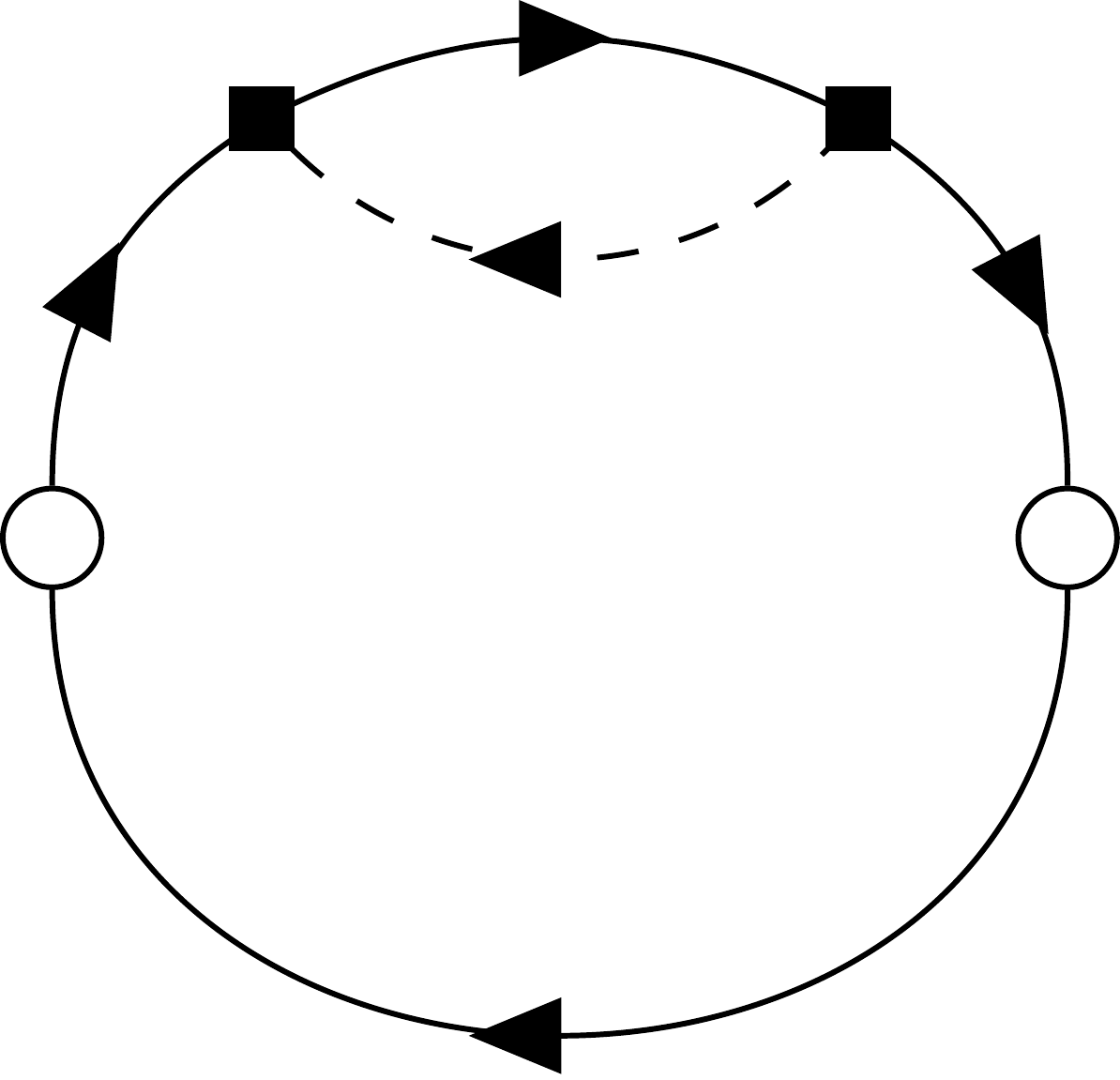}} ~~~
\subfloat[]{\includegraphics[width=0.12\textwidth]{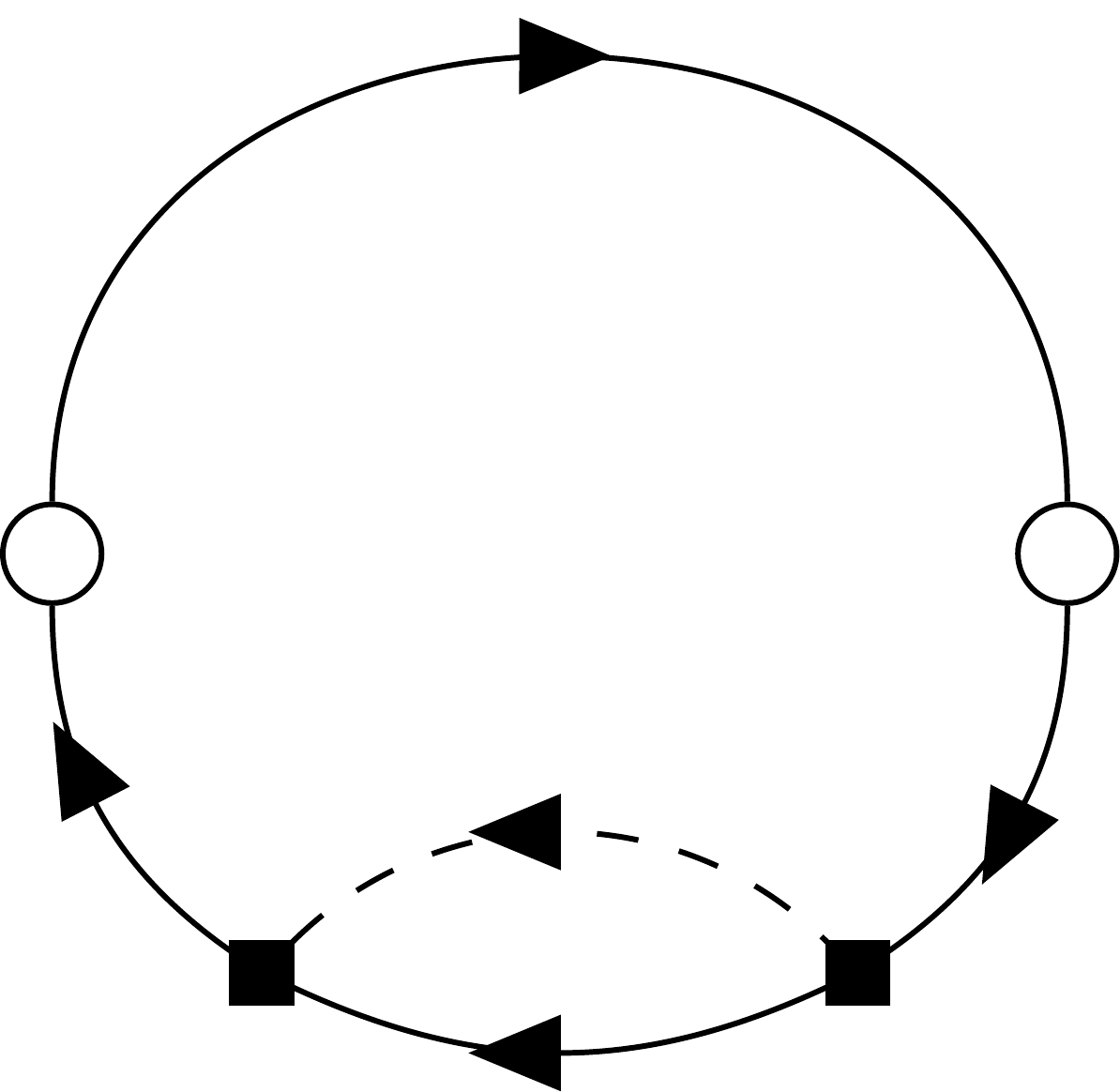}} ~~~
\subfloat[]{\includegraphics[width=0.12\textwidth]{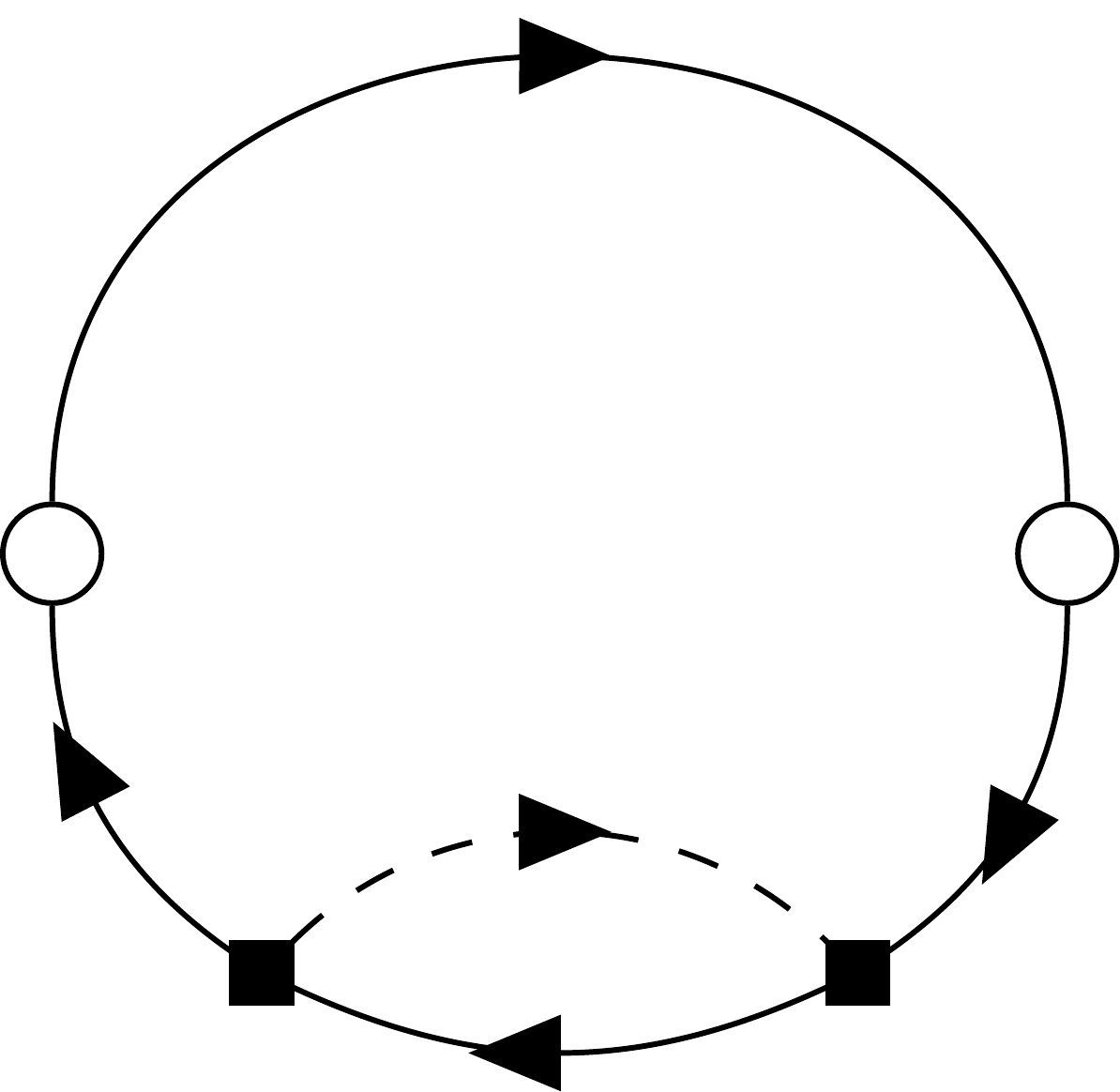}} ~~~
\subfloat[]{\includegraphics[width=0.12\textwidth]{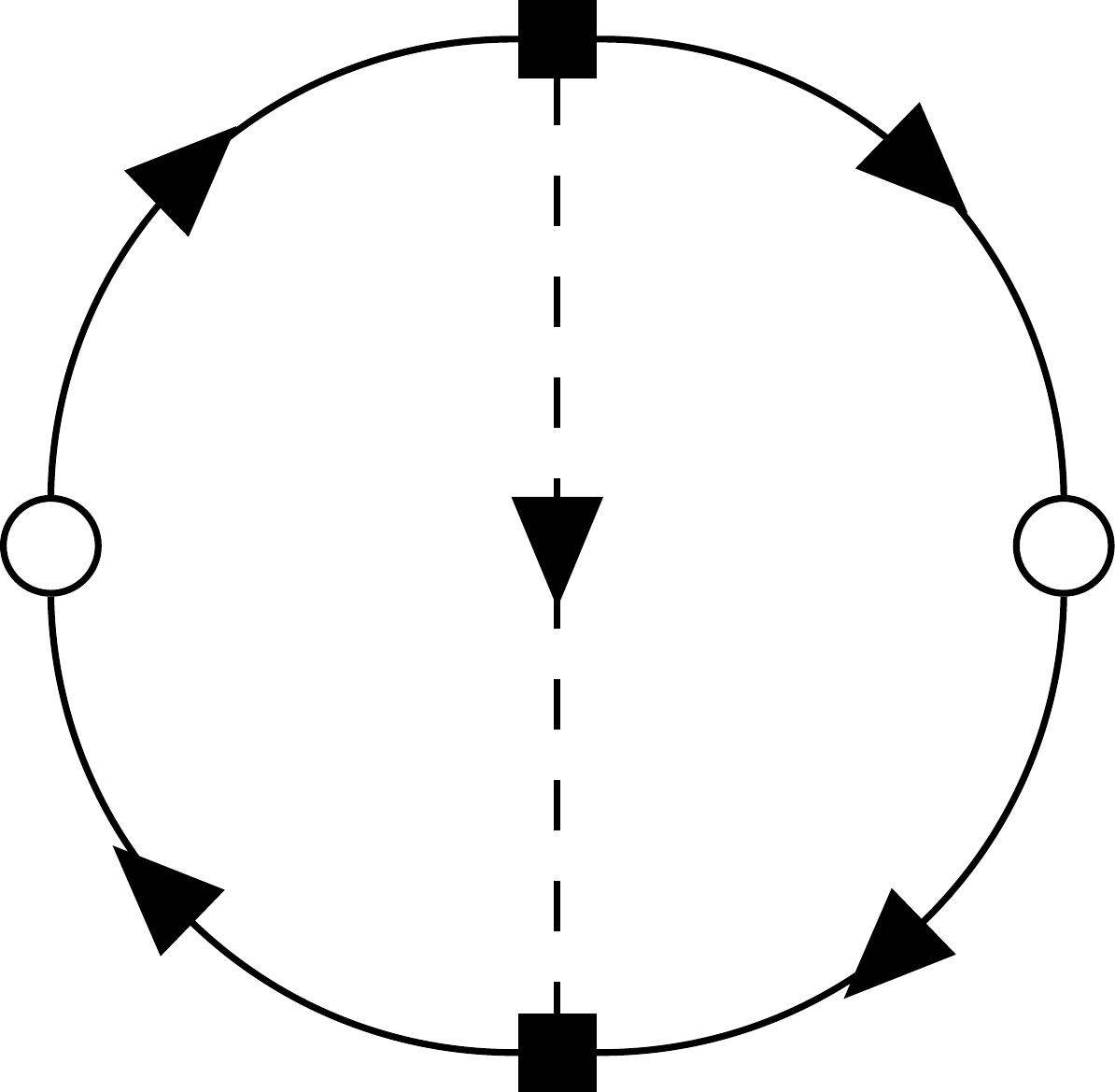}} ~~~
\subfloat[]{\includegraphics[width=0.12\textwidth]{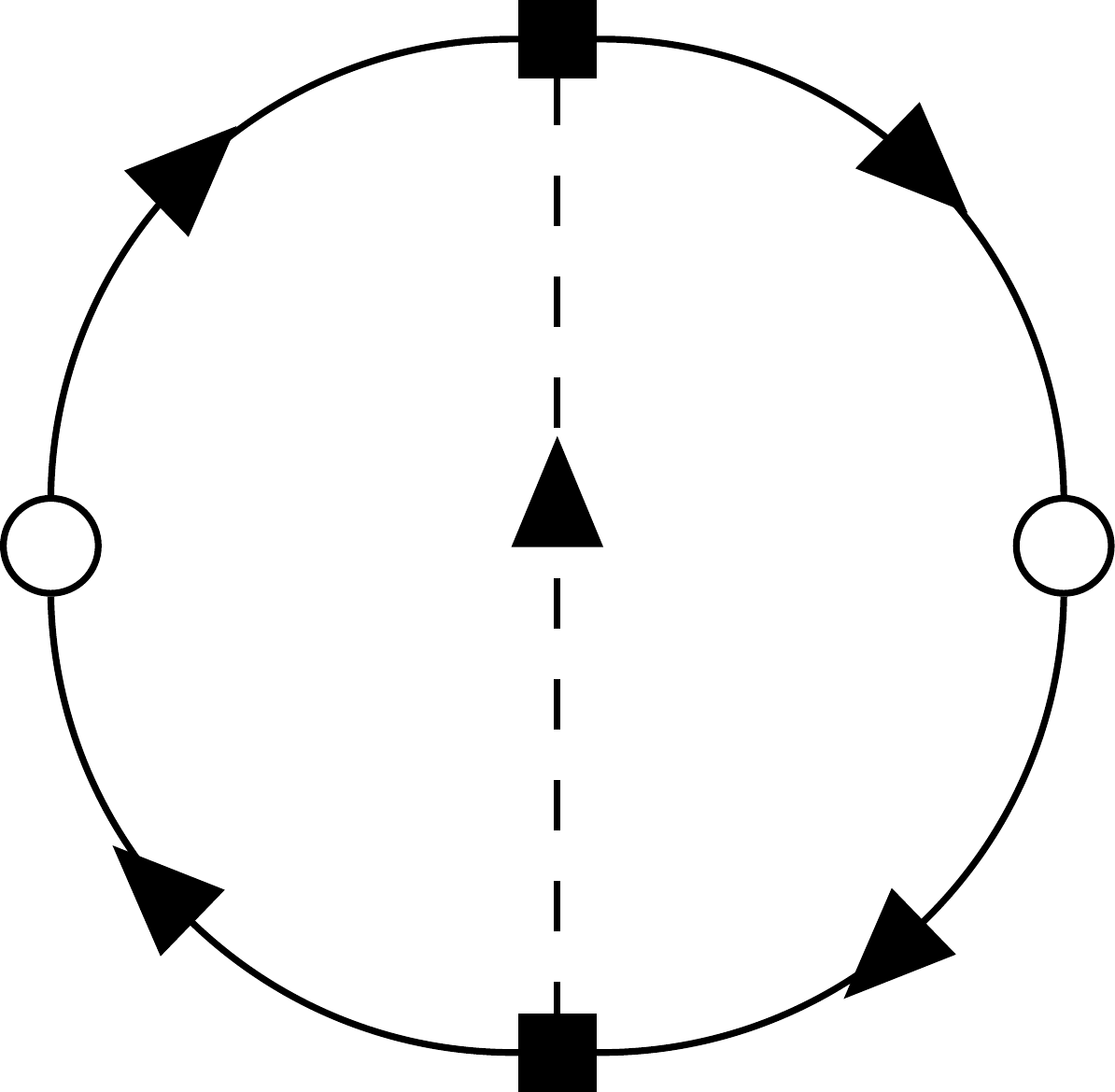}} \\
\caption{Diagrams contributing to the numerator, $N_{1}$ (Eq. (\ref{eq:n1})), of $\langle O_{1} \rangle = \langle S\up{a}(\tau) S\up{a}(0) \rangle$. Conventions as in Fig.~\ref{fig:denom}, and an open circle denotes the external $S\up{a}$ operator.} 
\label{fig:N1}
\end{figure}

We write the correlator as $\langle O_{1} \rangle = N_{1}/D$, and the perturbative expansions of the numerator and denominator are represented by the diagrams shown in Figs.~\ref{fig:denom} and \ref{fig:N1}. Note that these are not Feynman diagrams, and there is no Wick's theorem. The oriented line represents the worldline of the superspin, and the diagrams indicate the ordering of the operators whose traces are to be evaluated. The numerator and denominator have to be evaluated separately, and there is no automatic cancellation of disconnected contributions. The diagrams in Figs.~\ref{fig:denom} and \ref{fig:N1} yield
\begin{align}
%%%%
\label{eq:d}
D &= 1 + \gam\up{2} \lo \left( \Da + \Db + \Dc \right) + \gc\up{2} \lop \left( \Dap + \Dbp + \Dcp \right) \nonumber \\
&~~~~~~~~~~+ \gc\up{2} \lopp \left( \Dapp + \Dbpp + \Dcpp \right) \,, \\
%\end{align}
%\begin{align}
%%%%
\label{eq:n1}
N_{1} &= \lo + \gam\up{2} \left( \la\Da + \lb\Db + \lc\Dc \right) 
+ \gc\up{2} \left( \lap\Dap + \lbp\Dbp + \lcp\Dcp \right) \nonumber \\
&~~~~~~~~+ \gc\up{2} \left( \lapp\Dapp + \lbpp\Dbpp + \lcpp\Dcpp \right) \,,
\end{align}
where the average over the supergroup representation $\left\langle \mathcal{O} \right\rangle \equiv \left( \mbox{Tr} \mathcal{O} \right)/\left( \mbox{Tr} \mathbbm{1} \right)$, $ \mbox{Tr} \mathbbm{1} = \mathcal{D}(M,M',P)$, is carried out by the expressions
\begingroup
\allowdisplaybreaks
\begin{align}
\lo &= \left\langle S\up{a} S\up{a} \right\rangle = \frac{M+1}{2M} (M \mathcal{I}_{1,0} - \mathcal{I}_{2,0}) \,, \\
\lop &= \left\langle c_{\ell \alpha} c\up{\dagger}_{\ell \alpha}  \right\rangle = - \mathcal{I}_{1,1} + M \mathcal{I}_{0,1} \,, \\
\lopp &= \left\langle c\up{\dagger}_{\ell \alpha} c_{\ell \alpha}  \right\rangle = M' \mathcal{I}_{1,0} + \mathcal{I}_{1,1} \,, \\
\la &= \left\langle S\up{a} S\up{b} S\up{b} S\up{a}  \right\rangle = \frac{(M+1)\up{2}}{4M\up{2}}(M\up{2} \mathcal{I}_{2,0} 
+ \mathcal{I}_{4,0} - 2M \mathcal{I}_{3,0}) \,, \\
\lb &= \left\langle S\up{a} S\up{a} S\up{b} S\up{b}  \right\rangle = \frac{(M+1)\up{2}}{4M\up{2}}(M\up{2} \mathcal{I}_{2,0} 
+ \mathcal{I}_{4,0} - 2M \mathcal{I}_{3,0}) \,, \\
\lc &= \left\langle S\up{a} S\up{b} S\up{a} S\up{b}  \right\rangle = \frac{M+1}{4M\up{2}}(-M\up{3}\mathcal{I}_{1,0} + M\up{2}(M+2)\mathcal{I}_{2,0} 
-2M(M+1)\mathcal{I}_{3,0} \nonumber \\
&~~~~~~~~~~~~~~~~~~~~~~~~~+ (M+1)\mathcal{I}_{4,0}) \,,\\
\lap &= \left\langle S\up{a} c_{\ell \alpha} c_{\ell \alpha}\up{\dagger} S\up{a}  \right\rangle = -\frac{M+1}{2M}(2M\mathcal{I}_{2,1} 
- M\up{2} \mathcal{I}_{1,1} - \mathcal{I}_{3,1}) \,, \\
\lbp &= \left\langle S\up{a} S\up{a} c_{\ell \alpha} c\up{\dagger}_{\ell \alpha}  \right\rangle = -\frac{M+1}{2M}(2M\mathcal{I}_{2,1} 
- M\up{2} \mathcal{I}_{1,1} - \mathcal{I}_{3,1}) \,, \\
\lcp &= \left\langle S\up{a} c_{\ell \alpha} S\up{a} c\up{\dagger}_{\ell \alpha}  \right\rangle = -\frac{M+1}{2M}((2M-1)\mathcal{I}_{2,1} 
- M(M-1)\up{2} \mathcal{I}_{1,1} - \mathcal{I}_{3,1}) \,, \\
\lapp &= \left\langle S\up{a} c\up{\dagger}_{\ell \alpha} c_{\ell \alpha} S\up{a}  \right\rangle = \frac{M+1}{2M}(MM' \mathcal{I}_{2,0} 
-M' \mathcal{I}_{3,0} + M\mathcal{I}_{2,1} - \mathcal{I}_{3,1}) \,, \\
\lbpp &= \left\langle S\up{a} S\up{a} c\up{\dagger}_{\ell \alpha} c_{\ell \alpha}  \right\rangle = \frac{M+1}{2M}(MM' \mathcal{I}_{2,0} 
-M' \mathcal{I}_{3,0} + M\mathcal{I}_{2,1} - \mathcal{I}_{3,1}) \,, \\
\lcpp &= \left\langle S\up{a} c\up{\dagger}_{\ell \alpha} S\up{a} c_{\ell \alpha}  \right\rangle = \frac{M+1}{2M}(M'(M+1) \mathcal{I}_{2,0} 
-MM'\mathcal{I}_{1,0} - M' \mathcal{I}_{3,0} + (M+1)\mathcal{I}_{2,1} \nonumber \\
&~~~~~~~~~~~~~~~~~~~~~~~~~~~~~~- M\mathcal{I}_{1,1} -\mathcal{I}_{3,1}) \,. 
\end{align}
\endgroup
%%%%%

%%%%%%%%%%%%%%%%%%%%%%%%%%%%%%%%%%%%%%
%%%%%%%%%%%%%%%%%%%%%%%%%%%%%%%%%%%%%%

Also,
\begingroup
\allowdisplaybreaks
\begin{align}
`\Da &= \int_{0}\up{\tau} d\tau_{1} \int_{\tau_{1}}\up{\tau} d\tau_{2} G_{\phi} (\tau_{1} - \tau_{2})  
= - \frac{\widetilde{S}_{d+1} \tau\up{\ep}}{\ep (1-\ep)} \,, \\
\Db &= \int_{\tau}\up{\beta} d\tau_{1} \int_{\tau_{1}}\up{\beta} d\tau_{2} G_{\phi} (\tau_{1} - \tau_{2}) 
= - \frac{\widetilde{S}_{d+1} \tau\up{\ep}}{\ep (1-\ep)} \,, \\
\Dc &= \int_{0}\up{\tau} d\tau_{1} \int_{\tau}\up{\beta} d\tau_{2} G_{\phi} (\tau_{1} - \tau_{2})  
= \frac{2 \widetilde{S}_{d+1} \tau\up{\ep}}{\ep (1-\ep)} \,, \\
%\end{align}%
%
%\begin{align}
\Dap &= \int_{0}\up{\tau} d\tau_{1} \int_{\tau_{1}}\up{\tau} d\tau_{2} G_{\psi} (\tau_{2} - \tau_{1})  
= - \frac{\Gamma(r+1) \tau\up{2\rb}}{2\rb (1-2\rb)} \,, \\
\Dbp &= \int_{\tau}\up{\beta} d\tau_{1} \int_{\tau_{1}}\up{\beta} d\tau_{2} G_{\psi} (\tau_{2} - \tau_{1}) 
= - \frac{\Gamma(r+1) \tau\up{2\rb}}{2\rb (1-2\rb)} \,, \\
\Dcp &= \int_{0}\up{\tau} d\tau_{1} \int_{\tau}\up{\beta} d\tau_{2} G_{\psi} (\tau_{2} - \tau_{1}) 
= \frac{2 \Gamma(r+1) \tau\up{2\rb}}{2\rb (1-2\rb)} \,, \\
%%%%%%%%%%
\Dapp &= -\int_{0}\up{\tau} d\tau_{1} \int_{\tau_{1}}\up{\tau} d\tau_{2} G_{\psi} (\tau_{1} - \tau_{2})  
= - \frac{\Gamma(r+1) \tau\up{2\rb}}{2\rb (1-2\rb)} \,, \\
\Dbpp &= -\int_{\tau}\up{\beta} d\tau_{1} \int_{\tau_{1}}\up{\beta} d\tau_{2} G_{\psi} (\tau_{1} - \tau_{2}) 
= - \frac{\Gamma(r+1) \tau\up{2\rb}}{2\rb (1-2\rb)} \,, \\
\Dcpp &= -\int_{0}\up{\tau} d\tau_{1} \int_{\tau}\up{\beta} d\tau_{2} G_{\psi} (\tau_{1} - \tau_{2}) 
= - \frac{2\Gamma(r+1) \tau\up{2\rb}}{2\rb (1-2\rb)} \,.
\end{align}
\endgroup 
Note we evaluate the above integrals at $T=0$, by extending the integrals appropriately as explained in Ref.~\onlinecite{VBS2000}.
Here,
\begin{equation}
G_{\phi}(\tau) = \int \frac{d\up{d}k}{(2\pi)\up{d}} \frac{d\omega}{2\pi} \frac{e\up{-i\omega \tau}}{k\up{2} + \omega\up{2}} 
= \frac{\widetilde{S}_{d+1}}{|\tau|\up{d-1}} \,.
\end{equation}
Similarly,
\begin{align}
G_{\psi}(\tau) &= \int dk |k|\up{r} \int \frac{d\omega}{2\pi} \frac{e\up{-i\omega \tau}}{i\omega - k} \nonumber \\
&= \int dk |k|\up{r} \left[ -e\up{-k\tau} \left( \theta(k) \theta(\tau) - \theta(-k) \theta(-\tau) \right)  \right] \nonumber \\
&= \frac{\Gamma(1+r)}{|\tau|\up{1+r}} \left[ \theta(-\tau) - \theta(\tau) \right] \,.
\end{align}

From (\ref{eq:d}) and (\ref{eq:n1}) we obtain,
\begin{align}
\langle O_{1} \rangle = \frac{N_{1}}{D} &= \frac{\lo}{M} \bigg \lbrace 
1 + \gam\up{2} \left[ \left( \frac{\la}{\lo} - \frac{\lo}{M} \right) \Da 
+ \left( \frac{\lb}{\lo} - \frac{\lo}{M} \right) \Db 
+\left( \frac{\lc}{\lo} - \frac{\lo}{M} \right) \Dc \right]  \nonumber \\
&+ \gc\up{2} \left[ \left( \frac{\lap}{\lo} - \frac{\lop}{M} \right) \Dap 
+ \left( \frac{\lbp}{\lo} - \frac{\lop}{M} \right) \Dbp 
+\left( \frac{\lcp}{\lo} - \frac{\lop}{M} \right) \Dcp \right] \nonumber \\
&+ \gc\up{2} \left[ \left( \frac{\lapp}{\lo} - \frac{\lopp}{M} \right) \Dapp 
+ \left( \frac{\lbpp}{\lo} - \frac{\lopp}{M} \right) \Dbpp 
+\left( \frac{\lcpp}{\lo} - \frac{\lopp}{M} \right) \Dcpp \right]
\bigg \rbrace \,.
\end{align}
It is then straightforward to write, 
\begin{equation}
\label{eq:zs_m}
Z_{S} = 1 - \frac{\gamr\up{2}}{\ep} \Lgam - \frac{\gcr\up{2}}{2\rb} \Lg \,,
\end{equation}
where ,
\begin{align}
\Lgam &= \frac{\la + \lb -2\lc}{\lo} \,, \\
\Lg &= \frac{\lap+\lapp+\lbp+\lbpp-2\lcp-2\lcpp}{\lo} \,.
\end{align}
Note that for $M=2\,, M'=1$, we obtain $\Lgam=\Lg=2$ which agrees with the result that can be obtained from (\ref{eq:zs_zc}) and the results in Section~\ref{sec:rg}. 

\subsection{Electron correlator}

\begin{figure}[t]
\centering
\subfloat[]{\includegraphics[width=0.12\textwidth]{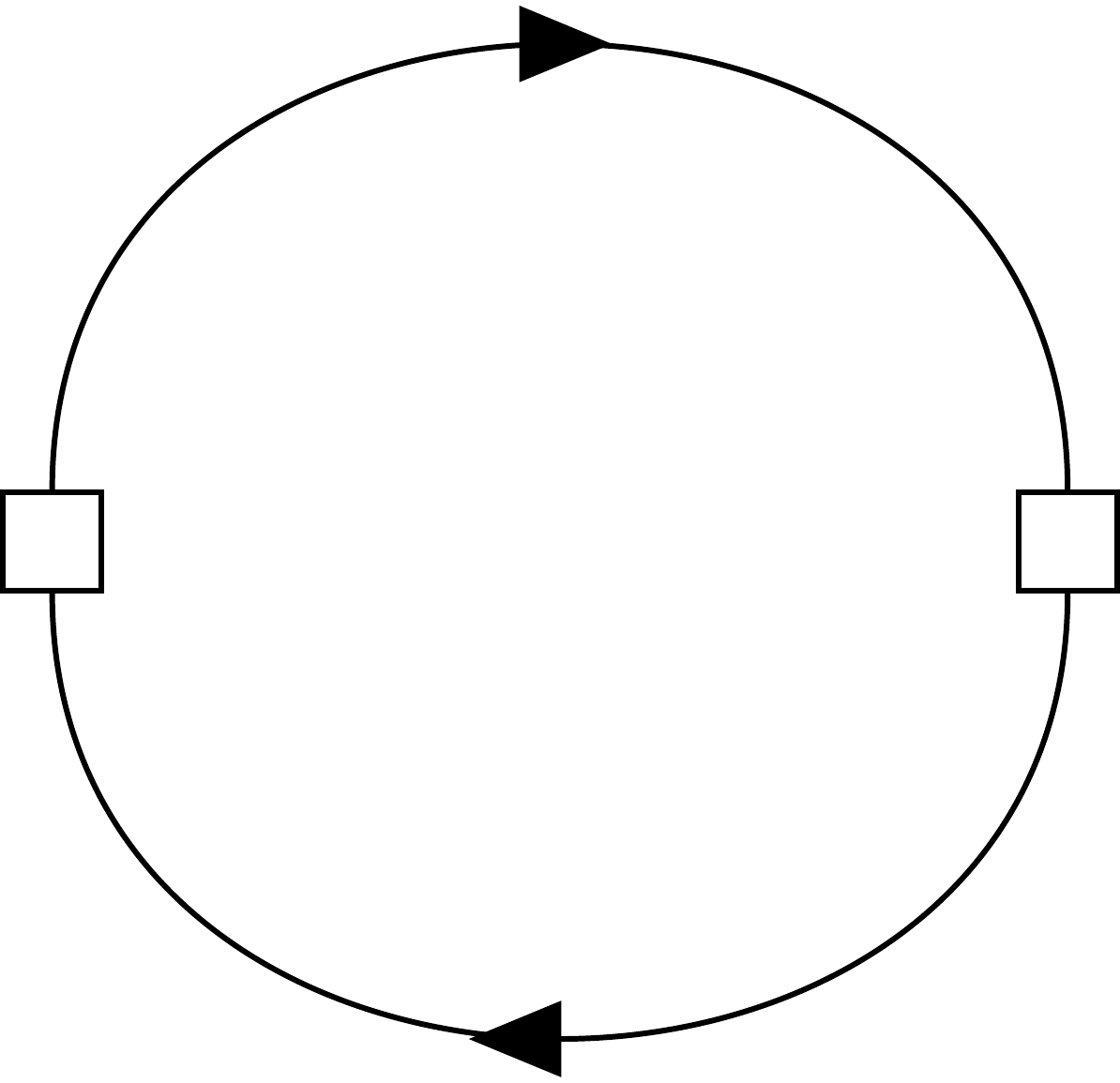}} ~~
\subfloat[]{\includegraphics[width=0.12\textwidth]{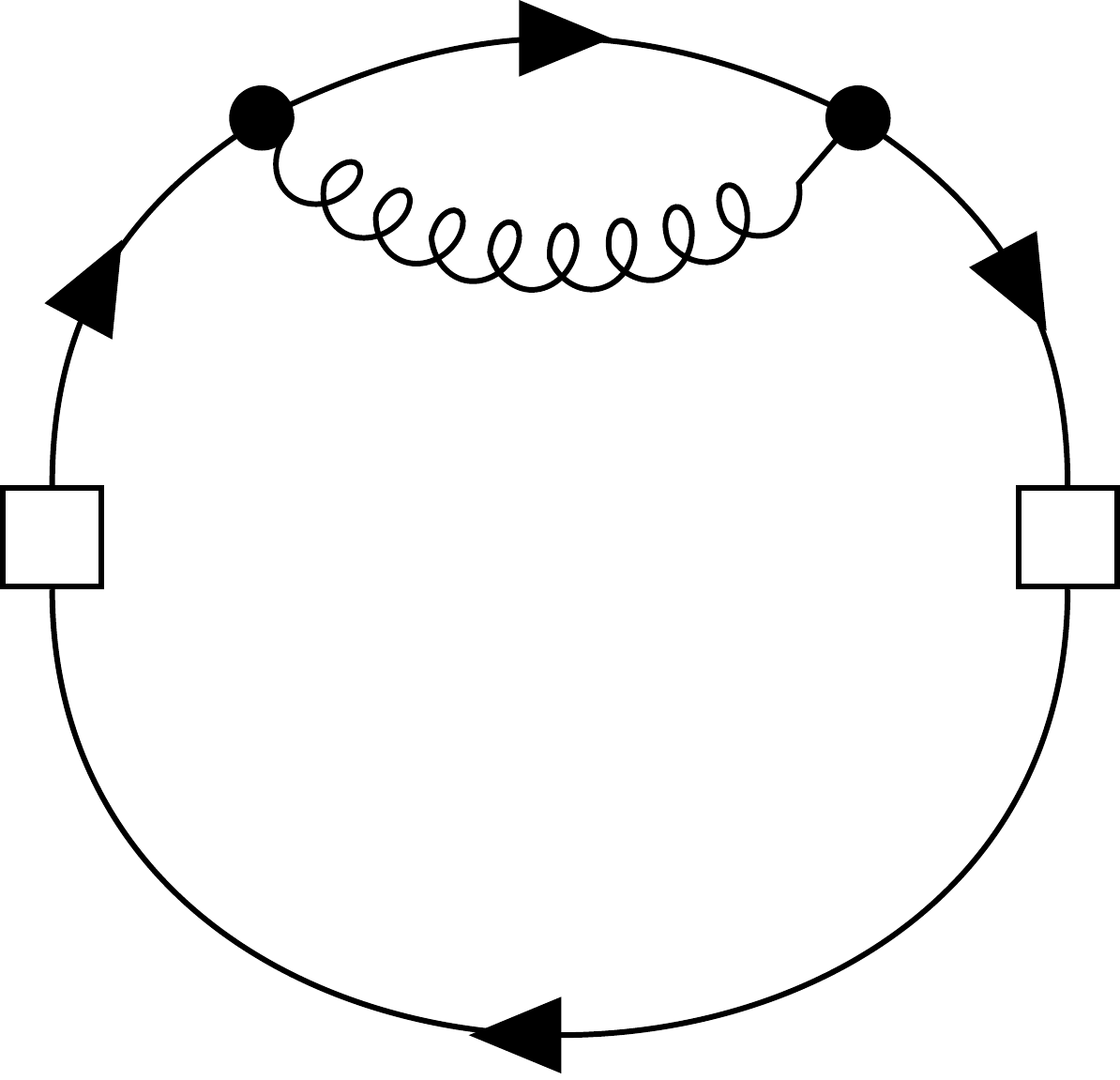}} ~~
\subfloat[]{\includegraphics[width=0.12\textwidth]{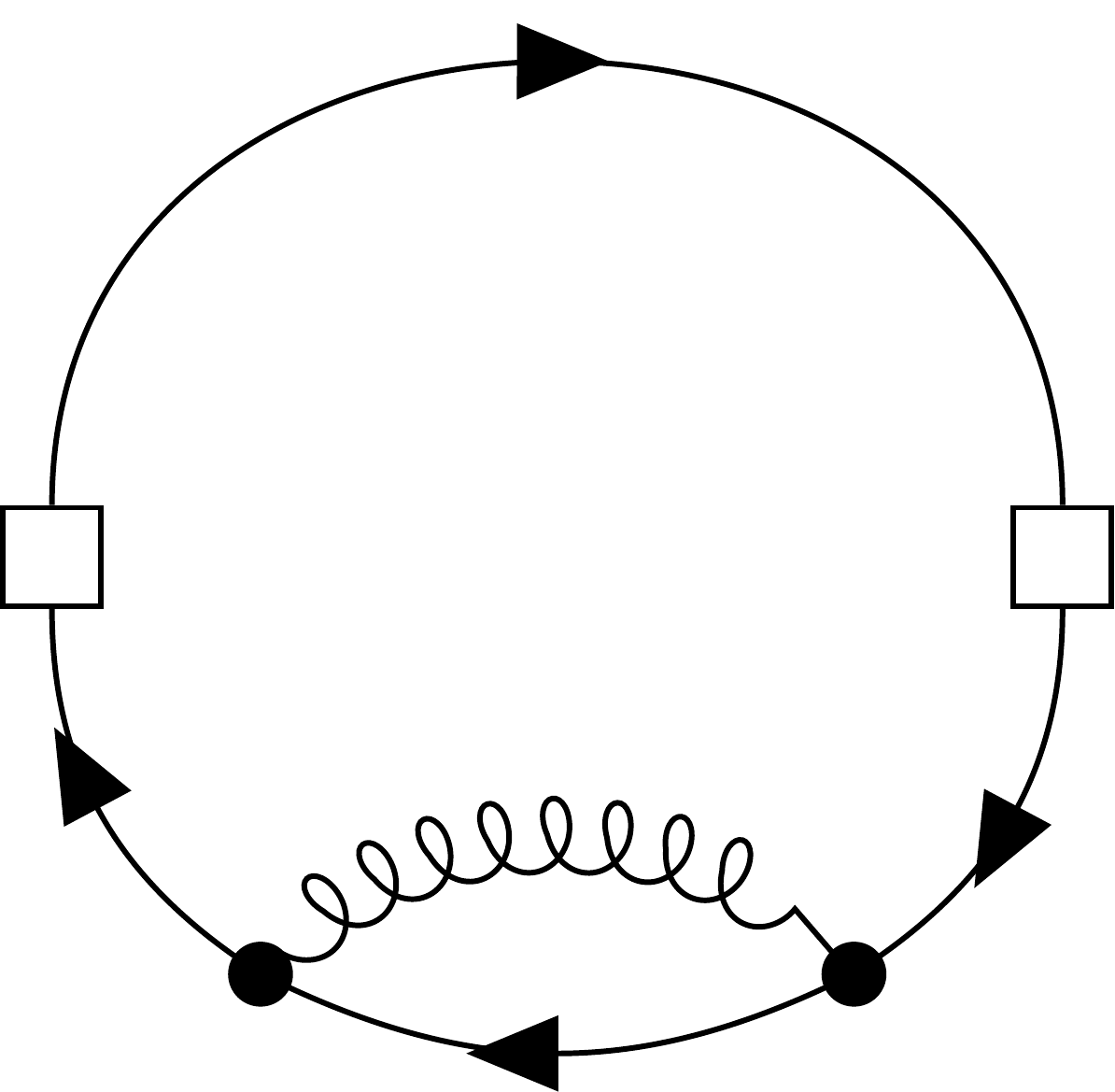}} ~~
\subfloat[]{\includegraphics[width=0.12\textwidth]{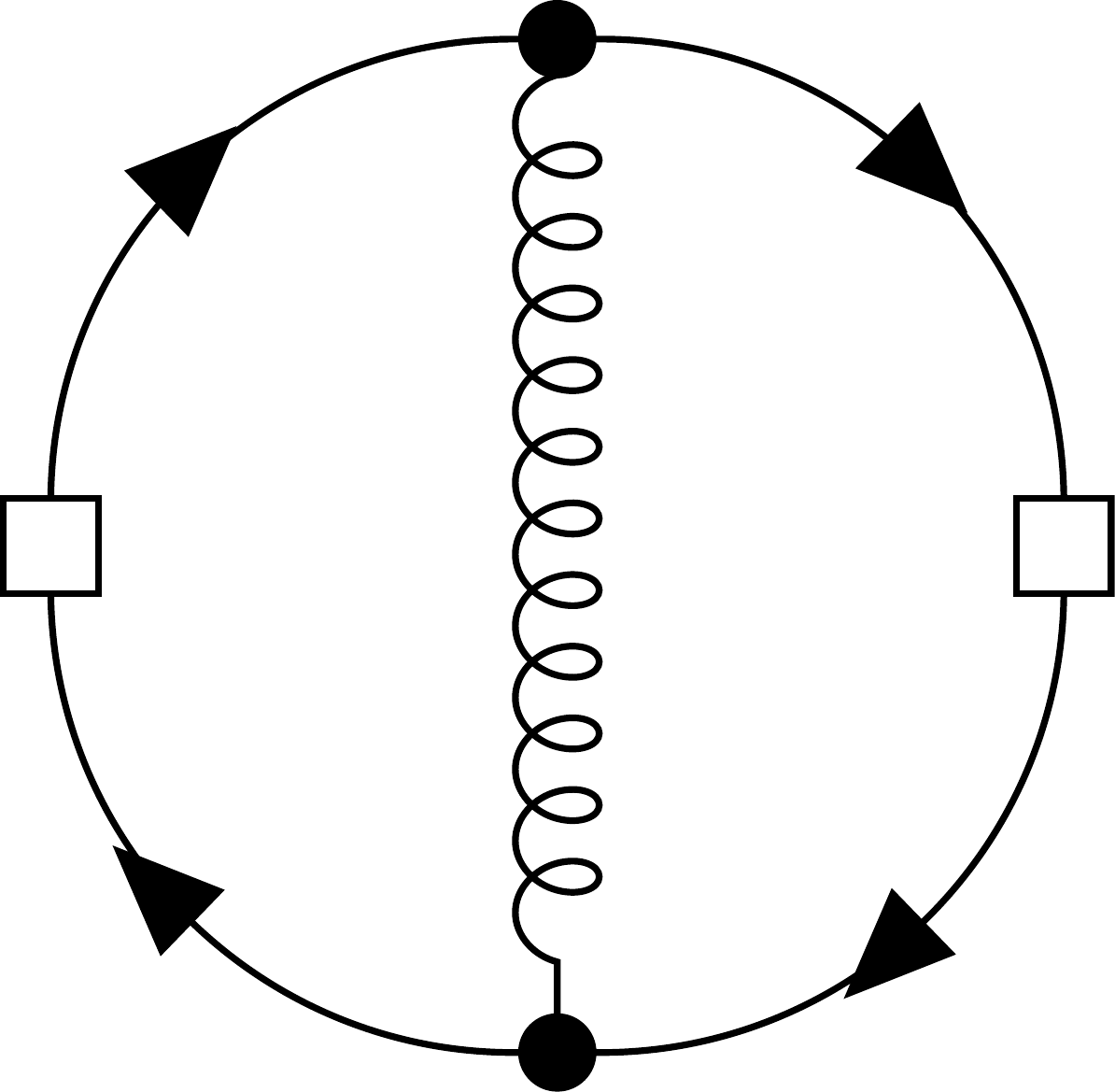}} ~~
\subfloat[]{\includegraphics[width=0.12\textwidth]{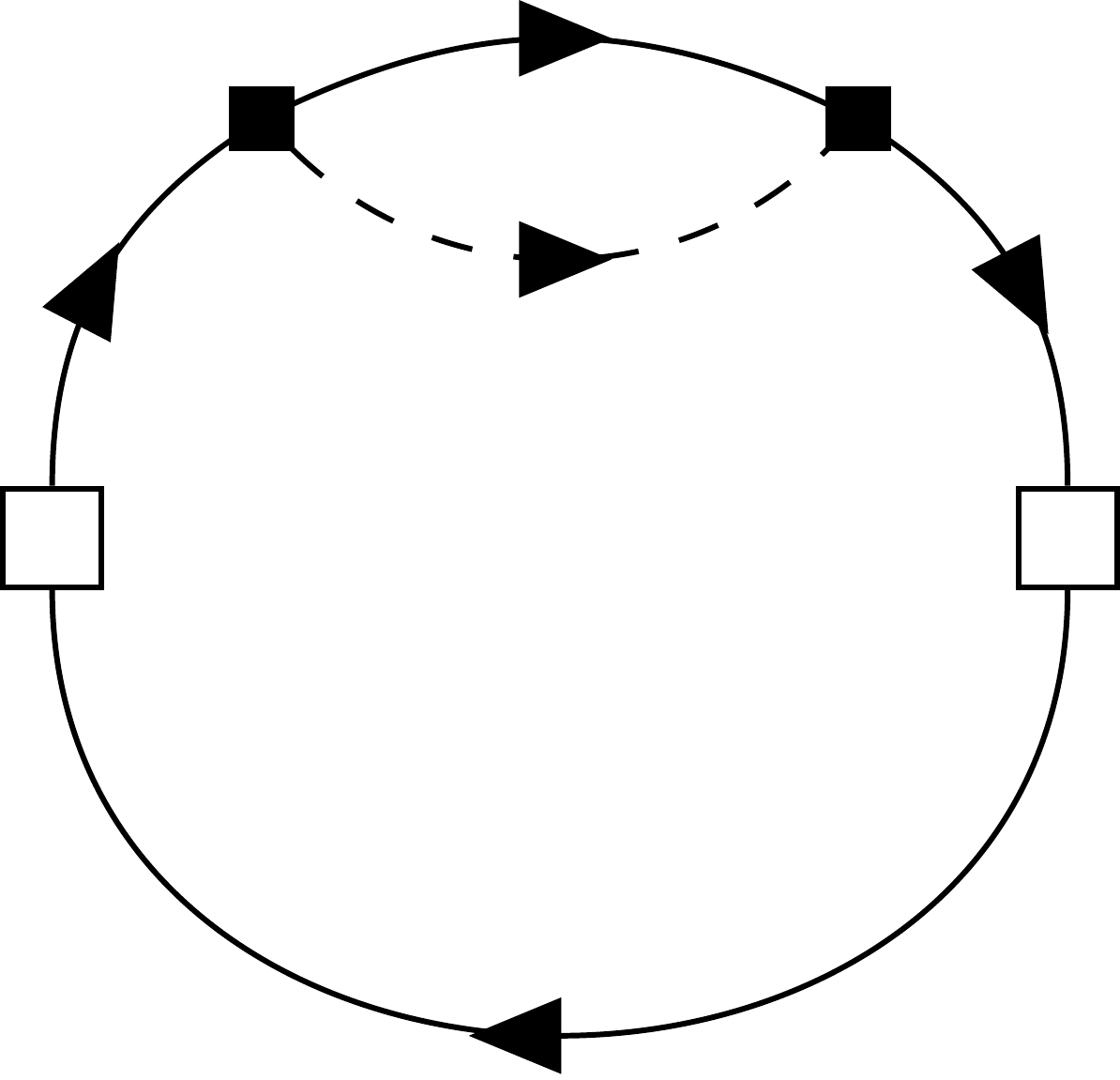}} \\
\subfloat[]{\includegraphics[width=0.12\textwidth]{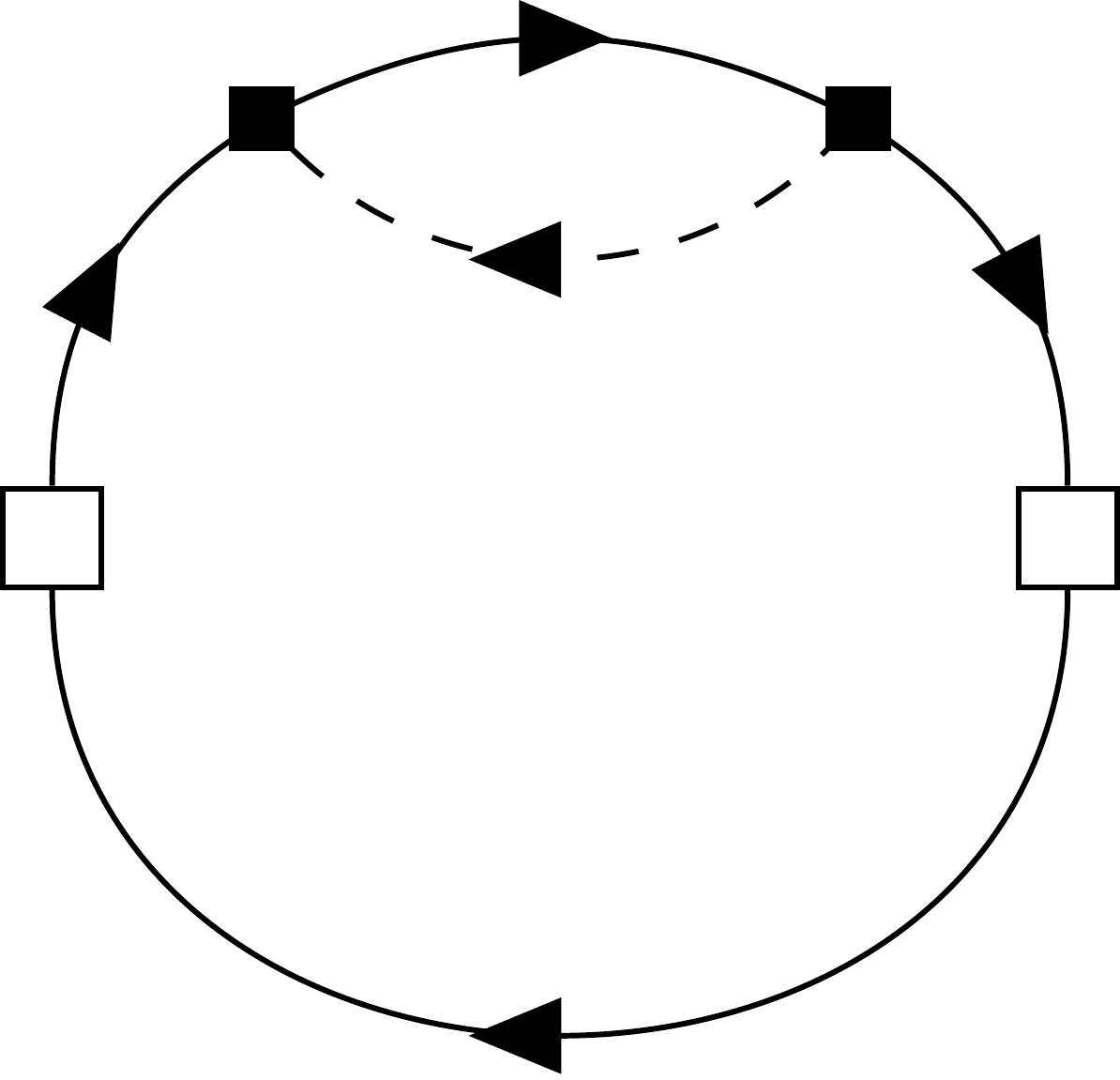}} ~~
\subfloat[]{\includegraphics[width=0.12\textwidth]{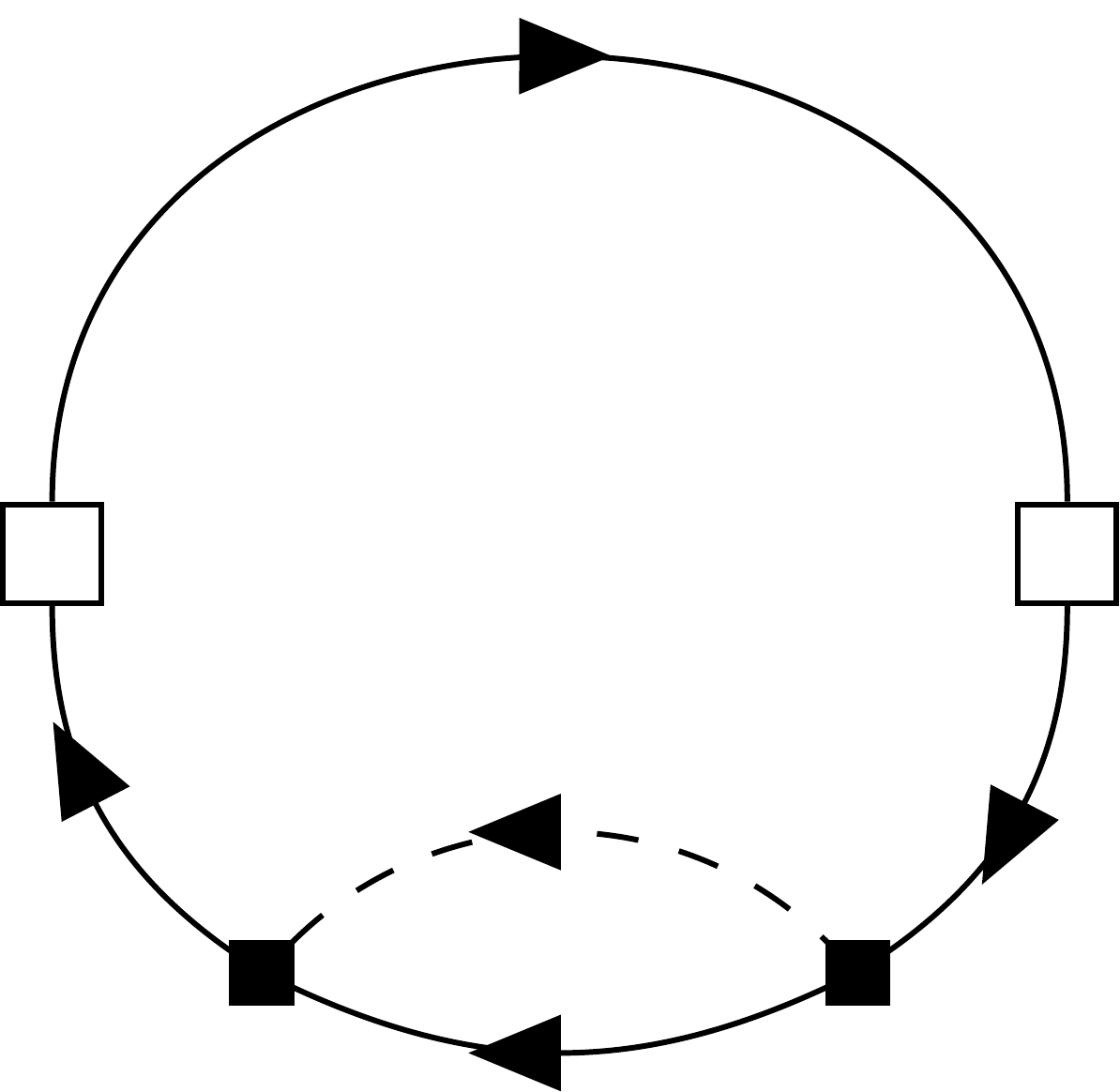}} ~~
\subfloat[]{\includegraphics[width=0.12\textwidth]{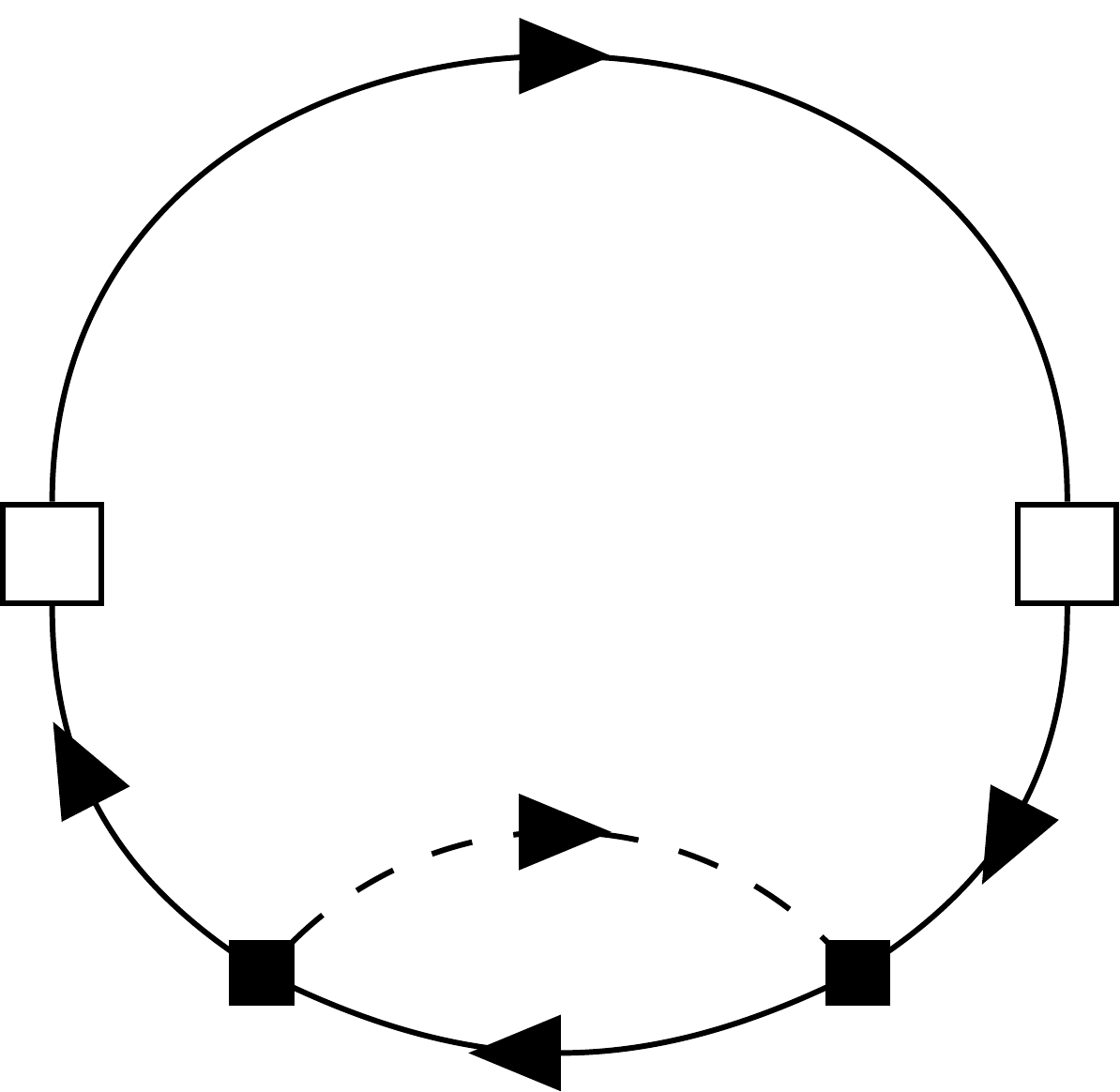}} ~~
\subfloat[]{\includegraphics[width=0.12\textwidth]{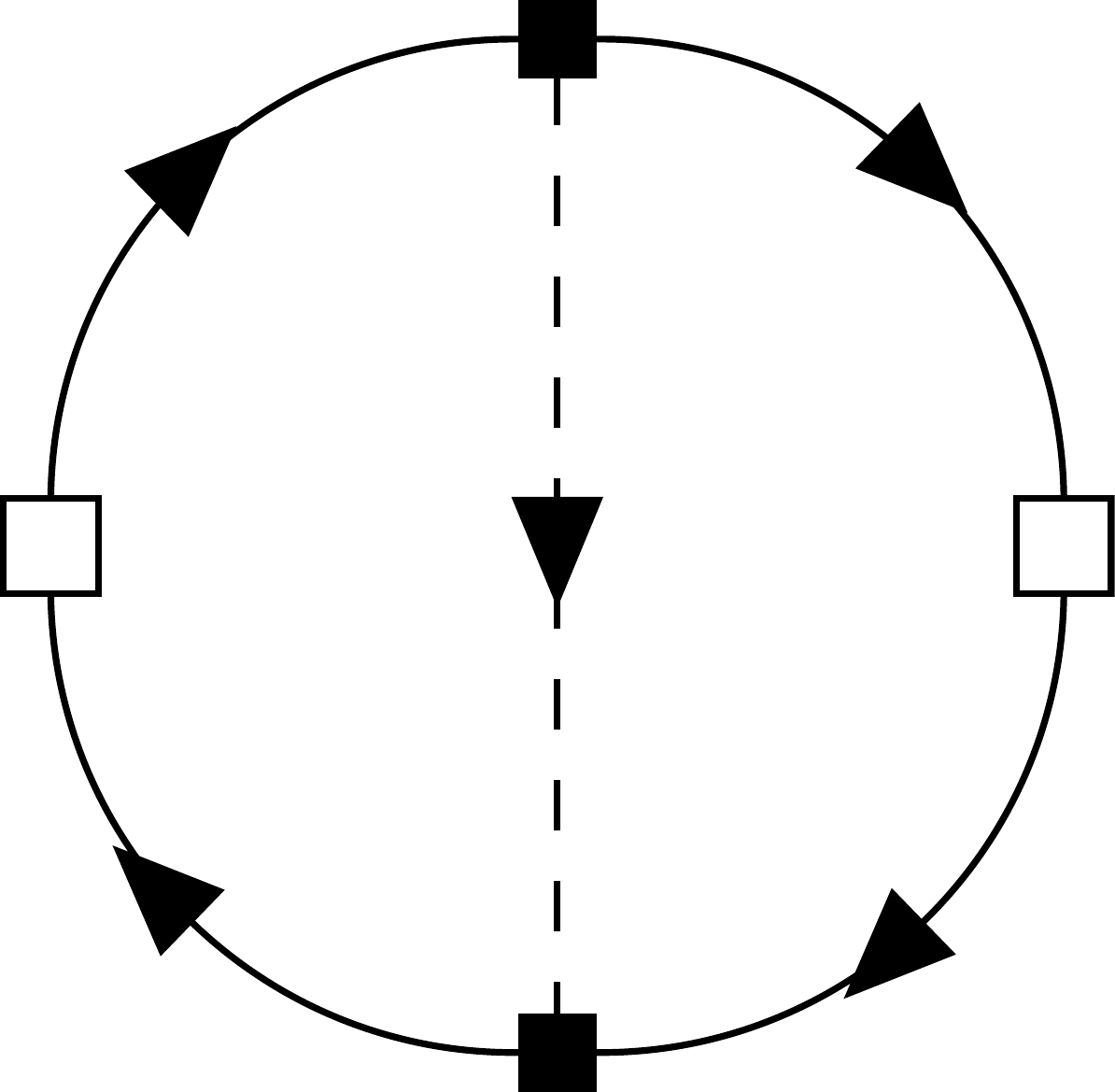}} ~~
\subfloat[]{\includegraphics[width=0.12\textwidth]{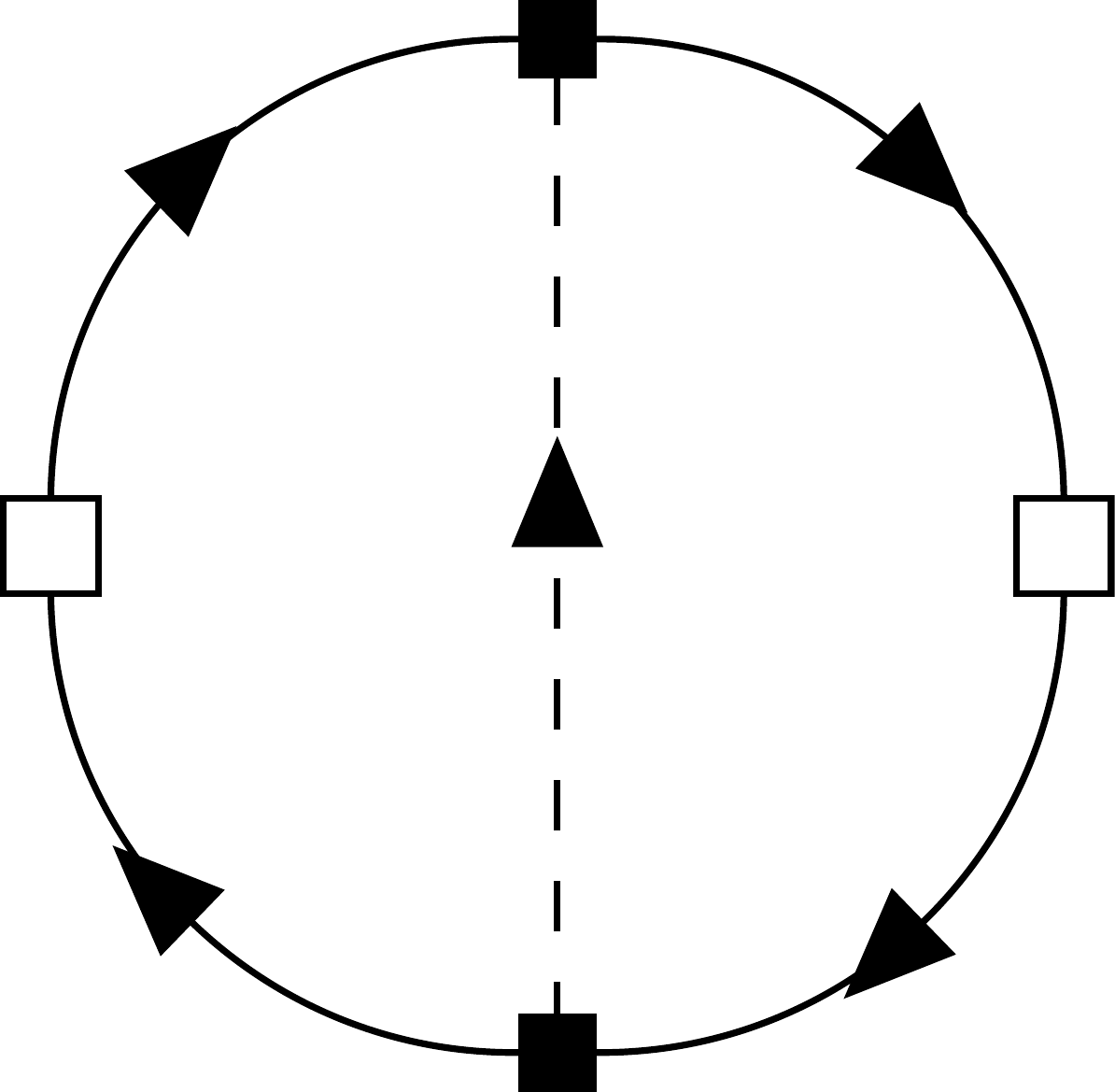}} \\
\caption{Diagrams contributing to the numerator, $N_{2}$ (Eq. (\ref{eq:n2})), of $\langle O_{2} \rangle = \langle c(\tau) c\up{\dagger}(0) \rangle$. Here conventionas are as in Fig.~\ref{fig:denom}, and open square denotes the external $c/c\up{\dagger}$ operator.} 
\label{fig:N2}
\end{figure}

Next we evaluate the electron correlation, $\langle O_{2} \rangle \equiv \langle c(\tau) c\up{\dagger}(0) \rangle = N_{2}/D$. The diagrams contributing to the numerator are shown in Fig. \ref{fig:N2}, while those contributing to the denominator have been already evaluated in (\ref{eq:d}). Thus we obtain,
%%%%%%%%%%%%%%%%%%%%%%%%%%%%%%%%%%%%%%
%%%%%%%%%%%%%%%%%%%%%%%%%%%%%%%%%%%%%%

\begin{align}
\label{eq:n2}
N_{2} &= \po + \gam\up{2} \left( \Pa\Da + \pb\Db + \pc\Dc \right) 
+ \gc\up{2} \left( \pap\Dap + \pbp\Dbp + \pcp\Dcp \right) \nonumber \\
&~~~~~~~+ \gc\up{2} \left( \papp\Dapp + \pbpp\Dbpp + \pcpp\Dcpp \right) \,, 
\end{align}
where,
\begingroup
\allowdisplaybreaks
\begin{align}
\po &= \left\langle c\up{\dagger}_{\ell \alpha} c_{\ell \alpha}\right\rangle = M'\mathcal{I}_{1,0} + \mathcal{I}_{1,1} \,,\\
\Pa &= \left\langle c\up{\dagger}_{\ell \alpha} S\up{a} S\up{a} c_{\ell \alpha}\right\rangle = \frac{M+1}{2M}(M'(M+2)\mathcal{I}_{2,0} 
-M'(M+1)\mathcal{I}_{1,0} - M'\mathcal{I}_{3,0} + (M+2)\mathcal{I}_{2,1} \nonumber \\
&~~~~~~~~~~~~~~~~~~~~~~~~~~~~~~- (M+1)\mathcal{I}_{1,1} - \mathcal{I}_{3,1}) \,, \\
\pb &= \left\langle c\up{\dagger}_{\ell \alpha} c_{\ell \alpha} S\up{a} S\up{a}\right\rangle = \frac{M+1}{2M}(MM'\mathcal{I}_{2,0} - M'\mathcal{I}_{3,0} 
+M\mathcal{I}_{2,1} -\mathcal{I}_{3,1}) \,, \\
\pc &= \left\langle c\up{\dagger}_{\ell \alpha} S\up{a} c_{\ell \alpha} S\up{a}  \right\rangle = \frac{M+1}{2M}(M'(M+1)\mathcal{I}_{2,0} - MM'\mathcal{I}_{1,0} 
-M'\mathcal{I}_{3,0} + (M+1)\mathcal{I}_{2,1} \nonumber \\
&~~~~~~~~~~~~~~~~~~~~~~~~~~~~~~- M\mathcal{I}_{1,1} -\mathcal{I}_{3,1}) \,, \\
\pap &= \left\langle c\up{\dagger}_{\ell \alpha} c_{\ell'\beta} c_{\ell'\beta}\up{\dagger} c_{\ell\alpha} \right\rangle = (M+1)M' \mathcal{I}_{1,0} - M' \mathcal{I}_{2,0} + (M+1)(M'+1)\mathcal{I}_{1,1}   \nonumber \\
&~~~~~~~~~~~~~~~~~~~~~~~~~+ (M+1)\mathcal{I}_{1,2} -(M'+1)\mathcal{I}_{2,1} - \mathcal{I}_{2,2}  \,, \\
\pbp &= \left\langle c\up{\dagger}_{\ell\alpha} c_{\ell\alpha} c_{\ell'\beta} c\up{\dagger}_{\ell'\beta} \right\rangle = M M' \mathcal{I}_{1,1} + M \mathcal{I}_{1,2} - M' \mathcal{I}_{2,1} - \mathcal{I}_{2,2}  \,, \\
\pcp &= \left\langle c\up{\dagger}_{\ell\alpha} c_{\ell'\beta} c_{\ell\alpha} c\up{\dagger}_{\ell'\beta} \right\rangle = - M M' \mathcal{I}_{1,1} -M \mathcal{I}_{1,2} + M' \mathcal{I}_{2,1} + \mathcal{I}_{2,2}  \,, \\
\papp &= \left\langle c\up{\dagger}_{\ell\alpha} c_{\ell'\beta}\up{\dagger} c_{\ell'\beta} c_{\ell\alpha} \right\rangle = M' (M'+1) (\mathcal{I}_{2,0}-\mathcal{I}_{1,0}) + (2M'+1) (\mathcal{I}_{2,1} -\mathcal{I}_{1,1})  %\nonumber \\
%&~~~~~~~~~~~~~~~~~~~~~~~~~~~~
- \mathcal{I}_{1,2} + \mathcal{I}_{2,2}  \,, \\
\pbpp &= \left\langle c\up{\dagger}_{\ell\alpha} c_{\ell\alpha} c\up{\dagger}_{\ell'\beta} c_{\ell'\beta}  \right\rangle =  M\up{\prime 2} \mathcal{I}_{2,0} + 2M' \mathcal{I}_{2,1} + \mathcal{I}_{2,2}  \,, \\
\pcpp &= \left\langle c\up{\dagger}_{\ell\alpha} c\up{\dagger}_{\ell'\beta} c_{\ell\alpha} c_{\ell'\beta}  \right\rangle = M'(M'+1)(\mathcal{I}_{1,0}-\mathcal{I}_{2,0}) + (2M'+1)(\mathcal{I}_{1,1}-\mathcal{I}_{2,1})  %\nonumber \\
%&~~~~~~~~~~~~~~~~~~~~~~~~~~~~
+ \mathcal{I}_{1,2} - \mathcal{I}_{2,2}  \,.
%%%%% 
\end{align}
\endgroup
Thus we have,
\begingroup
\allowdisplaybreaks
\begin{align}
\langle O_{2} \rangle = \frac{N_{2}}{D} &= \frac{\po}{M} \bigg \lbrace 
1 + \gam\up{2} \left[ \left( \frac{\Pa}{\po} - \frac{\lo}{M} \right) \Da 
+ \left( \frac{\pb}{\po} - \frac{\lo}{M} \right) \Db 
+\left( \frac{\pc}{\po} - \frac{\lo}{M} \right) \Dc \right]  \nonumber \\
&+ \gc\up{2} \left[ \left( \frac{\pap}{\po} - \frac{\lop}{M} \right) \Dap 
+ \left( \frac{\pbp}{\po} - \frac{\lop}{M} \right) \Dbp 
+\left( \frac{\pcp}{\po} - \frac{\lop}{M} \right) \Dcp \right] \nonumber \\
&+ \gc\up{2} \left[ \left( \frac{\papp}{\po} - \frac{\lopp}{M} \right) \Dapp 
+ \left( \frac{\pbpp}{\po} - \frac{\lopp}{M} \right) \Dbpp 
+\left( \frac{\pcpp}{\po} - \frac{\lopp}{M} \right) \Dcpp \right]
\bigg \rbrace \,.
\end{align}
\endgroup
Similarly, it is the straightforward to write, 
\begin{equation}
\label{eq:zc_m}
Z_{c} = 1 - \frac{\gamr\up{2}}{\ep} P_\gamma - \frac{\gcr\up{2}}{2\rb} P_g \,,
\end{equation}
where
\begin{align}
P_\gamma &= \frac{P_1+P_2 -2P_3}{P_0} \,, \\
P_g &= \frac{P_1'+P_2'-2P_3'+P_1''+P_2''-2P_3''}{P_0} \,.
\end{align}

Note that for $M=2\,, M'=1$, we obtain $\Pg=3$ and $\Pgam=3/4$ which agrees with the result that can be obtained from (\ref{eq:zs_zc}) and the results in Section~\ref{sec:rg}. 

\subsection{RG flow}

We are now in a position to write the beta functions for the coupling constants. Using (\ref{eq:renorm_fact_m}) we find two equations,
\begin{align}
\label{eq:beta1_m}
\frac{\ep}{2}\gamr Z_{S} + \left[ Z_{S} - \frac{\gamr}{2} \frac{\partial Z_{S}}{\partial \gamr} \right] \betgam 
- \frac{\gamr}{2} \frac{\partial Z_{S}}{\partial \gcr} \betg &= 0 \,, \\
%%%
\label{eq:beta2_m}
\rb\gcr Z_{c} + \left[ Z_{c} - \frac{\gcr}{2} \frac{\partial Z_{c}}{\partial \gcr} \right] \betg 
- \frac{\gcr}{2} \frac{\partial Z_{c}}{\partial \gamr} \betgam &= 0 \,.
\end{align}
We have used the exact relations $\widetilde{Z}_{\gcr}=\widetilde{Z}_{\gamr}=1$ in obtaining these equations. 
Solving these two equations and using the expressions for the renormalization factors found above we obtain the following one-loop beta functions,
\begin{align}
\label{eq:beta_g_m}
\betg &= -\rb \gcr + \frac{\Pg}{2} \gcr\up{3} + \frac{\Pgam}{2} \gcr \gamr\up{2} \,, \\
\label{eq:beta_gam_m}
\betgam &= -\frac{\ep}{2} \gamr + \frac{\Lgam}{2} \gamr\up{3} + \frac{\Lg}{2} \gamr \gcr\up{2} \,.
\end{align}
Recall that at $M=2, M'=1$, we have $\Pg=3$, $\Pgam=3/4$, $\Lg=2$, and $\Lgam=2$. With this the above expressions match the one-loop beta functions derived earlier in Sec. \ref{sec:rg}.

We can also calculate the anomalous dimension for the spin and electron operators, defined in  (\ref{eq:etasc2}). From (\ref{eq:beta1_m}) and (\ref{eq:beta2_m}) we obtain exactly the same equations derived before, {\it i.e.\/}, (\ref{eq:etas_eq}) and (\ref{eq:etac_eq}). Thus at the non-trivial fixed point where $\gamr*\neq0, \gcr*\neq0$ we obtain $\eta_{S}=\ep$ and $\eta_{c}=2\rb$ to all orders in $\ep$ and $\rb$. 

%%%%%%%%%%%%

\section{Large $M$ limit}
\label{app:largeM}

In this appendix we consider the large $M$ limit examined originally in the insulating spin model in Ref.~\onlinecite{SY92}. To extend the large $M$ limit to the $t$-$J$ model, we also need to
endow the electron with an additional orbital index $\ell = 1 \ldots M'$ as in (\ref{defc}), and take the large $M$ limit at fixed
\beq
k \equiv \frac{M'}{M}
\eeq
using SU($M'|M$) superspin formulation of Appendix~\ref{app:larger} while imposing the constraint (\ref{constraintM}) at $P=M/2$. Similar large $M$ limits were taken in particle-hole symmetric models in Refs.~\cite{Florens02,Florens04,Florens13,Fu2018} and for a non-random $t$-$J$ model in Refs.~\cite{Haule02,Haule03}. 

\begin{figure}
\begin{center}
\includegraphics[height=2.9in]{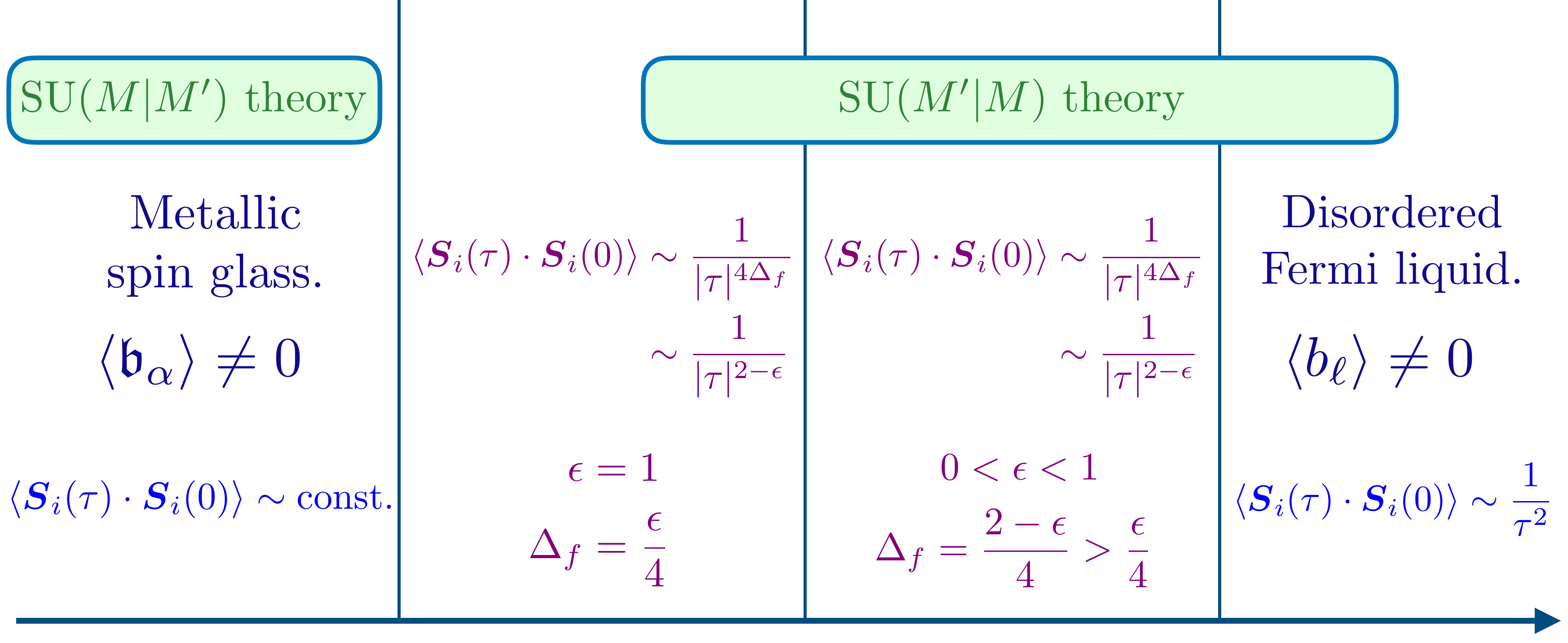} 
\end{center}
\caption{Schematic phase diagram in the large $M$ limit, with the phases displaying more rapidly decaying spin correlations as we move to the right. This appendix describes only the two intermediate critical phases in the SU($M'|M$) theory with bosonic holons and fermionic spinons. We propose in the present paper that for the case $M=2$, $M'=1$, these two critical phases reduce to a critical point with the characteristics of the $\epsilon=1$ phase, as shown in 
Fig.~\ref{fig:phasediag} . A description of the spin glass phase in the large $M$ limit requires fermionic holons and bosonic spinons in the SU($M|M'$) theory, which we do not examine here. We also note that for $M'>1$, the disordered Fermi liquid phase also has glassy correlations associated with the `orbital' index $\ell=1 \ldots M'$.}
\label{fig:phasediag:M}
\end{figure}

A sketch of our proposed large $M$ phase diagram is shown in Fig.~\ref{fig:phasediag:M}.
This applies to the theory obtained by taking the large $M$ limit of the path integral $\mathcal{Z}$ in (\ref{Z}) by inserting the SU($M'|M$) superspin
operators in (\ref{defc}) and (\ref{STM}), and rescaling $t^2 \rightarrow t^2/M$ and $J^2 \rightarrow J^2/M$.
See Ref.~\onlinecite{Fu2018} for details of a similar computation with particle-hole symmetry. In our case, both the boson and fermion Green's functions will have to be particle-hole asymmetric, as in Refs.~\cite{SY92,GPS00,GPS01,SS15}.

We will perform an analytic low energy analysis of the large $M$ equations, and find a critical solution which is in close correspondence with the RG fixed point $FP_4$ in Section~\ref{sec:rg}. However, the large $M$ solution appears to be present for a range of dopings, and not at a critical doping as in the RG analysis. We expect that either constraints from the higher energy structure of the large $M$ theory, or corrections higher order in $1/M$, will convert the critical phase to a critical point. It appears that the numerical studies of Haule {\it et al.\/} \cite{Haule02,Haule03} examined the finite temperature behavior about the critical phase described here as their model of the pseudogap. This contrasts with our model of the pseudogap in Fig.~\ref{fig:phasediag}
as a metallic spin glass flanking a critical point. We also note that the present large $M$ limit, with fermionic spinons, cannot obtain a metallic spin glass; instead we have to use the bosonic spinon approach outlined in Section~\ref{sec:pseudogap}.

\subsection{Green's functions}

We follow the condensed matter notation for Green's functions in which 
\bea
G_f (\tau) &=& - \langle T_\tau ( f(\tau) f^\dagger (0) ) \rangle, \nn \\
G_b (\tau) &=& - \langle T_\tau ( b(\tau) b^\dagger (0) ) \rangle.
\eea
We drop indices $\alpha,\ell$, and all Green's functions are diagonal in these indices.
It is useful to make ansatzes for the retarded Green's functions in the complex frequency plane, because then
the constraints from the positivity of the spectral weight are clear.
At the Matsubara frequencies, the Green's function is defined by
\beq
G(i \omega_n) = \int_0^{1/T} d \tau e^{i \omega_n \tau} G(\tau).
\eeq
So the bare Green's functions are
\bea
G_{f0} (i \omega_n) &=& \frac{1}{i \omega_n - \mu_f} \nn \\
G_{b0} (i \omega_n) &=& \frac{1}{i \omega_n - \mu_b}\,.
\eea 
The Green's functions are continued to all complex frequencies $z$ via
the spectral representation
\beq
G(z) = \int_{-\infty}^{\infty} \frac{d \Omega}{\pi} \frac{\rho (\Omega)}{z - \Omega}. \label{spec}
\eeq
For fermions, the spectral density obeys
\beq
\rho_f (\Omega ) > 0, \label{fcons}
\eeq
for all real $\Omega$ and $T$, and for bosons the constraint is
\beq
\Omega \, \rho_b (\Omega) > 0. \label{bcons}
\eeq
The retarded Green's function is $G^R (\omega) = G(\omega + i \eta)$ with $\eta$ a positive infinitesimal, while the advanced Green's function is $G^A (\omega) = G(\omega - i \eta)$.
It is also useful to tabulate the inverse Fourier transforms at $T=0$
\bea 
G(\tau) = \left\{
\begin{array}{ccc}
\displaystyle - \int_0^\infty \frac{d \Omega}{\pi} \rho (\Omega) e^{- \Omega \tau} &,& \mbox{for $\tau >0 $ and $T=0$} \\[1em] 
\displaystyle\int_0^\infty \frac{d \Omega}{\pi} \rho (-\Omega) e^{ \Omega \tau} &,& \mbox{for $\tau < 0$ and $T=0$}
\end{array} \right. . \label{ift}
\eea

\subsection{Saddle point and self-consistency equations}
\label{app:saddlepoint}

The saddle-point equations of the SU($M'|M$) form of (\ref{Z}), after rescaling $t^2 \rightarrow t^2/M$ and $J^2 \rightarrow J^2/M$, for the boson and fermion Green's functions are
\bea
G_b ( i \omega_n) &=& \frac{1}{i \omega_n + \mu_b - \Sigma_b (i \omega_n) } , \quad \Sigma_b (\tau) =  t^2 G_f(\tau) R(-\tau) \label{beqns} \\
G_f (i \omega_n) &=& \frac{1}{i \omega_n + \mu_f - \Sigma_f (i \omega_n)} , \quad \Sigma_f (\tau) =  J^2 G_f (\tau) Q(\tau)
- k \, t^2 R (\tau) G_b (\tau) \,,~~~~~~~~~ \label{feqns} 
\eea
while the self-consistency equations in (\ref{selfcons}) are
\beq
R(\tau) = - G_f (\tau) G_b (-\tau) \quad, \quad Q(\tau) = - G_f (\tau) G_f (-\tau)
\label{defRQ}
\eeq 
Here $\mu_f$ and $\mu_b$ are chemical potentials, determined by $\epsilon_0$ and the saddle point value of $\lambda$, and chosen to satisfy
\beq
\left\langle f^\dagger f \right\rangle = \frac{1}{2}-kp \quad, \quad  \left\langle b^\dagger b \right\rangle = p \,. \label{lutt1}
\eeq
Eqs.~(\ref{feqns},\ref{defRQ},\ref{lutt1}) are identical to the ones considered in Refs~\cite{Haule02,Haule03} as an EDMFT approximation to the non-random $t$-$J$ model. 
We will look for gapless critical solutions of the above equations. We expect that such solutions exist only for $p < p_c$, and that for $p > p_c$ the boson condenses, leading to a Fermi liquid solution.

\subsection{Low frequency ansatzes}
\label{app:ansatz}

For the fermion Green's function, we write at a complex frequency $|z| \ll J$
\beq
G_f (z) = C_f \frac{e^{-i (\pi \Delta_f + \theta_f)}}{z^{1-2 \Delta_f}} \quad, \quad \mbox{Im}(z) >0,
\label{Gfz}
\eeq
which is expressed in terms of three real parameters, $C_f$, $\Delta_f$ and $\theta_f$.
Then the constraint (\ref{fcons}) becomes
\beq
\sin ( \pi \Delta_f + \theta_f ) > 0 \quad, \quad \sin(\pi \Delta_f - \theta_f) > 0.
\label{signthetaf}
\eeq
The particle-hole symmetric value is $\theta_f = 0$.
Using (\ref{ift}) we obtain in $\tau$ space for $|\tau| \gg 1/J$
\bea 
G_f (\tau) = \left\{
\begin{array}{ccc}
\displaystyle - \frac{C_f \Gamma (2 \Delta_f) \sin (\pi \Delta_f + \theta_f)}{\pi |\tau|^{2 \Delta_f}} &,& \mbox{for $\tau > 0$ and $T=0$} \\[1em] 
\displaystyle \frac{C_f \Gamma (2 \Delta_f) \sin (\pi \Delta_f - \theta_f)}{\pi |\tau|^{2 \Delta_f}} &,& \mbox{for $\tau < 0$ and $T=0$}
\end{array} \right. . \label{Gftau}
\eea

We can also write corresponding ansatz for the fermionic correlator $R(\tau)$, which we will only need as a function of $\tau$
for $|\tau| \gg 1/J$
\bea 
R (\tau) = \left\{
\begin{array}{ccc}
\displaystyle - \frac{C_{+R}}{|\tau|^{2 (1 - \rb)}} &,& \mbox{for $\tau > 0$ and $T=0$} \\[1em] 
\displaystyle \frac{C_{-R}}{ |\tau|^{2 (1 - \rb)}} &,& \mbox{for $\tau < 0$ and $T=0$}
\end{array} \right. . \label{Rtau}
\eea
where $C_{+R} > 0$ and $C_{-R} > 0$, but they need not be equal. This corresponds to the ansatz in (\ref{QRpower}) which has $C_{+R} = C_{-R}$. We can allow these amplitudes to be distinct in the large $M$ limit. We also examined generalization of the RG analysis in Section~\ref{sec:rg} to the case $C_{+R} \neq C_{-R}$; we found that the perturbative RG then gave inconsistent renormalizations of the coupling $g$, and so $C_{+R} = C_{-R}$ in the context of the $\epsilon$ and $\rb$ expansion.

For the boson Green's function, we write at a complex frequency $|z| \ll J$
\beq
G_b (z) = C_b \frac{e^{-i (\pi \Delta_b + \theta_b)}}{z^{1-2 \Delta_b}} \quad, \quad \mbox{Im}(z) >0,
\label{Gbz}
\eeq
expressed in terms of the three real parameters, $C_b$, $\Delta_b$ and $\theta_b$. The constraint (\ref{bcons}) becomes
\beq
\sin ( \pi \Delta_b + \theta_b ) > 0 \quad, \quad \sin(\pi \Delta_b - \theta_b) < 0.
\label{signthetab}
\eeq
Using (\ref{ift}) we obtain in $\tau$ space for $|\tau| \gg 1/J$
\bea 
G_b (\tau) = \left\{
\begin{array}{ccc}
\displaystyle - \frac{C_b \Gamma (2 \Delta_b) \sin (\pi \Delta_b + \theta_b)}{\pi |\tau|^{2 \Delta_b}} &,& \mbox{for $\tau > 0$ and $T=0$} \\[1em] 
\displaystyle \frac{C_b \Gamma (2 \Delta_b) \sin (\pi \Delta_b - \theta_b)}{\pi |\tau|^{2 \Delta_b}} &,& \mbox{for $\tau < 0$ and $T=0$}
\end{array} \right. . \label{Gbtau}
\eea

Finally, we can write expressions similar to (\ref{Rtau}) for $Q(\tau)$
for $|\tau| \gg 1/J$
\bea 
Q (\tau) = \left\{
\begin{array}{ccc}
\displaystyle \frac{C_{Q}}{|\tau|^{2 - \epsilon}} &,& \mbox{for $\tau > 0$ and $T=0$} \\[1em] 
\displaystyle \frac{C_{Q}}{ |\tau|^{2 - \epsilon}} &,& \mbox{for $\tau < 0$ and $T=0$}
\end{array} \right. . \label{Qtau}
\eea
where $C_{Q} > 0$. This corresponds to the ansatz for $Q(\tau)$ in (\ref{QRpower})

We can relate the parameters in the ansatzes for the bosonic bath $Q(\tau)$ and the fermionic bath $R(\tau)$ to the parameters in the ansatzes for the Green's functions $G_f$ and $G_b$, by using the self-consistency conditions (\ref{defRQ}). This yields expressions for
 $\rb$, $\epsilon$, $C_{\pm R}$, and $C_{Q}$ in terms of $\Delta_{f,b}$, $\theta_{f,b}$, and $C_{f,b}$:
\bea
\rb &=& 1 - \Delta_f - \Delta_b \nn 
\epsilon &=& 2(1-2 \Delta_f) \nn 
C_{+R} &=& - \frac{C_f C_b \Gamma (2 \Delta_f) \Gamma (2 \Delta_b ) \sin (\pi \Delta_f + \theta_f) \sin (\pi \Delta_b - \theta_b)}{\pi^2} \nn 
C_{-R} &=& \frac{C_f C_b \Gamma (2 \Delta_f) \Gamma (2 \Delta_b ) \sin (\pi \Delta_f - \theta_f) \sin (\pi \Delta_b + \theta_b)}{\pi^2} \nn 
C_{Q} &=&  \frac{\left[C_f \Gamma (2 \Delta_f) \right]^2 \sin (\pi \Delta_f + \theta_f) \sin (\pi \Delta_f - \theta_f)}{\pi^2}\,.
\label{selfcon2}
\eea
However, in keeping with RG computation, we will defer application of the relations in (\ref{selfcon2}).

\subsection{Luttinger constraints}

The Luttinger constraints on the spectral asymmetries of the Green's functions are very similar to those in Refs.~\cite{GPS01,Davison17}, and can be obtained by the method of Ref.~\onlinecite{Gu2019}:
\begin{eqnarray}
 \frac{\theta_f}{\pi} + \left( \frac{1}{2} -\Delta_f \right) \frac{\sin (2 \theta_f)}{\sin (2\pi \Delta_f)} &=& kp \nonumber \\
\frac{\theta_b}{\pi} + \left( \frac{1}{2} -\Delta_b \right) \frac{\sin (2 \theta_b)}{\sin (2\pi \Delta_b)} &=& \frac{1}{2} + p \,. \nonumber \label{luttinger}
\end{eqnarray}
These constraints imply $- \pi \Delta_f < \theta_f < \pi \Delta_f$ and $\pi \Delta_b < \theta_b < \pi/2$. The determine the spectral asymmetry angles $\theta_{f,b}$ in terms of the density $p$ and the scaling dimensions $\Delta_{f,b}$.

\subsection{Self-energies}

We can now collect the corresponding expressions for the fermionic and bosonic self energies at $|\tau| \gg 1/J$ and $T=0$:
\bea 
\Sigma_f (\tau) &=& 
-\displaystyle k t^2 \left[ \frac{C_b C_{+R} \Gamma(2 \Delta_b)}{\pi} \right] 
 \frac{ \sin(\pi \Delta_b + \theta_b) }{|\tau|^{2\Delta_b + 2(1 - \rb)}} \nonumber \\
\displaystyle &~&~~~~~~~- J^2 \left[ \frac{C_f C_{Q} \Gamma(2 \Delta_f)}{\pi} \right] \frac{\sin (\pi \Delta_f + \theta_f)}{|\tau|^{2\Delta_f + 2 - \epsilon}}
 , \quad \mbox{ $\tau > 0$} 
 \label{Sftaup}
\eea
\bea 
\Sigma_f (\tau) &=& 
\displaystyle -k t^2 \left[ \frac{C_b C_{-R} \Gamma(2 \Delta_b))}{\pi} \right] 
 \frac{ \sin(\pi \Delta_b - \theta_b)}{|\tau|^{2 \Delta_b + 2 (1 - \rb)}} \nonumber \\
\displaystyle &~&~~~~~~~+ J^2 \left[ \frac{C_f C_{Q} \Gamma(2 \Delta_f)}{\pi} \right] \frac{\sin (\pi \Delta_f - \theta_f)}{|\tau|^{2 \Delta_f + 2 - \epsilon}}
 , \quad \mbox{$\tau < 0$} 
\label{Sftaum}
\eea
\bea 
\Sigma_b (\tau) &=& 
\displaystyle  -t^2 \left[ \frac{C_f C_{-R}  \Gamma(2 \Delta_f)}{\pi} \right] 
 \frac{ \sin(\pi \Delta_f + \theta_f) }{|\tau|^{2 \Delta_f + 2 (1 - \rb)}} , \quad \tau > 0\nonumber\\ \label{Sbtaup}
\eea
\bea 
\Sigma_b (\tau) &=& 
\displaystyle  -t^2 \left[ \frac{(C_f C_{+R} \Gamma(2 \Delta_f)}{\pi} \right] 
 \frac{ \sin(\pi \Delta_f - \theta_f)}{|\tau|^{2 \Delta_f + 2 (1 - \rb)}} , \quad \tau < 0\nonumber\\ \label{Sbtaum}
\eea
We also use the spectral representations for the self energies
\beq
\Sigma (z) = \int_{-\infty}^{\infty} \frac{d \Omega}{\pi} \frac{\sigma (\Omega)}{z - \Omega}. \label{sspec}
\eeq
Performing the inverse of the Laplace transforms in (\ref{ift}) we obtain at $T=0$ and $|\Omega| \ll J$
\bea 
\sigma_f (\Omega) &=& 
\displaystyle  \frac{\pi k t^2}{\Gamma(2 \Delta_b + 2 (1 - \rb))} \left[ \frac{C_b C_{+R}  \Gamma(2 \Delta_b)}{\pi} \right] 
 \frac{ \sin(\pi \Delta_b + \theta_b) }{|\Omega|^{1-2 \Delta_b - 2 (1 - \rb)}} \nonumber \\
\displaystyle &~&~~~~~~~+ \frac{\pi J^2}{\Gamma(2 \Delta_f + 2 - \epsilon)} \left[ \frac{C_f C_Q \Gamma(2 \Delta_f)}{\pi} \right] \frac{\sin (\pi \Delta_f + \theta_f) }{|\Omega|^{-1-2 \Delta_f  + \epsilon}}
 , \quad \mbox{ $\Omega > 0$} 
 \label{Sfomegap}
\eea
\bea 
\sigma_f (\Omega) &=& 
\displaystyle -\frac{\pi k t^2}{\Gamma (2 \Delta_b + 2(1 - \rb))}  \left[ \frac{C_b C_{-R} \Gamma(2 \Delta_b) }{\pi} \right] 
 \frac{ \sin(\pi \Delta_b - \theta_b)}{|\Omega|^{1-2 \Delta_b - 2 (1 - \rb)}} \nonumber \\
\displaystyle &~&~~~~~~~+ \frac{\pi J^2}{\Gamma(2 \Delta_f + 2 - \epsilon)}\left[ \frac{C_f C_Q \Gamma(2 \Delta_f)}{\pi} \right] \frac{\sin (\pi \Delta_f - \theta_f)}{|\Omega|^{-1-2 \Delta_f + \epsilon}}
 , \quad \mbox{$\Omega < 0$} 
\label{Sfomegam}
\eea
\bea 
\sigma_b (\Omega) &=& 
\displaystyle  \frac{\pi t^2}{\Gamma(2 \Delta_f + 2(1 - \rb))} \left[ \frac{C_f C_{-R} \Gamma(2 \Delta_f)}{\pi} \right] 
 \frac{\sin(\pi \Delta_f + \theta_f)}{|\Omega|^{1- 2 \Delta_f - 2 (1 - \rb)}} , \quad \Omega > 0\nonumber\\ \label{Sbomegap}
\eea
\bea 
\sigma_b (\Omega) &=& 
\displaystyle - \frac{\pi t^2}{\Gamma (2 \Delta_f + 2(1 - \rb))} \left[ \frac{C_f C_{+R} \Gamma(2 \Delta_f) C_b}{\pi} \right] 
 \frac{\sin(\pi \Delta_f - \theta_f)}{|\Omega|^{1-2 \Delta_f - 2 (1 - \rb)}} , \quad \Omega < 0\nonumber\\ \label{Sbomegam}
\eea

\subsection{Solution of the saddle point equations}

We will now determine the constraints placed by the saddle point equations (\ref{beqns}) and (\ref{feqns}) on the parameters of the low frequency ansatzes presented in Section~\ref{app:ansatz}. As in the body of the text, we will defer application of the self-consistency conditions (\ref{defRQ}), which led to the relations (\ref{selfcon2}). This  will allow us to make a more complete comparison of the results of the large $M$ theory with that in Section~\ref{sec:rg}.

From (\ref{Gbz}) we have
\beq
\Sigma_b (z) = -\frac{1}{C_b} e^{i (\pi \Delta_b + \theta_b)} z^{(1 - 2 \Delta_b)} . \label{Sbz}
\eeq
Comparing (\ref{Sbomegap}), (\ref{Sbomegam}) and (\ref{Sbz}) we have the exponent identity
\beq
\Delta_f  + \Delta_b  = \rb \,. \label{e5}
\eeq
In the present large $M$ limit, the electron Green's function is given by 
\beq
G_c (\tau) = - G_f (\tau) G_b (-\tau)\,, \label{Gc1}
\eeq
and so the anomalous dimension of the electron operator is
\beq
\eta_c = 2 (\Delta_f + \Delta_b)\,. 
\label{etaclargeM}
\eeq
Now we see that the large $M$ result (\ref{e5}) is precisely the result (\ref{etacrb}) for the electron anomalous dimension obtained to all orders in the $\epsilon$ and $\rb$ expansions.

Comparing the amplitudes of (\ref{Sbomegap}), (\ref{Sbomegam}) and (\ref{Sbz}) we obtain
\beq
\displaystyle  \frac{\pi t^2}{\Gamma(2 \Delta_f + 2(1 - \rb))} \left[ \frac{C_f C_{-R} \Gamma(2 \Delta_f)}{\pi} \right] 
 \sin(\pi \Delta_f + \theta_f) = \frac{\sin (\pi \Delta_b + \theta_b)}{C_b} \label{c1}
\eeq
\beq
\displaystyle  \frac{\pi t^2}{\Gamma(2 \Delta_f + 2(1 - \rb))} \left[ \frac{C_f C_{+R} \Gamma(2 \Delta_f)}{\pi} \right] 
 \sin(\pi \Delta_f - \theta_f) = - \frac{\sin (\pi \Delta_b - \theta_b)}{C_b} \label{c1a}
\eeq
From (\ref{Gfz}) we have
\beq
\Sigma_f (z) = -\frac{1}{C_f} e^{i (\pi \Delta_f + \theta_f)} z^{(1 - 2 \Delta_f)} . \label{Sfz}
\eeq
The comparison of this with (\ref{Sfomegap}), (\ref{Sfomegam}) leads to two possible solutions, appearing as the two intermediate critical phases in Fig.~\ref{fig:phasediag:M}.

\subsubsection{$\Delta_f > \epsilon/4 $}
The second $J^2$ terms in (\ref{Sfomegap}) and (\ref{Sfomegam}) are much smaller than the $t^2$ terms
when $\Delta_f - \epsilon/2 > \Delta_b - \rb$; using (\ref{e5}), we obtain the condition $\Delta_f > \epsilon/4$. So the $J^2$ terms can be neglected. Indeed, the low energy solution is then entirely independent of the strength of the exchange interaction, which is rather different from the structure of the $FP_4$ fixed point in Section~\ref{sec:rg} with both $\gcr*$ and $\gamr*$ non-zero. Instead, it is the $FP_3$ fixed point, with $\gamr* =0$, which matches the structure of the present large $M$ solution, and this fixed point was found to be unstable in the RG analysis for $M=2$, $M'=1$. 
We will therefore only write down the saddle point equations here, and not consider this case further. 

From (\ref{Sfz}), (\ref{Sfomegap}) and (\ref{Sfomegam}) we have
\beq
\displaystyle \frac{\pi k t^2}{\Gamma(2 \Delta_b + 2 (1 - \rb))} \left[ \frac{C_b C_{+R} \Gamma(2 \Delta_b)}{\pi} \right] 
  \sin(\pi \Delta_b + \theta_b) = \frac{\sin(\pi \Delta_f + \theta_f) }{C_f} \label{c2}
\eeq
\beq
\displaystyle \frac{\pi k t^2}{\Gamma(2 \Delta_b + 2 (1 - \rb))} \left[ \frac{C_b C_{-R} \Gamma(2 \Delta_b)}{\pi} \right] 
  \sin(\pi \Delta_b - \theta_b) = -\frac{\sin(\pi \Delta_f - \theta_f) }{C_f} \label{c2a}
\eeq
The combination of (\ref{c1}), (\ref{c2a}) or (\ref{c1a}), (\ref{c2}) yields
\beq
\frac{\Gamma(2 \Delta_f) \sin(\pi \Delta_f + \theta_f) \sin(\pi \Delta_f - \theta_f)}{ \Gamma(2 \Delta_b) \sin(\pi \Delta_b + \theta_b) \sin(\pi \Delta_b - \theta_b) } = -k
\eeq
We also have from (\ref{c1}), (\ref{c1a}) or (\ref{c2}), (\ref{c2a})
\beq
\frac{C_{+R}}{C_{-R}} = - \frac{\sin(\pi \Delta_f + \theta_f) \sin(\pi \Delta_b - \theta_b)}{\sin(\pi \Delta_f - \theta_f) \sin(\pi \Delta_b + \theta_b)} \label{crcr}
\eeq
which is consistent with the relations in (\ref{selfcon2}).

We can also apply the self-consistency relations in (\ref{selfcon2}) to the exponents, and obtain $\rb = 1/2$ and $\Delta_f = (2-\epsilon)/4$.

\subsubsection{$\Delta_f =\epsilon/4 $}

Now the $t^2$ and $J^2$ terms  in (\ref{Sfomegap}) and (\ref{Sfomegam}) are equally important, and we will see that the structure of this large $M$ solution is very similar to that of the critical point found in the RG analysis in Section~\ref{sec:rg}. 

Solving (\ref{e5}) and $\Delta_f -\epsilon/2 = \Delta_b - \rb $ we obtain the scaling dimensions.
\beq
\Delta_f = \frac{\epsilon}{4}
\label{Deltafe}
\eeq
\beq
\Delta_b = \rb - \frac{\epsilon}{4}
\label{Deltabe}
\eeq
In the large $M$ limit, the spin correlator is given by
\beq
\left\langle \vec{S}(\tau) \cdot \vec{S} (0) \right\rangle \sim - G_f (\tau) G_f (-\tau)\,
\label{SSGfGf}
\eeq
and so the anomalous dimension of the spin operator is
\beq
\eta_S = 4 \Delta_f
\label{etaSDeltaf}
\eeq
We now see that the spin anomalous dimension implied by the large $M$ equations (\ref{Deltafe}) and (\ref{etaSDeltaf}) 
is consistent with the result (\ref{etaSep}) obtained to all orders in the $\epsilon$ and $\rb$ expansion.

Comparison of the amplitude of (\ref{Sfomegap}) with (\ref{Sfz}) yields
\bea 
\displaystyle &&  k t^2 \frac{C_b C_{+R} \Gamma(2 \rb -\epsilon/2)}{\Gamma(2-\epsilon/2)}
\sin(\pi \Delta_b + \theta_b)  \nonumber \\
&&~~~~~~~~~~+  J^2 \frac{C_f C_Q \Gamma(\epsilon/2)}{\Gamma(2-\epsilon/2)} \sin (\pi \Delta_f + \theta_f)  = \frac{ \sin (\pi \Delta_f + \theta_f)}{C_f}
 \label{c3}
\eea
while the comparison of (\ref{Sfomegam}) with (\ref{Sfz}) yields
\bea 
\displaystyle &&  -k t^2 \frac{C_b C_{-R} \Gamma(2 \rb -\epsilon/2 )}{\Gamma(2-\epsilon/2)}
\sin(\pi \Delta_b - \theta_b)  \nonumber \\
&&~~~~~~~~~~+  J^2 \frac{C_f C_Q \Gamma(\epsilon/2)}{\Gamma(2-\epsilon/2)}  \sin (\pi \Delta_f - \theta_f)  = \frac{\sin (\pi \Delta_f - \theta_f)}{C_f}
 \label{c3a}
\eea
Comparison of (\ref{c3}) and (\ref{c3a}) again yields (\ref{crcr}), which is consistent with (\ref{selfcon2}).

Let us combine the saddle point equations (\ref{c1}), (\ref{c1a}), (\ref{Deltafe}), (\ref{Deltabe}), (\ref{c3}), and (\ref{c3a}) with self-consistency equations (\ref{selfcon2}), 
and collect all the equations which determine the parameters $\Delta_{f,b}$, $\theta_{f,b}$, and $C_{f,b}$ in the low frequency ansatzes for $G_f$ and $G_b$ in (\ref{Gfz}) and (\ref{Gbz}). All these equations reduce to $\epsilon=1$, $\rb = 1/2$, and the following independent equations
\bea
&& ~~~~~~~~~\Delta_f = \frac{1}{4} \nonumber \\
&& ~~~~~~~~~\Delta_b = \frac{1}{4} \nonumber \\
&& ~~t^2 C_f^2 C_b^2 \cos(2 \theta_f) = \pi \nonumber \\
&& J^2 C_f^4 \cos(2 \theta_f) - k t^2 C_f^2 C_b^2 \cos(2 \theta_b)  = \pi \,. \label{alleqns}
\eea
Note that the values of $\Delta_f$ and $\Delta_b$ above, combined with (\ref{etaclargeM})
and (\ref{etaSDeltaf}) yield the self-consistent values in (\ref{etaSetac}).
For the last two equations in (\ref{alleqns}), notice the bounds $|\theta_f| < \pi/4$
and $\pi/4 < \theta_b < \pi/2$ below (\ref{luttinger}); so all the co-efficients on the left hand sides of (\ref{alleqns}) are positive, and the last two equations determine the values of $C_f$ and $C_b$. The values of $\theta_f$ and $\theta_b$ are then determined by the particle density $p$ from (\ref{luttinger}). So this low energy solution can exist at a variable particle density, and the present low energy $M=\infty$ theory describes a potential critical phase, rather than a critical point.

Finally, let us note the form of the electron Green's function from (\ref{Gc1})
\bea 
G_c (\tau) = \left\{
\begin{array}{ccc}
\displaystyle  \frac{C_f C_b \sin (\pi/4 + \theta_f) \sin (\pi/4 - \theta_b)}{\pi |\tau|} &,& \mbox{for $\tau > 0$ and $T=0$} \\[1em] 
\displaystyle \frac{C_f C_b  \sin (\pi/4 - \theta_f) \sin (\pi/4 + \theta_b)}{\pi |\tau|} &,& \mbox{for $\tau < 0$ and $T=0$}
\end{array} \right. . \label{Gc2}
\eea
The exponent and signs of (\ref{Gc2}) agree with the self-consistent electron Green's function obtained in (\ref{Gc3}) (recall (\ref{signthetaf}) and (\ref{signthetab})), but it appears that the magnitudes of the amplitudes in (\ref{Gc2}) can be different between $\tau > 0$ and $\tau < 0$. This is a subtle feature of the large $M$ theory which is not reproduced by the $\epsilon$ and $\rb$  expansion in the body of the paper. This is related to the discussion below (\ref{Rtau}).

Also note that the $1/\tau$ decay of (\ref{Gc2}) is similar to that of a Fermi liquid. Nevertheless, this state is not a Fermi liquid because the spin correlator in (\ref{SSGfGf}) decays as $1/|\tau|$, in contrast to the $1/\tau^2$ decay expected in a Fermi liquid. The exponents in (\ref{SSGfGf}) and (\ref{Gc2}) can be understood together in a picture of fractionalization of the electron into spinons  and holons, where the spinon and holon correlators both decay as $1/\sqrt{\tau}$.

%%%%%%%%%%%%%%%%%%%%%%%%%%%%%%%%%%%%%%%%%%%%%%%%%%%%%%%%%%%%%%%%%%%%%%%%%%%%
\section{RG details}
\label{app:2loop}
This appendix contains further details on the RG computation of Section~\ref{sec:rg}.

%%%%%%%%%%%%%%%%%%%%%%%%%%%%%%
The beta functions are defined as follows:
\begin{equation}
\betg = \rgs \frac{d\gcr}{d\rgs}|_{\gc} \,; ~~~~~  \betgam = \rgs \frac{d\gamr}{d\rgs}|_{\gam}
\label{eq:beta_def}
\end{equation}
To begin,
\begin{equation}
\rgs \frac{d\gc}{d\rgs} = 0 = \rb \frac{\rgs\up{\rb} \zg \gcr}{\sqrt{\zf \zb}} + \rgs \frac{\rgs\up{\rb} \gcr}{\sqrt{\zf \zb}} \frac{d\zg}{d\rgs} + \rgs \frac{\rgs\up{\rb} \zg}{\sqrt{\zf \zb}} \frac{d\gcr}{d\rgs} 
- \frac{\rgs}{2} \frac{\rgs\up{\rb} \zg \gcr}{\sqrt{\zf \zb}} \left[\frac{1}{\zf} \frac{d\zf}{d\rgs} + \frac{1}{\zb} \frac{d\zb}{d\rgs} \right] \,,
\end{equation}
which gives us,
\begin{align}
\rb \gcr \zg \zf \zb &+ \betg \left[ \zg \zf \zb  + \gcr \zf \zb \frac{\partial \zg}{\partial \gcr} 
- \frac{\gcr}{2} \left( \zg \zb \frac{\partial \zf}{\partial \gcr} + \zg \zf \frac{\partial \zb}{\partial \gcr} \right) \right] 
\nonumber \\
&+ \betgam \left[\gcr \zf \zb \frac{\partial \zg}{\partial \gamr}
 -\frac{\gcr}{2} \left( \zg \zb \frac{\partial \zf}{\partial \gamr} + \zg \zf \frac{\partial \zb}{\partial \gamr} \right) \right] = 0 \,.
\label{eq:beta1}
\end{align}
Similarly, we have
\begin{equation}
\frac{\ep}{2} \gamr \zgam \zf + \betg \gamr \left[\zf \frac{\partial \zgam}{\partial \gcr} - \zgam \frac{\partial \zf}{\partial \gcr} \right] 
+ \betgam \left[\gamr \zf \frac{\partial \zgam}{\partial \gamr} + \zf \zgam  - \gamr \zgam \frac{\partial \zf}{\partial \gamr} \right] = 0 \,.
\label{eq:beta2}
\end{equation}

\subsection{Flow away from criticality}

For the flow equation of $\kinr$ at one-loop, we will follow the momentum-shell RG procedure, where the cut-off $D$ is kept explicitly. In this case, we introduce masses for bosons and fermions, but keeping in mind that only their difference is physically relevant. To this end we consider the Fourier-transformed action,
\begin{align}
S(D-\delta D) &= S_{\psi} (D- \delta D) + S_{\gcr} (D - \delta D) + S_{\gamr} (D- \delta D) + S_{\phi} (D - \delta D) \nonumber \\
&+ \frac{1}{\beta} \sum_{i\omega_n, \alpha} \fd_{\alpha} \left[ -i\omega_n + \lam + m_f + \Sigma_{F} \right] \fa_{\alpha} 
+ \frac{1}{\beta} \sum_{i\omega_n} \bd \left[ -i\omega_n + \lam + m_b + \Sigma_{B} \right] \ba \,,
\end{align}
where the self energies are evaluated as follows:
\begin{align}
\Sigma_{a} &= \gc\up{2} \int_{D-\delta D}\up{D} dk \frac{k\up{r}}{i\omega_n - \lam - k -mb} =
-\gc\up{2} D\up{r-1} \left( \delta D + (i\omega_n - \lam - m_b) \frac{\delta D}{D} \right)  \nonumber \\
&= -\gcr\up{2} \left( \delta D + (i\omega_n - \lam - m_b) \frac{\delta D}{D} \right) 
~~~~~~~~~~~~~~(\text{with}~~ \gcr\up{2} \equiv \gc\up{2} D\up{r-1}) \,, \\
\Sigma_{b} &= \gam\up{2} \frac{3}{4} \frac{S_{d}}{2} \int_{D - \delta D}\up{D} dk \frac{k\up{d-2}}{i\omega_n - \lam - k - m_f}
= -\frac{3}{4} \gam\up{2} \frac{S_d}{2} D\up{d-3} \left( \delta D + (i\omega_n - \lam - m_f) \frac{\delta D}{D} \right) \nonumber \\
&= -\frac{3}{4} \gamr\up{2} \frac{S_d}{2 \widetilde{S}_{d+1}} \left( \delta D + (i\omega_n - \lam - m_f) \frac{\delta D}{D} \right) ~~~~~~~~~~~~~~(\text{with}~~ \gamr\up{2} \equiv \gam\up{2} D\up{d-3} \widetilde{S}_{d+1})  \,, \\
\Sigma_{c} &= 2\gc\up{2} \int_{D-\delta D}\up{D} dk \frac{k\up{r}}{i\omega_n - \lam - k -mb} =
-2\gc\up{2} D\up{r-1} \left( \delta D + (i\omega_n - \lam - m_b) \frac{\delta D}{D} \right)  \nonumber \\
&= -2\gcr\up{2} \left( \delta D + (i\omega_n - \lam - m_b) \frac{\delta D}{D} \right) 
~~~~~~~~~~~~~~(\text{with}~~ \gcr\up{2} \equiv \gc\up{2} D\up{r-1}) \,,
\end{align}
with $\Sigma_{F} = \Sigma_{a} + \Sigma_{b}$ and $\Sigma_{B} = \Sigma_{c}$. The scaling factor is $l = 1 + \delta D/D$ such that under the scaling $k'=lk$ and $i\omega' = l i\omega$. Thus we have,
\begin{align}
S'(D) &= l\up{-3-r} (S'_{c} (D) + S'_{\gcr} (D) ) + l\up{-1} S'_{\gamr} (D) + l\up{-d+1} S'_{\phi} (D) \nonumber \\
&+ \frac{l\up{-2}}{\beta}  \sum_{i\omega'_{n}} \fd \left[ (-i\omega'_n + \lam)(1+ \gcr\up{2}\frac{\delta D}{D} + \frac{3}{4}\gamr\up{2}\frac{\delta D}{D}) + l m_{f} (1 + \frac{3}{4}\gamr\up{2}\frac{\delta D}{D}) - l \gcr\up{2} \delta D 
+ l \gcr\up{2} m_{b} \frac{\delta D}{D} - l \frac{3}{4} \gamr\up{2} \delta D
\right] \fa  \nonumber \\
&+ \frac{l\up{-2}}{\beta}  \sum_{i\omega'_{n}} \bd \left[ (-i\omega'_n + \lam)(1+ 2\gcr\up{2}\frac{\delta D}{D} ) + l m_{b} 
 - l 2\gcr\up{2} \delta D 
+ l 2\gcr\up{2} m_{f} \frac{\delta D}{D} 
\right] \ba \,.
\end{align}
Thus we have the following expressions for the renormalized masses:
\begin{align}
\label{eq:massf}
m'_{f} &= \left( 1 + \frac{\delta D}{D} - \gcr\up{2} \frac{\delta D}{D} \right) m_f 
- \left(\gcr\up{2} + \frac{3}{4} \gamr\up{2} \right) \frac{\delta D}{D} + \gcr\up{2} m_{b} \frac{\delta D}{D} \,, \\
\label{eq:massb}
m'_{b} &= \left( 1 + \frac{\delta D}{D} - 2\gcr\up{2} \frac{\delta D}{D} \right) m_b - 2\gcr\up{2} \frac{\delta D}{D} + 2\gcr\up{2} m_{f} \frac{\delta D}{D} \,.
\end{align}
Note that along with this the fermionic and bosonic operators, bosonic field and the coupling constants are also renormalized. For instance, $\fa' = l\up{-1 + \gcr\up{2}/2 + 3\gamr\up{2}/8} \fa$ and $\ba' = l\up{-1 + \gcr\up{2}} \ba$. In addition to the self-energy corrections there is also a vertex correction to $\gamr$ at this order. However, this does not influence the mass renormalization and thus we can already proceed to calculate the flow equation for the mass.
In our notation introduced earlier, $\kinr \equiv m_{f} - m_{b}$. Now,
\begin{equation}
\label{eq:beta_s}
\bets \equiv -D\frac{\delta \kinr}{\delta D} = -\kinr + 3 \gcr\up{2} \kinr - \gcr\up{2} + \frac{3}{4} \gamr\up{2}  \,.
\end{equation}
We can compute the relevant eigenvalue associated with the flow of $s$ at the fixed points of the beta functions, and find
\begin{equation}
    \lambda_s = 1 + \frac{3\ep-16\rb}{6} \,,
    \label{lambdas}
\end{equation}
at the non-trivial fixed point $FP_{4}$. At the self-consistent values, {\it i.e.\/}, $\ep=1$ and $\rb=1/2$ we have $\lambda_s = 1/6$, although we cannot trust the result at such large values of $\ep$ and $\rb$. Similarly, at $FP_{1}$, $FP_{2}$, and $FP_{3}$ we find $\lambda_s$ to be $1$, $1$, and $1-2\rb$ respectively.

Within the momentum-shell RG, we get the same beta functions for $\gcr$ and $\gamr$, after considering the vertex correction as well.%, as quoted above in Eqs. \ref{eq:beta_g} and \ref{eq:beta_gam}.

%%%%%%%%%%%%%%%%%%%%%%%%%%%%%%%%%%%%%%%%%%%%%%%%%%%%%%%%%%%%%%%%%%%%%%%%%

\subsection{Particle density}
\label{app:particle}

%%%%%%%%%%%%%%%%
\begin{figure}[h]
\centering
\subfloat[]{\includegraphics[width=0.15\textwidth]{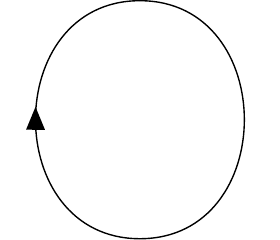}} ~~~~~~
\subfloat[]{\includegraphics[width=0.15\textwidth]{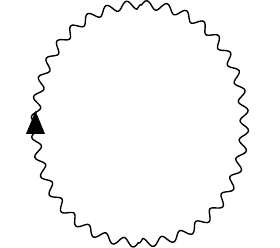}} 
\caption{Lowest order diagrams for particle densities. There are corrections, which are expected to vanish at $T=0$.}
\label{fig:part}
\end{figure}
%%%%%%%%%%%%%%%%

We can also calculate the particle densities ($n_{f/b}$). This can be done at any $\kinr$. We will first make the following identification:
$\kinr = \tilde{\kinr}/\beta$, such that $\tilde{\kinr}$ is small. This will facilitate us to do a small $\tilde{\kinr}$ expansion. From the diagrams we find,
\begin{align}
\charg_f (\lam)&= 2 e\up{-\beta (\lam + \kinr)} \sim 2 e\up{-\beta \lam} (1-\tilde{\kinr}) \,, \\
\charg_b (\lam) &= e\up{-\beta \lam} \,.
\end{align}
$\charg(\lam) = \charg_{f} (\lam) + \charg_{b} (\lam)$, such that,
\begin{equation}
n_{f/b} = \lim_{\lam \rightarrow \infty} \frac{\charg_{f/b}}{\charg} \,.
\end{equation}
So, we have,
\begin{align}
n_{f} &= \frac{2-2\tilde{\kinr}}{3- 2\tilde{\kinr}} \sim \frac{2}{3} (1-\tilde{\kinr})(1+\frac{2}{3}\tilde{\kinr}) 
\sim \frac{2}{3} - \frac{2}{9} \tilde{\kinr} \,, \\
n_{b} &= \frac{1}{3- 2\tilde{\kinr}} \sim \frac{1}{3} (1+\frac{2}{3}\tilde{\kinr}) 
\sim \frac{1}{3} + \frac{2}{9} \tilde{\kinr} \,.
\end{align}
Note that we still satisfy the particle density constraint $n_{f} + n_{b}=1$ exactly. It is also interesting to note that $n_b = 1/3$ at zeroth order, which corresponds to $p_c = 1/3$.

%%%%%%%%%%%%%%%%%%%%%%%%%%%%%%%%%%%%%%%%%%%%%%%%%%%%%%%%%%%%%%%%%%%%%%%%%

\subsection{Two-loop self energy}

%%%%%%%%%%%%%%%%
\begin{figure}[]
\centering
\subfloat[]{\includegraphics[width=0.2\textwidth]{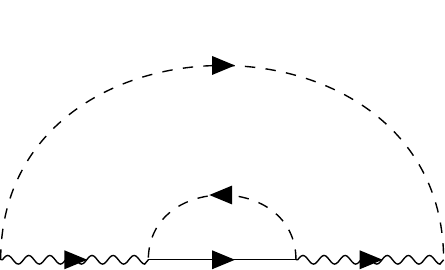}} ~~~~~~
\subfloat[]{\includegraphics[width=0.2\textwidth]{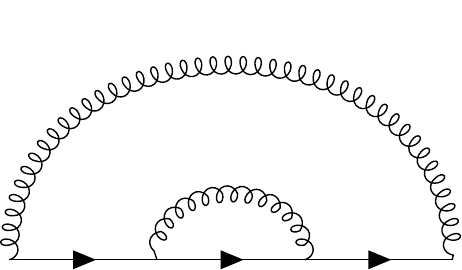}} ~~~~~~
\subfloat[]{\includegraphics[width=0.2\textwidth]{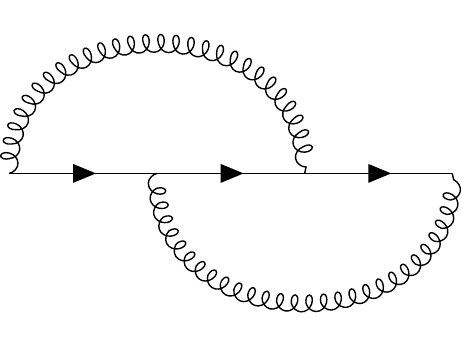}} ~~~~~~
\subfloat[]{\includegraphics[width=0.2\textwidth]{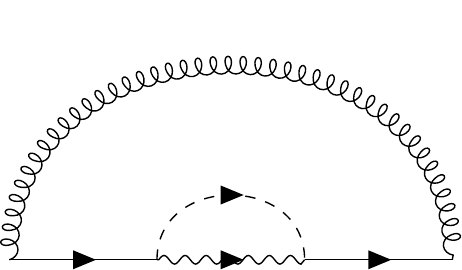}} 
\caption{Two-loop diagrams for fermion self energy.
In these diagrams, solid line is for $f$ propagator,  dashed line is for $\psi$ propagator, wavy is for $b$ propagator, and spiral is for $\phi$ propagator.}
\label{fig:dia2f}
\end{figure}
%%%%%%%%%%%%%%%%
We first evaluate the fermionic self energies to two-loop order. The relevant Feynman diagrams are shown in Fig. \ref{fig:dia2f}. 
\begin{align}
\label{eq:fse2a}
\Sigma\up{f}_{\ref{fig:dia2f} (a)} (i\nu) &= -2 \frac{\gc\up{4}}{\beta\up{2}} \sum_{i\omega_{1n},i\omega_{2n}} \int dk_{1} dk_{2}
\frac{|k_{1}|\up{r}}{i\omega_{1n}-k_{1}} 
\frac{|k_{2}|\up{r}}{i\omega_{2n}-k_{2}} \frac{1}{(i\nu - i\omega_{1n} -\lam)\up{2}} 
\frac{1}{i\nu - i\omega_{1n} + i\omega_{2n} - \lam} 
\nonumber \\
%&\text{Factor of 2 comes due to internal f line} \nonumber \\
&= 2 \gc\up{4} \int dk_{1} dk_{2} 
\frac{|k_{1}|\up{r} |k_{2}|\up{r} \theta(k_{1}) \theta(-k_{2})}{(i\nu - k_{1} -\lam)\up{2} (i\nu - k_{1} + k_{2} - \lam)} 
\nonumber \\
&= 2 \gc\up{4} (i\nu-\lam)\up{-1+2r} \left[\frac{1}{8 \rb^2} - \frac{(2 \pi i - 1)}{4 \rb} \right]
\,, 
\end{align}
\begin{align}
%%%%%%
\label{eq:fse2b}
\Sigma\up{f}_{\ref{fig:dia2f} (b)} (i\nu) &= 
\frac{9}{16} \frac{\gam\up{4}}{\beta\up{2}} \sum_{i\omega_{1n},i\omega_{2n}} \int d\up{d}k_{1} d\up{d}k_{2} 
\frac{1}{\omega_{1n}\up{2} + k_{1}\up{2}} \frac{1}{\omega_{2n}\up{2} + k_{2}\up{2} } 
\frac{1}{(i\nu + i\omega_{1n} - \lam)\up{2}} \frac{1}{i\nu + i\omega_{1n} + i\omega_{2n} - \lam} \nonumber \\
&= \frac{9}{16} \gam\up{4} \int d\up{d}k_{1} \frac{d\up{d}k_{2} }{4 k_{1} k_{2}} 
\frac{1}{(i\nu - k_{1} - \lam)\up{2}} \frac{1}{i\nu - k_{1} - k_{2} - \lam}  \nonumber \\
&=\frac{9}{16} \gam\up{4} (i\nu-\lam)\up{-5+2d} 
\left[ \frac{1}{2 \ep^2}-\frac{-1+\No + 2 i \pi }{2 \ep} \right]
\,,
\end{align}
\begin{align}
%%%%%%
\label{eq:fse2c}
\Sigma\up{f}_{\ref{fig:dia2f} (c)} (i\nu) &= 
\frac{-3}{16} \frac{\gam\up{4}}{\beta\up{2}} \sum_{i\omega_{1n},i\omega_{2n}} \int d\up{d}k_{1} d\up{d}k_{2} 
\frac{1}{\omega_{1n}\up{2} + k_{1}\up{2}} \frac{1}{\omega_{2n}\up{2} + k_{2}\up{2} } 
\frac{1}{i\nu + i\omega_{1n} - \lam} \frac{1}{i\nu + i\omega_{1n} + i\omega_{2n} - \lam} 
\frac{1}{i\nu + i\omega_{2n} - \lam} \nonumber \\
&= \frac{-3}{16} \gam\up{4} \int d\up{d}k_{1} \frac{d\up{d}k_{2} }{4 k_{1} k_{2}} 
\frac{1}{i\nu - k_{1} - \lam} \frac{1}{i\nu - k_{2} - \lam} \frac{1}{i\nu - k_{1} - k_{2} - \lam} \nonumber \\
&=\frac{-3}{16} \gam\up{4} (i\nu-\lam)\up{-5+2d} 
\left[-\frac{1}{\ep^2}-\frac{1- 2\No - 4 i \pi }{2 \ep} \right]
\,, 
\end{align}
\begin{align}
%%%%%%
\label{eq:fse2d}
\Sigma\up{f}_{\ref{fig:dia2f} (d)} (i\nu) &= 
-\frac{3}{4} \frac{\gam\up{2} \gc\up{2}}{\beta\up{2}} \sum_{i\omega_{1n},i\omega_{2n}} \int d\up{d}k_{1} dk_{2} 
|k_{2}|\up{r}
\frac{1}{\omega_{1n}\up{2} + k_{1}\up{2}} 
\frac{1}{(i\nu + i\omega_{1n} - \lam)\up{2}} \frac{1}{i\omega_{2n} - k_{2}} 
\frac{1}{i\nu + i\omega_{1n} - i\omega_{2n} - \lam} \nonumber \\
&= \frac{3}{4}  \gam\up{2} \gc\up{2} \int d\up{d}k_{1} dk_{2} \frac{|k_{2}|\up{r}}{2 k_{1}} \theta(k_{2}) 
\frac{1}{(i\nu - k_{1} - \lam)\up{2}} \frac{1}{i\nu - k_{1} - k_{2} - \lam} \nonumber \\
&= \frac{3}{4}  \gam\up{2} \gc\up{2} (i\nu-\lam)\up{-3+d+r}
\left[ \frac{1}{2 \rb (\ep+ 2\rb)}+\frac{-2 i\pi (\ep+2\rb) - \No \ep + 4\rb}{4 \rb (\ep+2\rb)} \right]
\,.
\end{align}

%%%%%%%%%%%%%%%%
\begin{figure}[]
\centering
\subfloat[]{\includegraphics[width=0.2\textwidth]{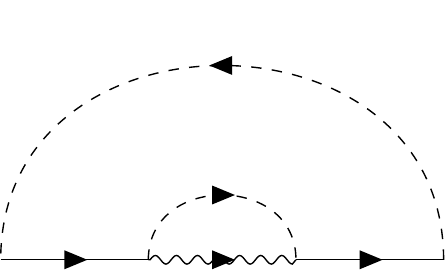}} ~~~~~~
\subfloat[]{\includegraphics[width=0.2\textwidth]{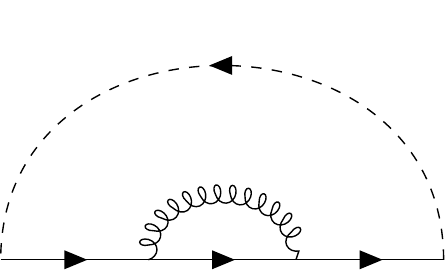}} 
\caption{Two-loop diagrams for boson self energy.
In these diagrams, solid line is for $f$ propagator,  dashed line is for $\psi$ propagator, wavy is for $b$ propagator, and spiral is for $\phi$ propagator.}
\label{fig:dia2b}
\end{figure}
%%%%%%%%%%%%%%%%

We will now evaluate the bosonic self energies to two-loop order. The relevant Feynman diagrams are shown in Fig. \ref{fig:dia2b}.
\begin{align}
\label{eq:bse2a}
\Sigma\up{b}_{\ref{fig:dia2b} (a)} (i\nu) &= -2 \frac{\gc\up{4}}{\beta\up{2}} \sum_{i\omega_{1n},i\omega_{2n}} \int dk_{1} dk_{2}
\frac{|k_{1}|\up{r}}{i\omega_{1n}-k_{1}} 
\frac{|k_{2}|\up{r}}{i\omega_{2n}-k_{2}} \frac{1}{(i\nu + i\omega_{1n} -\lam)\up{2}} 
\frac{1}{i\nu + i\omega_{1n} - i\omega_{2n} - \lam} 
\nonumber \\
%&\text{Factor of 2 due to internal f line} \nonumber \\
&=2 \gc\up{4} \int dk_{1} dk_{2} \frac{\theta(-k_{1}) \theta(k_{2}) |k_{1}|\up{r} |k_{2}|\up{r}}{(i\nu + k_{1} -\lam)\up{2} (i\nu + k_{1} - k_{2} - \lam)} 
\nonumber \\
&= 2 \gc\up{4} (i\nu-\lam)\up{-1+2r}
\left[ \frac{1}{8 \rb^2} - \frac{2i\pi -1}{4 \rb} \right]
\,,
\end{align}
\begin{align}
%%%%%%%%%%%%
\label{eq:bse2b}
\Sigma\up{b}_{\ref{fig:dia2b} (b)} (i\nu) &= 
2 \frac{3}{4} \frac{\gc\up{2} \gamr\up{2}}{\beta\up{2}} \sum_{i\omega_{1n},i\omega_{2n}} \int dk_{1} d\up{d}k_{2}
\frac{|k_{1}|\up{r}}{i\omega_{1n}-k_{1}} 
\frac{1}{\omega\up{2}_{2n} + k_{2}\up{2}} \frac{1}{(i\nu + i\omega_{1n} -\lam)\up{2}} 
\frac{1}{i\nu + i\omega_{1n} + i\omega_{2n} - \lam} 
\nonumber \\
&=2 \frac{3}{4} \gc\up{2} \gamr\up{2} \int dk_{1} d\up{d}k_{2} 
\frac{\theta(-k_{1})}{2k_{2}} \frac{|k_{1}|\up{r}}{(i\nu + k_{1} -\lam)\up{2} (i\nu + k_{1} - k_{2} - \lam)}  \nonumber \\ 
&=2 \frac{3}{4} \gc\up{2} \gamr\up{2} (i\nu-\lam)\up{-3+d+r}
\left[ \frac{1}{\ep (\ep+2\rb)}+\frac{-2 i \pi (\ep+2\rb) -\No \ep + 2\ep }{2 \ep (\ep+2\rb)} \right]
\,.
\end{align}

%%%%%%%%%%%%%%%%%%%%%%%%%%%%%%%%%%%%%%%%%%%%%%%%%%%%%%%%%%%%%%%%%%%%%%%%%

\subsection{Two-loop vertex corrections}

%%%%%%%%%%%%%%%%
\begin{figure}[]
\centering
\subfloat[]{\includegraphics[width=0.25\textwidth]{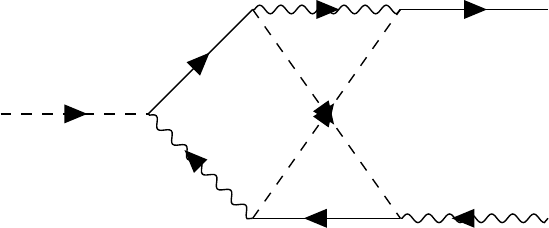}} ~~~~~~
\subfloat[]{\includegraphics[width=0.25\textwidth]{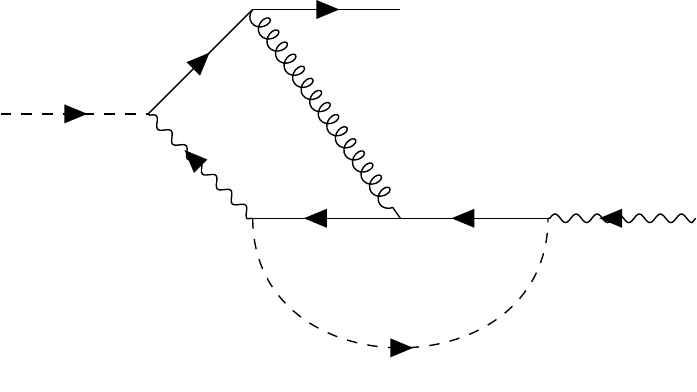}} ~~~~~~
\subfloat[]{\includegraphics[width=0.25\textwidth]{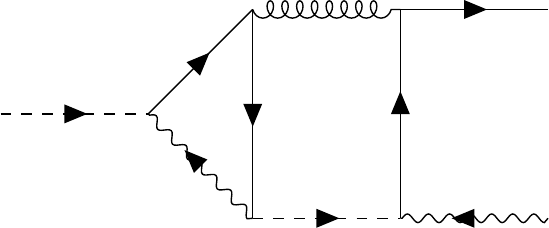}}  
\caption{Two-loop diagrams for vertex corrections to $\gc$.
In these diagrams, solid line is for $f$ propagator,  dashed line is for $\psi$ propagator, wavy is for $b$ propagator, and spiral is for $\phi$ propagator.}
\label{fig:dia2vg}
\end{figure}
%%%%%%%%%%%%%%%%

Let us first evaluate the vertex correction to the fermionic bath coupling $\gc$ at two-loop level. The relevant Feynman diagrams are shown in Fig. \ref{fig:dia2vg},.
\begin{align}
\label{eq:ver2ga}
\Gamma\up{g}_{\ref{fig:dia2vg} (a)} &= \frac{\gc\up{5}}{\beta\up{2}} \sum_{i\omega_{1n},i\omega_{2n}} \int dk_{1} dk_{2}
\frac{|k_{1}|\up{r}}{i\Omega_{1n}-i\omega_{1n} - k_{1}} \frac{|k_{2}|\up{r}}{i\omega_{2n} -i\Omega_{2n} - k_{2}} \nonumber \\
&~~~~~~~~~~~~~~~~~~~~\times 
\frac{1}{i\omega_{1n}-\lam} \frac{1}{i\omega_{2n}-\lam} \frac{1}{i\omega_{2n}-i\Omega_{1n}+i\omega_{1n}-\lam} 
\frac{1}{i\omega_{2n}-i\Omega_{2n}+i\omega_{1n}-\lam} \nonumber \\
&= - \gc\up{5} \int dk_{1} dk_{2} 
\frac{|k_{1}|\up{r} \theta(k_{1})}{i\Omega_{1n}-k_{1}-\lam} \frac{|k_{2}|\up{r} \theta(-k_{2})}{i\Omega_{2n} +k_{2}-\lam} 
\frac{1}{i\Omega_{2n} +k_{2}-k_{1}-\lam} \frac{1}{i\Omega_{1n}+k_{2}-k_{1}-\lam} \nonumber
\end{align}
We set the external frequency of $c$ particle to zero. In other words, $\Omega_{1n}=\Omega_{2n}$. Then
\begin{align}
\Gamma\up{g}_{\ref{fig:dia2vg} (a)} &=- \gc\up{5} \int dk_{2} \frac{|k_{2}|\up{r} \theta(-k_{2})}{i\Omega_{1n} +k_{2}-\lam} 
\bigg[ \frac{\pi  \csc (\pi  r) (l-i \Omega_{1n})^r}{k_{2}^2} 
-\frac{\pi  (\lam-i\Omega_{1n}) \csc (\pi  r) (-k_{2}+\lam-i\Omega_{1n})^{r-1}}{k_{2}^2} \nonumber \\
&~~~~~~~~~~~~~~~~~~~~~~~~~-\frac{\pi  (r-1) \csc (\pi  r) (-k_{2}+\lam-i \Omega_{1n})^{r-1}}{k_{2}^2}
\bigg] \nonumber \\
&=- \gc A_{\mu} \gcr\up{4} 
\left[ \left(\frac{1}{4\rb^2} - \frac{2 i \pi }{2\rb} \right) 
-\left( \frac{1}{4\rb^2} + \frac{\gamma_{E} -2 i \pi +\log (4)+\psi ^{(0)}\left(\frac{1}{2}\right)}{2\rb} \right)
+\frac{1}{4 \rb} \right] \nonumber \\
&=- \gc A_{\mu} \gcr\up{4} \left[ \frac{1}{4 \rb} \right] 
 \,, 
\end{align}
where we used $\gamma_{E} +\log (4)+\psi ^{(0)}\left(\frac{1}{2}\right)=0$.
\begin{align}
%%%%%%%%%%%%%%
\label{eq:ver2gb}
\Gamma\up{g}_{\ref{fig:dia2vg} (b)} &= \gc C_{s1} \frac{\gc\up{2} \gam\up{2}}{\beta\up{2}} \sum_{i\omega_{1n},i\omega_{2n}} \int dk_{1} d\up{d}k_{2} 
\frac{|k_{1}|\up{r}}{i\omega_{1n} - k_{1}} \frac{1}{\omega_{2n}\up{2} + k_{2}\up{2}} \frac{1}{i\Omega_{1n}+i\omega_{1n}-\lam} 
\frac{1}{i\Omega_{1n}+i\omega_{1n}+i\omega_{2n}-\lam}  \nonumber \\
&~~~~~~~~~~~~~~~~~~~~~~~\times\frac{1}{i\Omega_{1n}+i\omega_{2n}-\lam} \frac{1}{i\Omega_{2n}+i\omega_{2n}-\lam} \nonumber \\
&\text{Note that $i\Omega_{1n,2n}$ is external bosonic (fermionic) freq.,
$i\omega_{1n, 2n}$ is internal fermionic (bosonic) freq. } \nonumber \\
&= \gc C_{s1} \gc\up{2} \gam\up{2} \int dk_{1} \frac{d\up{d}k_{2}}{2k_{2}} 
\frac{|k_{1}|\up{r} \theta(-k_{1})}{i\Omega_{1n}+k_{1}-\lam} \frac{1}{i\Omega_{1n}-k_{2}-\lam} \frac{1}{i\Omega_{2n}-k_{2}-\lam} 
\frac{1}{i\Omega_{1n}+k_{1}-k_{2}-\lam} \,, 
\end{align}
\begin{align}
%%%%%%%%%%%%%%
\label{eq:ver2gc}
\Gamma\up{g}_{\ref{fig:dia2vg} (c)} &= \gc C_{s2} \frac{\gc\up{2} \gam\up{2}}{\beta\up{2}} \sum_{i\omega_{1n},i\omega_{2n}} \int dk_{1} d\up{d}k_{2} 
\frac{|k_{1}|\up{r}}{i\omega_{1n} - i\Omega_{1n} - k_{1}} \frac{1}{(\omega_{1n}-\Omega_{2n})\up{2} + k_{2}\up{2}} 
\frac{1}{i\omega_{1n}-\lam} \frac{1}{i\omega_{2n}-i\Omega_{1n}+i\Omega_{2n}-\lam}  \nonumber \\
&~~~~~~~~~~~~~~~~~~~~~~~\times\frac{1}{i\omega_{1n}+i\omega_{2n}-i\Omega_{1n}-\lam} \frac{1}{i\omega_{2n}-\lam} \nonumber \\
&= \gc C_{s2} \frac{\gc\up{2} \gam\up{2}}{\beta} \sum_{i\omega_{1n}} \int dk_{1} d\up{d}k_{2} 
\frac{|k_{1}|\up{r}}{i\omega_{1n} - i\Omega_{1n} - k_{1}} \frac{1}{(\omega_{1n}-\Omega_{2n})\up{2} + k_{2}\up{2}} 
\frac{1}{i\omega_{1n}-\lam} \times 0 = 0 \,.
\nonumber \\
&\text{Note that $i\Omega_{1n,2n}$ is external bosonic (fermionic) freq.,
$i\omega_{1n, 2n}$ is internal fermionic (bosonic) freq. }
\end{align}
However, $C_{s1}=0$. So there is only one non-zero contribution to the $\gc$ vertex correction.

%%%%%%%%%%%%%%%%
\begin{figure}[]
\centering
\subfloat[]{\includegraphics[width=0.25\textwidth]{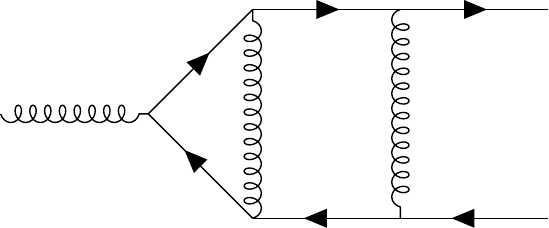}} ~~~~~~
\subfloat[]{\includegraphics[width=0.25\textwidth]{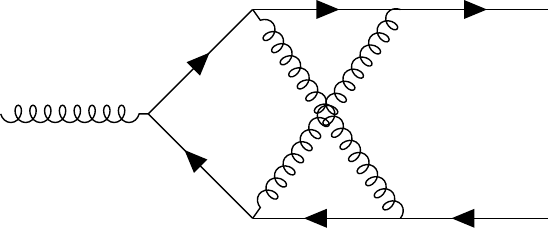}} ~~~~~~
\subfloat[]{\includegraphics[width=0.25\textwidth]{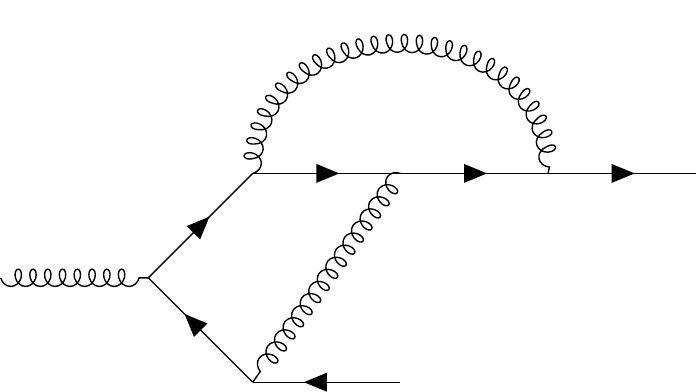}} \\ %~~~~~~
\subfloat[]{\includegraphics[width=0.25\textwidth]{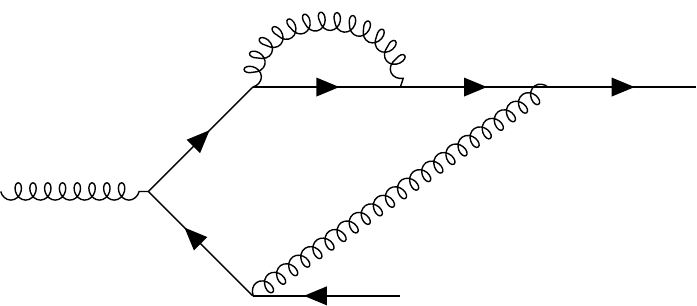}} ~~~~~~
\subfloat[]{\includegraphics[width=0.25\textwidth]{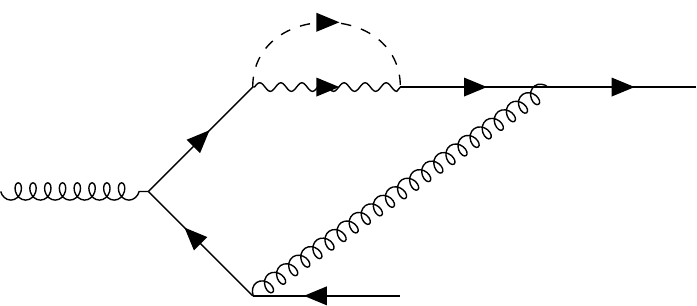}} ~~~~~~
\subfloat[]{\includegraphics[width=0.25\textwidth]{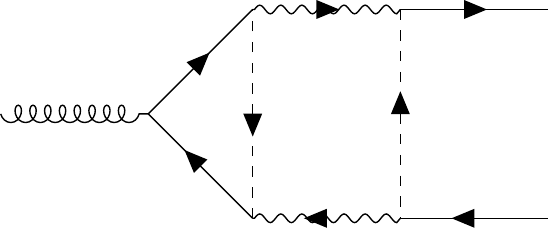}}
\caption{Two-loop diagrams for vertex corrections to $\gam$.
In these diagrams, solid line is for $f$ propagator,  dashed line is for $\psi$ propagator, wavy is for $b$ propagator, and spiral is for $\phi$ propagator.}
\label{fig:dia2vgam}
\end{figure}
%%%%%%%%%%%%%%%%

We will now evaluate the two-loop vertex correction to the bosonic bath coupling $\gam$. The corresponding diagrams are shown in Fig. \ref{fig:dia2vgam}.
\begin{align}
\label{eq:ver2gama}
\Gamma\up{\gamma}_{\ref{fig:dia2vgam} (a)} &= \gam \frac{1}{16} \frac{ \gam\up{4}}{\beta\up{2}} \sum_{i\omega_{1n},i\omega_{2n}} 
\int d\up{d}k_{2} d\up{d}k_{1} 
\frac{1}{\omega_{1n}\up{2} + k_{1}\up{2}} \frac{1}{\omega_{2n}\up{2} + k_{2}\up{2}} \frac{1}{i\Omega_{1n}+i\omega_{1n}-\lam} 
\frac{1}{i\Omega_{1n}+i\omega_{1n}+i\omega_{2n}-\lam}  \nonumber \\
&~~~~~~~~~~~~~~~~~~~~~~~\times\frac{1}{i\Omega_{2n}+i\omega_{1n}+i\omega_{2n}-\lam}  
\frac{1}{i\Omega_{2n}+i\omega_{1n}-\lam} \nonumber \\
&= \gam \frac{1}{16} \gam\up{4} \int \frac{d\up{d}k_{2} d\up{d}k_{1} }{4 k_{1} k_{2}}
\frac{1}{i\Omega_{1n} -k_{1}-k_{2}-\lam} \frac{1}{i\Omega_{2n}-k_{1}-k_{2}-\lam} \frac{1}{i\Omega_{1n}-k_{1}-\lam} 
\frac{1}{i\Omega_{2n}-k_{1}-\lam}  \nonumber \\
&= \gam \frac{1}{16} B_{\mu} \gamr\up{4} \left( \sdd \right)\up{2}
\frac{\pi  (-1)^{2 d-6} (d-2) \csc (\pi  d) \Gamma (6-2 d) \Gamma (d-1)}{\Gamma (5-d)} \nonumber \\
&= \gam \frac{1}{16} B_{\mu} \gamr\up{4} 
\left[ \frac{1}{2 \ep^2} - \frac{3+\No +2 i \pi }{2 \ep}  \right] \,, 
\end{align}
\begin{align}
%%%%%%%%%%%
\label{eq:ver2gamb}
\Gamma\up{\gamma}_{\ref{fig:dia2vgam} (b)} &= \gam \frac{5}{16} \frac{ \gam\up{4}}{\beta\up{2}} \sum_{i\omega_{1n},i\omega_{2n}} 
\int d\up{d}k_{2} d\up{d}k_{1} 
\frac{1}{\omega_{1n}\up{2} + k_{1}\up{2}} \frac{1}{\omega_{2n}\up{2} + k_{2}\up{2}} \frac{1}{i\Omega_{1n}+i\omega_{1n}-\lam} 
\frac{1}{i\Omega_{1n}+i\omega_{1n}+i\omega_{2n}-\lam}  \nonumber \\
&~~~~~~~~~~~~~~~~~~~~~~~\times\frac{1}{i\Omega_{2n}+i\omega_{1n}+i\omega_{2n}-\lam}  
\frac{1}{i\Omega_{2n}+i\omega_{2n}-\lam} \nonumber \\
&= \gam \frac{5}{16} \gam\up{4} \int \frac{d\up{d}k_{2} d\up{d}k_{1} }{4 k_{1} k_{2}}
\frac{1}{i\Omega_{1n} -k_{1}-k_{2}-\lam} \frac{1}{i\Omega_{2n}-k_{1}-k_{2}-\lam} \frac{1}{i\Omega_{1n}-k_{1}-\lam} 
\frac{1}{i\Omega_{2n}-k_{2}-\lam} \nonumber \\
&=\gam \frac{5}{16} B_{\mu} \gamr\up{4} \left( \sdd \right)\up{2} 
\frac{\pi  (-1)^{2 d-6} \csc (\pi  d) ((d-1) \Gamma (6-2 d) \Gamma (d-3)+\pi  \csc (\pi  d) \Gamma (3-d))}{\Gamma (3-d)} 
\nonumber \\ 
&=\gam \frac{5}{16} B_{\mu} \gamr\up{4} 
\left[ \frac{1}{2\ep} \right] \,, %\\
\end{align}
\begin{align}
%%%%%%%%%%%
\label{eq:ver2gamc}
\Gamma\up{\gamma}_{\ref{fig:dia2vgam} (c)} &= \gam 2 \frac{1}{16} \frac{ \gam\up{4}}{\beta\up{2}} \sum_{i\omega_{1n},i\omega_{2n}} 
\int d\up{d}k_{2} d\up{d}k_{1} 
\frac{1}{\omega_{1n}\up{2} + k_{1}\up{2}} \frac{1}{\omega_{2n}\up{2} + k_{2}\up{2}} \frac{1}{i\Omega_{1n}+i\omega_{1n}-\lam} 
\frac{1}{i\Omega_{1n}+i\omega_{1n}+i\omega_{2n}-\lam}  \nonumber \\
&~~~~~~~~~~~~~~~~~~~~~~~\times\frac{1}{i\Omega_{1n}+i\omega_{2n}-\lam}  
\frac{1}{i\Omega_{2n}+i\omega_{2n}-\lam} \nonumber \\
&\text{Factor of 2 accounts for similar graph with internal $\phi$ line on lower edge} \nonumber \\
&= \gam 2 \frac{1}{16} \gam\up{4} \int \frac{d\up{d}k_{2} d\up{d}k_{1} }{4 k_{1} k_{2}}
\frac{1}{i\Omega_{1n} -k_{1}-k_{2}-\lam} \frac{1}{i\Omega_{2n}-k_{2}-\lam} \frac{1}{i\Omega_{1n}-k_{1}-\lam} 
\frac{1}{i\Omega_{1n}-k_{2}-\lam} \nonumber \\
&= \gam 2 \frac{1}{16} B_{\mu} \gamr\up{4} \left( \sdd \right)\up{2} 
\left[ \pi ^2 (-1)^{2 d-6} (d-3) \csc ^2(\pi  d) 
+ \frac{\pi  (-1)^{2 d-6} \csc (\pi  d) \Gamma (6-2 d) \Gamma (d-2)}{\Gamma (4-d)}\right] \nonumber \\
&= \gam 2 \frac{1}{16} B_{\mu} \gamr\up{4}
\left[ -\frac{1}{\ep} 
+ \frac{1}{2 \ep^2} - \frac{\No +2 i \pi }{2 \ep} \right] \,, 
\end{align}
\begin{align}
%%%%%%%%%%%
\label{eq:ver2gamd}
\Gamma\up{\gamma}_{\ref{fig:dia2vgam} (d)} &= \gam 2 \left(\frac{-3}{16}\right) \frac{ \gam\up{4}}{\beta\up{2}} \sum_{i\omega_{1n},i\omega_{2n}} 
\int d\up{d}k_{2} d\up{d}k_{1} 
\frac{1}{\omega_{1n}\up{2} + k_{1}\up{2}} \frac{1}{\omega_{2n}\up{2} + k_{2}\up{2}} 
\frac{1}{(i\Omega_{1n}+i\omega_{2n}-\lam)\up{2}} 
\frac{1}{i\Omega_{1n}+i\omega_{1n}+i\omega_{2n}-\lam}  \nonumber \\
&~~~~~~~~~~~~~~~~~~~~~~~\times\frac{1}{i\Omega_{2n}+i\omega_{2n}-\lam}   \nonumber \\
&\text{Factor of 2 is for similar graph with internal $\phi$ line on lower edge} \nonumber \\
&= \gam 2 \left(\frac{-3}{16}\right) \gam\up{4} \int \frac{d\up{d}k_{2} d\up{d}k_{1} }{4 k_{1} k_{2}}
\frac{1}{i\Omega_{1n} -k_{1}-k_{2}-\lam} \frac{1}{i\Omega_{2n}-k_{2}-\lam} 
\frac{1}{(i\Omega_{1n}-k_{2}-\lam)\up{2}} \nonumber \\
&= \gam 2 \left(\frac{-3}{16}\right) B_{\mu} \gamr\up{4}  \left( \sdd \right)\up{2} 
\left[ \frac{\pi  (-1)^{2 d-5} \csc (\pi  d) \Gamma (6-2 d) \Gamma (d-1)}{\Gamma (5-d)} \right]  \nonumber \\
&= \gam 2 \left(\frac{-3}{16}\right) B_{\mu} \gamr\up{4} 
\left[ -\frac{1}{2 \ep^2} - \frac{-2-\No -2 i \pi }{2 \ep} \right] \,, 
%%%%%%%%%%%
\end{align}
\begin{align}
\label{eq:ver2game}
\Gamma\up{\gamma}_{\ref{fig:dia2vgam} (e)} &= -\gam 2\left(\frac{-1}{4}\right) \frac{\gc\up{2} \gam\up{2}}{\beta\up{2}} \sum_{i\omega_{1n},i\omega_{2n}} 
\int dk_{2} d\up{d}k_{1} 
\frac{|k_{2}|\up{r}}{i\omega_{2n} - k_{2}} \frac{1}{\omega_{1n}\up{2} + k_{1}\up{2}} \frac{1}{i\Omega_{2n}+i\omega_{1n}-\lam} 
\frac{1}{i\Omega_{1n}+i\omega_{1n}-i\omega_{2n}-\lam}  \nonumber \\
&~~~~~~~~~~~~~~~~~~~~~~~\times\frac{1}{(i\Omega_{1n}+i\omega_{1n}-\lam)\up{2}}  \nonumber \\
&\text{Factor of two is for similar graph with boson line on lower edge} \nonumber \\
&= \gam 2\left(\frac{-1}{4}\right) \gc\up{2} \gam\up{2} \int dk_{2} d\up{d}k_{1} \frac{|k_{2}|\up{r} \theta(k_{2})}{2k_{1}} 
\frac{1}{i\Omega_{1n} -k_{1}-k_{2}-\lam} \frac{1}{i\Omega_{2n}-k_{1}-\lam} \frac{1}{(i\Omega_{1n}-k_{1}-\lam)\up{2}} \,, 
\nonumber \\
&= \gam 2\left(\frac{-1}{4}\right) \gcr\up{2} \gamr\up{2} \sdd 
\left[ -\frac{\pi  \Gamma (d-1) \csc (\pi  r) \Gamma (-d-r+4) (l-i \Omega_{1n})^{d+r-4}}{\Gamma (3-r)} \right] \nonumber \\
&= \gam 2\left(\frac{-1}{4}\right) A_{\mu} B_{\mu} \gcr\up{2} \gamr\up{2} 
\left[ 
-\frac{1}{2 \rb (\ep +2 \rb)}+\frac{2 i \pi (\ep+2\rb) + \No \ep + 2 \ep + 4 \rb}{4 \rb (\ep+2 \rb)}
\right] \,, %\\
\end{align}
\begin{align}
%%%%%%%%%%%
\label{eq:ver2gamf}
\Gamma\up{\gamma}_{\ref{fig:dia2vgam} (f)} &= -\gam C_{sf} \frac{\gc\up{4}}{\beta\up{2}} \sum_{i\omega_{1n},i\omega_{2n}} 
\int dk_{1} dk_{2} 
\frac{|k_{1}|\up{r}}{i\Omega_{1n} - i\omega_{1n} - k_{1}} \frac{|k_{2}|\up{r}}{i\omega_{2n} - k_{2}} 
\frac{1}{i\omega_{1n}-\lam} \frac{1}{i\omega_{1n}+i\omega_{2n}-\lam}  \nonumber \\
&~~~~~~~~~~~~~~~~~~~~~~~\times\frac{1}{i\Omega_{2n}-i\Omega_{1n}+i\omega_{1n}+i\omega_{2n}-\lam} 
\frac{1}{i\Omega_{2n}-i\Omega_{1n}+i\omega_{1n}-\lam} 
\nonumber \\
&= \gam C_{sf} \gc\up{4} \int dk_{1} dk_{2} |k_{1}|\up{r} |k_{2}|\up{r} \theta(k_{1}) \theta(-k_{2})  
\frac{1}{i\Omega_{1n} -k_{1}+k_{2}-\lam} \frac{1}{i\Omega_{2n}-k_{1}+k_{2}-\lam} \nonumber \\
&~~~~~~~~~~~~~~~~~~~~~~~~~~~~~~~~~~~\times \frac{1}{i\Omega_{1n}-k_{1}-\lam} 
\frac{1}{i\Omega_{2n}-k_{1}-\lam} \nonumber \\
&= \gam C_{sf} \gcr\up{4} 
\left[ \frac{\pi  r \csc (\pi  r) \Gamma (2-2 r) \Gamma (r+1) (l-i \Omega_{1n})^{2 r-2}}{\Gamma (3-r)} \right] \nonumber \\
&= \gam C_{sf} \gcr\up{4} A_{\mu}\up{2}
\left[ \frac{1}{8 \rb^2}-\frac{3+2 i \pi }{4 \rb} \right] 
\,.
\end{align}
We get $C_{sf}=0$. Using the results in this appendix we obtain the renormalization factors and beta functions 
in Sec. \ref{sec:rg}.
%%%%%%%%%%%%%%%%%%%%%%%%%%%%%%%%%%%%%%%%%%%%%%%%%%%%%%%%%%%%%%%%%%%%%%%%%

\bibliography{dqcp}

\end{document}